\documentclass[twoside,12pt]{article}
\usepackage{lineno}
\usepackage{epsfig}
\usepackage{graphicx}
\usepackage{float}
\usepackage{amsmath,amssymb}

\usepackage{color}
\usepackage{subfloat}
\usepackage{hyperref, url}

\usepackage[normalem]{ulem}

\definecolor{dgreen}{cmyk}{1.,0.,1.,0.2}
\definecolor{orange}{cmyk}{0.,0.353,1.,0.}

\newcommand{\red}[1]{{\color{red}{#1}}}
\newcommand{\mean}[1]{{\langle{#1}\rangle}}

\def \deta {\Delta\eta} 
\def \dphi {\Delta\phi}

\def \CP {${\cal CP}$}
\def \P {${\cal P}$}
\def \pt {p_T}

\def \ask#1 {\red{[#1]} }

\def\NPA{{\em Nucl. Phys.} A}

\newcommand{\be}{\begin{equation}}
\newcommand{\ee}{\end{equation}}
\newcommand{\bea}{\begin{eqnarray}}
\newcommand{\eea}{\end{eqnarray}}

\begin{document}

%\linenumbers
%\modulolinenumbers[2]

\title{ \vspace{1cm} Chiral Magnetic and Vortical Effects \\ in High-Energy Nuclear Collisions \\ --- A Status Report}

\author{D.E.\ Kharzeev,$^{1,2}$ J.\ Liao,$^{3,4}$ S.A.\ Voloshin,$^5$ G.\
Wang,$^{6}$
\\
$^1$Department of Physics and Astronomy, Stony Brook University, \\Stony Brook, New York 11794-3800, USA\\
$^2$Department of Physics and RIKEN-BNL Research Center, \\Brookhaven National Laboratory,  Upton, New York 11973-5000, USA\\
$^3$Physics Department and Center for Exploration of Energy and Matter, \\
Indiana University, 727 E Third Street, Bloomington, IN 47405, USA\\
$^4$RIKEN BNL Research Center, Bldg. 510A,    Brookhaven National Laboratory, \\Upton, NY 11973, USA\\
$^5$Department of Physics and Astronomy, Wayne State University, \\ 666 W. Hancock, Detroit, Michigan 48201\\
$^6$ Department of Physics and Astronomy, University of California, \\ Los Angeles, CA 90095, USA
}

\maketitle

\begin{abstract} 
The interplay of quantum anomalies with magnetic field and vorticity results in a variety of 
novel non-dissipative transport phenomena in systems with chiral fermions, including the quark-gluon plasma. Among them is the Chiral Magnetic Effect (CME) -- the generation of electric current
along an external magnetic field induced by chirality
imbalance.  Because
the chirality imbalance is related to the global topology of gauge fields, the
CME current is topologically protected and hence non-dissipative even in the
presence of strong interactions. As a result, the CME and related quantum
phenomena affect the hydrodynamical and transport behavior of strongly coupled quark-gluon plasma, and can be studied in relativistic heavy ion collisions where strong magnetic fields are created by the colliding ions. Evidence for the CME and related phenomena has been reported by the STAR Collaboration at Relativistic Heavy Ion Collider at BNL, and by the ALICE Collaboration at the Large Hadron Collider at CERN. 
The goal of the present review is to provide an elementary introduction into the physics of anomalous chiral effects, to describe the current status of experimental studies in heavy ion physics, and to outline the future work, both in experiment and theory, needed to eliminate the existing uncertainties in the interpretation of the data. 
\\
\\
\\

%{Color codes: \new{added}, \old{old}, \mod{modified}, \ask{to-do/questions} }

\end{abstract}
%\eject
%\tableofcontents

 %%%  Here we include the various chapters %%%%%%

%%%%%%%%%%%%%%
%COPY BELOW
%%%%%%%%%%%%%%%

\section{Introduction}
\label{sec:1}

Quantum Chromodynamics (QCD) presents a remarkable example of a  theory with known symmetries and well established elementary constituents, but with emergent behavior that remains mysterious to this day.  In spite of forty years of intense effort, it is still not clear how the asymptotic states of QCD perturbation theory -- colored quarks and gluons -- transform into the asymptotic states actually observed in experiment, the color-singlet hadrons. 
Because the confinement of color is not present in perturbation theory, its mechanism has to arise from a non-perturbative dynamics. 

It is widely perceived that such non-perturbative dynamics originates in the topological sector of QCD. The main topic of this review, namely the Chiral Magnetic Effect (CME), is motivated by attempts to find  observable manifestations of the topological structure of the theory. In this review we will provide a simple intuitive introduction into CME and other similar anomalous transport effects, discuss the  phenomenology of these effects for heavy ion collisions, and review the current status of experimental search. For the rest of the Introduction, however, let us first discuss the theoretical foundations and the distinctive features of the CME phenomenon.

A salient feature of QCD ignored in the perturbative approach is the compactness of 
the non-Abelian gauge group. This may well be at the origin of the difficulties of perturbative QCD in describing the ground state of the theory. Indeed, the compact gauge group SU(3) allows for topologically nontrivial configurations of the gluon field. 
The existence of these configurations in the ground state of the theory essentially modifies the vacuum structure  -- 
a superposition of an infinite set of topologically distinct states connected by tunneling instanton  transitions \cite{Belavin:1975fg} becomes the  ``$\theta$-vacuum" of the theory \cite{Callan:1976je,'tHooft:1976fv}.  It is likely that topological effects in QCD are responsible for the  chiral symmetry breaking (see \cite{Schafer:1996wv} for a review) and possibly for confinement (see \cite{Kharzeev:2015xsa} for a recent proposal).

The crucial importance of the compactness of the gauge group for the structure of the ground state can be illustrated by using electrodynamics of superconductors as an example. The U(1) gauge group of electrodynamics with elements $e^{i \varphi}$ can be treated both as a compact (i.e. defined on a circle, with identification $\varphi \to \varphi + 2 \pi n$) or a non-compact (i.e. defined on an infinite line) group.  The Abrikosov vortex in a type II superconductor \cite{Abrikosov:1956sx} corresponds to the  continuous circle onto circle $S_1 \to S_1$ mapping from the azimuthal angle of space onto the phase angle of the electromagnetic order field, and its existence is thus linked to the compactness of U(1). One would not be able to understand the existence of the ground state of a superconductor using any finite order computation in U(1) perturbation theory. 

The example of the Abrikosov vortex is suggestive since it emerges as a crucial ingredient of confinement mechanism proposed for QCD by 't Hooft \cite{'tHooft:1977hy}, Mandelstam \cite{Mandelstam:1974pi}, and Nambu \cite{Nambu:1975ba}. Indeed, if magnetic monopoles existed, a pair of magnetic monopole and anti-monopole inside a superconductor would be connected by the Abrikosov vortex, since magnetic field is expelled from the bulk of the material due to Meissner effect. As the vortex possesses a fixed amount of energy per unit of length, the monopole and anti-monopole would be bound by a linear confining potential. In the dual picture with magnetic and electric charges exchanged, the opposite electric charges become  connected by an confining electric flux tube expelled from the bulk due to the condensation of 
magnetic charges (dual to the condensate of electrically charged Cooper pairs).  

Superconductors also demonstrate the deep link between topology and non-dissipative currents. Since this link is of crucial importance for our discussion, let us elaborate on it by using superconductor as an example. Around the Abrikosov vortex, there exists a supercurrent that screens the magnetic field of the vortex in the bulk. The corresponding physics is captured by the London relation between the electric current and gauge potential ($\nabla \cdot \vec{\bf A} = 0$):
\be\label{london}
\vec{\bf J} = - \lambda^{-2} \vec{\bf A} \,\, . 
\ee
Integrating Eq.(\ref{london}) along the circle around the vortex and using the Stokes theorem, we get
\be\label{line}
\int_{\partial S} \vec{\bf J} = -\lambda^{-2}  \int_{\partial S} \vec{\bf A} 
= - \lambda^{-2}  \int_{S} B = -\lambda^{-2}  \Phi_B  \,\, ,
\ee
where $\Phi_B$ is the magnetic flux; 
from Eq.(\ref{line}) one can deduce the existence of the supercurrent around the vortices filled by magnetic flux. Due to the compactness of U(1) group, the magnetic flux through an Abrikosov vortex is quantized. Therefore, the amount of current circulating around the vortex is determined by its magnetic flux which in turn is fixed by topology of the $S_1 \to S_1$ mapping. This means that the circulating supercurrent is topologically protected -- it is not allowed to dissipate, or else the (conserved) $S_1 \to S_1$ winding number would have to change.

In the absence of electric charges, $\vec{\bf E} = - \dot{\vec{\bf A}}$ and so the London relation Eq.(\ref{london}) yields 
\be
\vec{\bf E} = \lambda^2 \dot{\vec{\bf J}} \,\, . 
\ee
This means that a constant electric field induces an electric current growing with time, so the charges accelerate as if they met no obstacles, following the Newton's law $e \vec{\bf E} = m \dot{\vec{\bf v}}$ -- in other words, we get a superconducting current. Note that the motion of dual confining strings can be described by using the London theory as well \cite{Nambu:1974zg}, suggesting connection between the mechanism of confinement and the non-dissipative currents. 

It is instructive to consider the relation Eq.(\ref{london}) as an analog of Ohm's law $\vec{\bf J} =  \sigma \vec{\bf E}$ for superconductors. Let us first examine the discrete symmetries of quantities on both sides of the Ohm's law. Both electric current and electric field are even under parity transformation, so the ohmic conductivity $\sigma$ has to be $\cal P$-even as well. Under time reversal, the electric current is $\cal T$-odd -- if we film the propagation of current and then watch the film backwards in time, the current would flow in the opposite direction. The electric field however is $\cal T$-even (due to the time derivative in its definition), and so the Ohmic conductivity has to be odd under $\cal T$-reversal for the Ohm's law to make sense. This is natural since the ohmic conductivity describes processes of dissipation that produce entropy, and entropy production by the second law of thermodynamics is an irreversible process -- it generates an arrow of time. Because of this, almost all of the known transport coefficients in fact are odd under $\cal T$-reversal. 

Let us however examine the discrete symmetries of the quantities entering the London relation Eq.(\ref{london}) that replaces the Ohm's law for superconductors. Both $\vec{\bf J}$ and $\vec{\bf A}$ are odd under $\cal T$-reversal, so the coefficient relating them has to be $\cal T$-even, and no current dissipation is allowed!  Note that in a superconductor this has been achieved due to the spontaneous breaking of gauge invariance (or, to be more precise, due to the loss of invariance with respect to the U(1) rotations in the ground state). The non-dissipative chiral magnetic currents discussed in this review, as we shall see, are also topologically protected, but do not require the loss of U(1) invariance. They appear in every system possessing chiral fermions in the presence of chirality imbalance.   

How can a chirality imbalance arise in quark-gluon matter? The compact nature of SU(3) gauge group implies the existence of topological solutions, similar to the case of Abrikosov vortices arising from the compact U(1) gauge group. If these solutions are chiral, they can transfer chirality to quarks through the chiral anomaly \cite{Adler:1969gk,Bell:1969ts}, creating an imbalance between the numbers of left- and right-handed fermions. In fact, the Atiyah-Singer index theorem \cite{Atiyah:1984tf} implies that the extended topological configurations of gluon fields support the chiral zero modes of fermions. Since the fermions of QCD -- the quarks -- also possess electric charges, their topology can be effectively 
probed by a background Abelian magnetic field. The topology of zero modes of chiral fermions in a magnetic field leads to a number of surprising novel phenomena some of which are reviewed below; see the volume \cite{Kharzeev:2013jha} and Refs. \cite{Kharzeev:2013ffa,Liao:2014ava,Kharzeev:2015kna,Shovkovy:2012zn,Andersen:2014xxa,Rebhan:2014rxa,Shuryak:2014zxa} for complementary reviews of topics not covered here in detail. 

A particularly interesting phenomenon stemming from the interplay of chirality, magnetic field and the chiral anomaly is the Chiral Magnetic Effect (CME). 
This term \cite{Kharzeev:2007jp,Fukushima:2008xe} refers to 
 the generation of electric current induced by the 
chirality imbalance in the presence of an external magnetic field:
\be\label{cme_current}
\vec{\bf J} = \sigma_5\ \vec{\bf B}  \,\, ,
\ee
where $\sigma_5 = e^2/(2 \pi^2)\ \mu_5$ is the chiral magnetic conductivity expressed in terms of the chiral chemical potential $\mu_5$. For QCD with $N_c$ colors and $N_f$ dynamical quarks of charges $Q_f e$, one has to sum over the quark colors and flavors:  $\sigma_5 = N_c \sum_f Q_f^2 e^2 / (2 \pi^2)\ \mu_5$. In heavy ion collisions, the CME leads to the event-by-event fluctuations of electric dipole moment of the quark-gluon plasma \cite{Kharzeev:2004ey,Kharzeev:2007tn} -- the effect that will be the main focus of our discussion in this review. 

Similarly to superconductivity discussed above, the CME is a macroscopic quantum effect - it is a manifestation of the chiral anomaly creating a collective motion in the Dirac sea. Analogously to the London relation, the quantities ${\vec {\bf J}}$ and ${\vec {\bf B}}$ on both sides of Eq.(\ref{cme_current}) are $\cal T$-odd, so the chiral magnetic conductivity is $\cal T$-even, and no dissipation is allowed \cite{Kharzeev:2011ds}. The same conclusion can also be reached \cite{Kharzeev:2013ffa} by using the Onsager relations of non-equilibrium statistical mechanics. However, unlike superconductivity, the CME does not require a spontaneous symmetry breaking, or formation of a condensate -- it is driven solely by the chirality imbalance. 
Because the chirality imbalance is related to the global topology of gauge fields, the CME current is topologically protected \cite{Kharzeev:2009fn} and thus 
{\it non-dissipative} even in the presence of strong interactions. As a result, the CME and related quantum phenomena affect the hydrodynamical 
and transport behavior of systems possessing chiral fermions, from the quark-gluon plasma to Dirac semimetals.

The persistence of CME at strong coupling and small frequencies makes the hydrodynamical description of the effect possible, and indeed it arises naturally within the relativistic hydrodynamics as shown by Son and Surowka \cite{Son:2009tf}.  The quantum anomalies in general have been found to modify hydrodynamics in a significant way, see \cite{Erdmenger:2008rm,Sadofyev:2010pr,Kalaydzhyan:2011vx,Kharzeev:2011ds,Lublinsky:2009wr,Landsteiner:2011cp,Neiman:2010zi,Jensen:2012kj,Basar:2013qia}, and
\cite{Zakharov:2012vv} for a review.  The principle of ``no entropy production from ${\cal P}$-odd and ${\cal T}$-even anomalous terms" \cite{Kharzeev:2011ds} can be used to constrain the relativistic conformal hydrodynamics at second order  in the derivative expansion, where it allows to compute analytically 13 out of 18 anomalous transport coefficients. 

Anomalous hydrodynamics has been found to possess a novel gapless collective excitation -- the ``chiral magnetic wave" \cite{Kharzeev:2010gd}, see also \cite{Newman:2005hd}. It is analogous to sound, but in strong magnetic field propagates along the direction of the field with its wave speed  reaching the speed of light \cite{Kharzeev:2010gd}. The chiral magnetic wave is the hydrodynamical mechanism of transporting the CME current; it transforms an initial chiral or electric charge fluctuation into a macroscopic observable asymmetry in the distribution of electric charge \cite{CMW_calc_Burnier,Hirono:2014oda}. 

The anomaly-induced effects away from equilibrium can be described by using the chiral kinetic theory \cite{Stephanov:2012ki,Son:2012wh,Chen:2014cla}. In particular, the chiral kinetic theory can be used to derive the chiral magnetic wave \cite{Stephanov:2014dma}. In the presence of chirality-flipping transitions, the chiral magnetic wave at frequencies smaller than the transition rate has been found to give rise to a diffusive vector mode  \cite{Stephanov:2014dma}. The chiral kinetic description of transport in systems with finite chemical potential has been recently studied in \cite{Monteiro:2015mea}.

Since topological fluctuations 
are accompanied by the changes in net chirality, QCD matter
subjected to an external magnetic field will develop fluctuations of the electric dipole moment \cite{Kharzeev:2007jp,Fukushima:2008xe,Kharzeev:2004ey,Kharzeev:2007tn}. A clean and explicitly soluble example is the electric dipole moment of the QCD instanton in a strong magnetic field \cite{Basar:2011by}. Similar effect can be induced by rotation, due to the coupling of vorticity to the spin. The corresponding ``chiral vortical effect" can be disentangled from the CME due to the difference in the coupling of quarks to magnetic field and vorticity \cite{Kharzeev:2010gr}. 
The physics of CME and related phenomena has been recently reviewed in \cite{Kharzeev:2013ffa,Liao:2014ava,Kharzeev:2015kna} which contain an extensive set of references. 

After the recent discovery of Dirac and Weyl semimetals, the studies of CME in condensed matter has begun. The experimental observation of CME in Dirac semimetal ${\rm ZrTe_5}$ has been reported  in \cite{Li:2014bha}. Very recently, the effect has been also observed in ${\rm Na_3Bi}$ \cite{Xiong2015}, ${\rm TaAs}$ \cite{Huang2015}, and ${\rm TaP}$ \cite{Shekhar2015}.

%%%%%%%%%%%%
Under extreme conditions of high temperature and/or high baryon density, the vacuum of QCD changes its properties, and deconfinement and chiral symmetry restoration take place. This deconfinement transition is accompanied by the rapid change in the rate and nature of topological transitions connecting different topological sectors. 
The heavy ion program opens a possibility to study these phenomena.
Moreover, since the colliding ions create strong magnetic fields $eB \sim {\cal O}(10\ m_{\pi}^2)$ \cite{Kharzeev:2007jp} (see \cite{Tuchin:2013ie} for review), the interplay of QCD topology with an Abelian magnetic background can be studied experimentally.  %%%%%
%\vskip0.3cm

In this review  we will focus our attention on the discussion of recent CME-related developments in  heavy ion physics, and the plans for the future experiments aimed at establishing (or falsifying) the presence of CME in heavy ion data. 
The paper is organized as follows. In section 2 we give an elementary introduction into the physics of anomaly-induced transport in systems with chiral fermions, using the quark-gluon plasma as an example. In section 3 we discuss the theoretical predictions of these anomalous chiral effects in relativistic heavy ion collisions. In section 4 we present the current status of the experimental studies of anomalous chiral effects at Relativistic Heavy Ion Collider (RHIC) at BNL and the Large Hadron Collider (LHC) at CERN. We describe the experimental observables proposed for detecting the signal and the experimental evidences for the anomalous chiral effects, as well as the backgrounds and uncertainties. We also outline the future measurements that will allow to reach a definite conclusion on the occurrence of anomalous chiral effects in the quark-gluon plasma produced in heavy ion collisions.

\section{Anomalous Chiral Effects for Pedestrians}
\label{sec:2}

The purpose of this Section is mainly pedagogical, aiming to provide
a simple and intuitive way of understanding the physics of anomalous chiral effects. 
For a more detailed and technical discussion, we refer the reader to the previous reviews \cite{Kharzeev:2013ffa,Liao:2014ava,Kharzeev:2015kna,Flow_CME,Miransky:2015ava,Huang:2015oca}. 

To set the stage of subsequent discussions, let us consider the
quark-gluon plasma (QGP) with restored chiral symmetry for light
quarks. For each specific flavor of these chiral fermions, one can
introduce the corresponding vector current ${J}^\mu$ and axial current
${ J}^\mu_5$:
\begin{eqnarray} \label{eq_currents}
{ J}^\mu = \langle \bar \Psi \gamma^\mu \Psi \rangle \,\,\, , \,\,\,
{J}_5^\mu=\langle \bar \Psi \gamma^\mu \gamma^5 \Psi \rangle.
\end{eqnarray} 
By virtue of the chiral symmetry, one can also separately introduce
the right-handed (RH) and left-handed (LH) fermions, and relate them
to the above currents via ${J}_{R/L}^\mu= ({J}^\mu \pm {J}^\mu_5)/2$.
Equivalently one has ${J}^\mu = {J}^\mu_R+{J}^\mu_L$ and ${J}^\mu_5 =
{J}^\mu_R - {J}^\mu_L$.  The thermodynamic states of such QGP can be
specified by, in addition to the temperature $T$, the vector chemical
potential $\mu$ (pertinent to the vector number density $J^0$) and the
axial chemical potential $\mu_5$ (pertinent to the axial number
density $J^0_5$). The quantity $\mu_5$ characterizes the imbalance of
RH and LH fermions in the system, and the QGP with nonzero $\mu_5$ is
a {\it chiral medium}. Such a chiral QGP may be created locally in
heavy-ion collisions through a variety of mechanisms (e.g. topological
fluctuations in the gluonic sector, glasma flux tubes, or simply
fluctuations in the quark
sector~\cite{Kharzeev:2007jp,Kharzeev:2004ey,Kharzeev:2007tn,Kharzeev:2001ev,Iatrakis:2014dka,Fukushima:2010vw})
on an event-by-event basis.  Again one may also introduce
correspondingly the RH and LH chemical potentials $\mu_{R,L}=\mu \pm
\mu_5$.

For simplicity we will discuss the anomalous chiral effects with the
single-fermion-species example in this Section. The generalization to
the multi-flavor and multi-color case would be straightforward.
Experimental measurements often concern  the
electromagnetic charge or baryonic charge rather than the quark-level
currents. The conserved charge currents can be constructed from those
of quarks by summing over relevant flavors and colors, e.g.
\begin{eqnarray} \label{eq_Q_B}
{J}^\mu_Q = N_c \sum_f e Q_f  {J}^\mu_f \,\,\, , \,\,\, {J}^\mu_B = N_c
\sum_f B_f {J}^\mu_f,
\end{eqnarray}
where $Q_f$ and $B_f$ are the electric and baryonic charges of a given
flavor, respectively, e.g. for $(u,d,s)$ flavors,
$Q_f=(2/3,-1/3,-1/3)$ and $B_f=(1/3,1/3,1/3)$.

\subsection{The Chiral Magnetic Effect}

A powerful way of probing properties of matter, known since the time
of Ohm and still widely used today, is to apply external
electromagnetic fields and examine the responses of matter. For
example, in an electrically conducting medium, an electric current can
be generated in the presence of an external electric field
\begin{eqnarray} \label{eq_Ohm}
\vec{\mathbf J} = \sigma \vec{\mathbf E},
\end{eqnarray} 
which is the famous Ohm's law, with $\sigma$ being the electric
conductivity characterizing the vector charge transport property of
matter. (Note we have ``hidden'' an electric charge $Qe$ factor on
both sides in the above.) Obviously the QGP with electrically charged
quarks roaming around is a conductor.

There are however more interesting questions one may ask regarding the
QGP transport. What would happen if one uses an external magnetic
field $\vec{\mathbf B}$ as a probe, instead? Can a vector current be
generated similarly to that in Eq.~(\ref{eq_Ohm})? Normally this is
{\it forbidden} by the symmetry argument: $\vec{\mathbf J}$ is a
${\cal P}$-odd vector quantity while $\vec{\mathbf B}$ is a ${\cal
  P}$-even axial vector quantity. But the situation is different if
the underlying medium itself is {\it chiral}, such as a
chiral QGP with {\it nonzero $\mu_5$} whose parity ``mirror image''
has an opposite $\mu_5$. 
As already discussed in the Introduction section, in such case the Chiral Magnetic Effect (CME)   ~\cite{Kharzeev:2007jp,Kharzeev:2004ey,Kharzeev:2007tn} predicts the generation of a vector current $\vec{\mathbf J} = \sigma_5    \vec{\mathbf B}$ in response to the $\vec{\mathbf B}$ field, as given in Eq.(\ref{cme_current}).  The
$\sigma_5=\frac{Q e }{2\pi^2} \mu_5$ is a chiral magnetic conductivity. (Again if
one wants to specifically consider the electric current, then
$\vec{\mathbf J} \to Qe \vec{\mathbf J} = (Qe)^2/(2\pi^2) \mu_5
\vec{\mathbf B}$.) 

The generation of a vector current in the presence of chirality imbalance was 
first discussed by Vilenkin \cite{Vilenkin:1979ui}. However chirality imbalance itself is a necessary but not sufficient ingredient of the CME -- the corresponding current does not vanish only when the chiral charge is not conserved, i.e. in the presence of chiral anomaly. The detailed discussion of this issue, and additional references to earlier work, can be found in \cite{Kharzeev:2013ffa}. Because both the chirality imbalance and the chiral anomaly are involved in the CME and related phenomena, we will refer to them  as ``anomalous chiral effects" in this review.

\begin{figure}[!hbt]
\begin{center}
\includegraphics[width=0.9\textwidth]{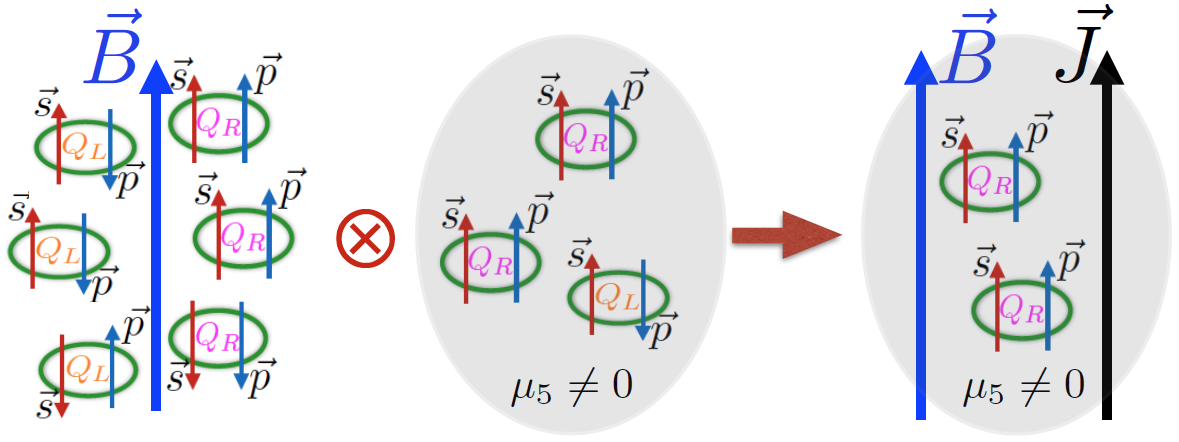}
\end{center}
\caption{(Color online) Illustration of the Chiral Magnetic Effect. To
  be specific, the illustration is for just one kind of massless
  quarks with positive electric charge $Q>0$ and for the case of
  $\mu_5>0$. For quarks with negative electric charge the quark
  current $\vec{\mathbf J}$ is generated in the opposite direction
  (owing to the opposite spin polarization) but their contribution to
  the electric current would be the same as that from positively
  charged quarks. For $\mu_5<0$ the current will flip direction.}
\label{fig_CME}
\end{figure}

Intuitively the CME may be understood in the following way, as
illustrated in Fig.~\ref{fig_CME}.  The magnetic field leads to a spin
polarization (i.e. ``magnetization'') effect, with quarks' spins
preferably aligned along the $\vec{\mathbf B}$ field direction, which
implies $\langle \vec s \rangle \propto (Qe) \vec{\mathbf B}$. Quarks
with specific chirality have their momentum $\vec p$ direction
correlated with spin $\vec s$ orientation: $\vec p \, || \vec s$ for RH
quarks, while $\vec p \, || (-\vec s)$ for LH ones. In the presence of
chirality imbalance, i.e. $\mu_5\neq 0$, there will be a net
correlation between average spin and momentum $\langle \vec p \rangle
\propto \mu_5 \langle \vec s \rangle$. For example, if $\mu_5>0$ there
are more RH quarks, and the momentum is preferably in parallel to
spin. It is therefore evident that $\langle \vec p \rangle \propto (Qe)
\mu_5 \vec{\mathbf B}$, which implies a vector current of these quarks
$\vec{\mathbf J} \propto \langle \vec p \rangle \propto (Qe) \mu_5
\vec{\mathbf B}$.

Of course, the precise coefficient of the chiral magnetic conductivity
$\sigma_5$ has to be determined dynamically. Remarkably, computations
in various systems ranging from free gas to infinitely strongly
coupled field theories, have inevitably found the same universal value
independent of dynamical details (see e.g. the reviews in \cite{Kharzeev:2013jha} and further references therein). This points to a certain deep origin
of the CME, and indeed this coefficient is entirely dictated by the
{\em chiral anomaly}. A most elaborative way to manifest this profound
connection is perhaps through the following derivation (see
e.g.~\cite{Fukushima:2008xe}). Let us assume a CME-induced electric
current $(Qe)\vec {\mathbf J} = (Qe) \sigma_5  \vec{\mathbf
  B}$. To probe the existence of such a current we turn on an
arbitrarily small auxiliary electric field $\vec{\mathbf E}\, || \,
\vec{\mathbf B}$ and examine the energy changing rate of the
system. The straightforward electrodynamic way of computation
``counts'' the work per unit time (i.e. power) done by such an
electric field $P=\int_{\vec{\mathbf x}} \vec{\mathbf J}\cdot
\vec{\mathbf E} = \int_{\vec{\mathbf x}} [(Qe) \sigma_5 
]\vec{\mathbf E}\cdot \vec{\mathbf B}$.  Alternatively for this system
of chiral fermions, the (electromagnetic) chiral anomaly suggests the
generation of axial charges at the rate $dQ_5/dt=\int_{\vec{\mathbf
    x}} C_A \vec{\mathbf E}\cdot \vec{\mathbf B}$ with
$C_A=(Qe)^2/(2\pi^2)$ the universal anomaly coefficient. Now a nonzero
axial chemical potential $\mu_5\neq 0$ implies an energy cost for
creating each unit of axial charge, thus the energy changing rate via
anomaly counting would give the power $P=\mu_5 (dQ_5/dt) =
\int_{\vec{\mathbf x}} [C_A \mu_5] \vec{\mathbf E}\cdot \vec{\mathbf
  B}$. These reasonings therefore lead to the following
identification:
\begin{eqnarray}
 \int_{\vec{\mathbf x}} [(Qe) \sigma_5  ]\vec{\mathbf E}\cdot
 \vec{\mathbf B} = \int_{\vec{\mathbf x}} [C_A \mu_5] \vec{\mathbf
   E}\cdot \vec{\mathbf B}
\end{eqnarray}
 for any auxiliary $\vec{\mathbf E}$ field. Thus the $\sigma_5$ must
 take the universal value $\frac{C_A \mu_5}{Qe}=\frac{Q e }{2\pi^2} \mu_5$ that is completely fixed by the chiral anomaly.

The transport phenomenon in Eq.~(\ref{cme_current}) bears a distinctive
feature that is intrinsically different from Eq.~(\ref{eq_Ohm}). The
chiral magnetic conductivity $\sigma_5$ is a ${\cal T}$-even transport
coefficient while the usual conductivity $\sigma$ is ${\cal
  T}$-odd~\cite{Kharzeev:2011ds}. That is, the CME current can be
generated as an equilibrium current without producing entropy, while
the usual conducting current is necessarily dissipative.

\subsection{The Chiral Separation Effect}

By reminding ourselves of the axial counterpart in
Eq.~(\ref{eq_currents}) of the vector current, which we have discussed
so far, it may be natural to ask: could axial current also be
generated under certain circumstances in response to external probe
fields? The answer is positive. A complementary transport phenomenon
to the CME has been found and named the Chiral Separation Effect
(CSE)~\cite{son:2004tq,Metlitski:2005pr}:
 \begin{eqnarray} \label{eq_CSE}
\vec{\mathbf J}_5 = \sigma_{s}   \vec{\mathbf B}   \,\, .
\end{eqnarray}
It states that an axial current is generated along an external
$\vec{\mathbf B}$ field, with its magnitude in proportion to the
system's (nonzero) vector chemical potential $\mu$ as well as the
field magnitude. The coefficient (which may be called the CSE conductivity)
is given by $\sigma_s=\frac{Qe}{2\pi^2}\mu$.

\begin{figure}[!hbt]
\begin{center}
\includegraphics[width=0.9\textwidth]{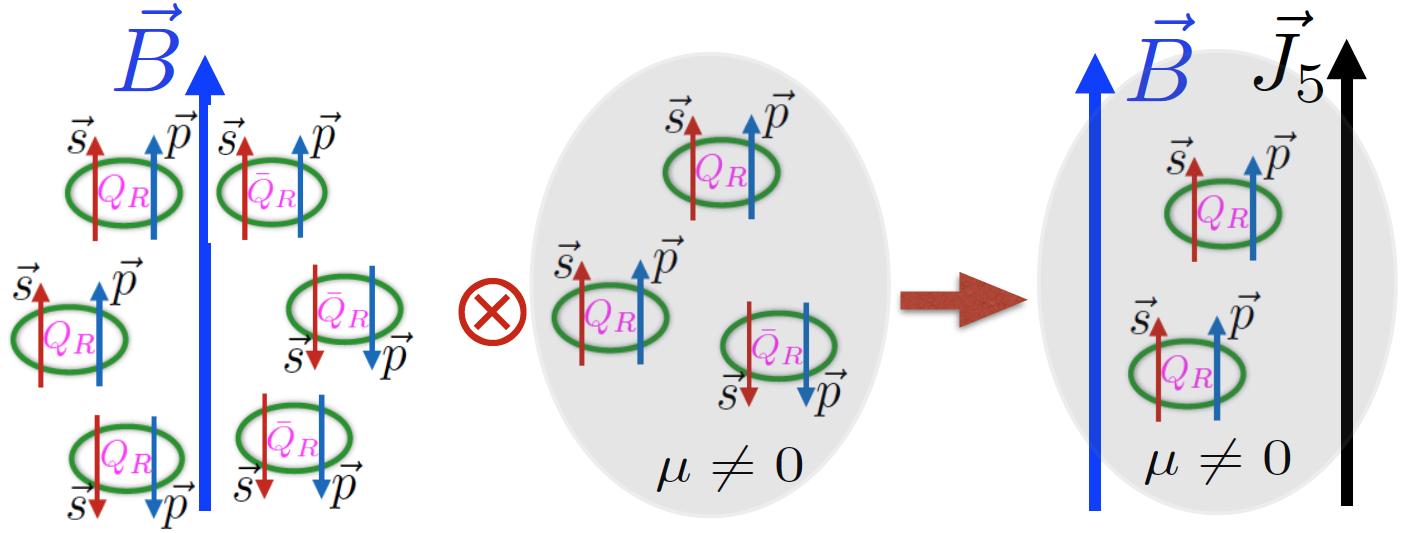}
\end{center}
\caption{(Color online) Illustration of the Chiral Separation
  Effect. To be specific, the illustration is for just one kind of
  right-handed (RH) quarks(with $Q>0$) and their antiquarks (with
  $Q<0$) and for the case of $\mu>0$ (i.e. more quarks than
  antiquarks). For left-handed (LH) quarks (and anti-quarks) the LH
  quarks' current is generated in the opposite direction but their
  contribution to the axial current $\vec{\mathbf J}_5$ would be the
  same as that of RH quarks. For $\mu<0$ the current will flip
  direction.}
\label{fig_CSE}
\end{figure}

Intuitively the CSE may be understood in the following way, as
illustrated in Fig.~\ref{fig_CSE}.  The magnetic field leads to a spin
polarization (i.e. ``magnetization'') effect, with $\langle \vec s
\rangle \propto (Qe) \vec{\mathbf B}$. This effect implies that the
positively charged quarks have their spins preferably aligned along
the $\vec{\mathbf B}$ field direction, while the negatively charged
anti-quarks have their spins oppositely aligned. Now RH quarks and
antiquarks (with $\vec p || \vec s$) will have opposite average
momentum $\langle \vec p \rangle \propto \langle \vec s \rangle
\propto (Qe) \vec{\mathbf B}$, i.e. with more RH quarks/antiquarks
moving in the direction parallel/antiparallel to $\vec{\mathbf B}$. Furthermore with
nonzero $\mu\neq 0$ (e.g. considering $\mu>0$) there would then be a
net current of RH quarks/antiquarks $\vec{\mathbf J}_R \propto \langle
\vec p \rangle (n_Q-n_{\bar Q}) \propto (Qe) \mu \vec{\mathbf B}$. The
LH quarks/antiquarks would form an opposite current $\vec{\mathbf J}_L
\propto -(Qe) \mu \vec{\mathbf B}$ but contribute the same as the RH
quarks/antiquarks to form together an axial current along the magnetic
field: $\vec{\mathbf J}_5 \propto (Qe) \mu \vec{\mathbf B}$.

It is instructive to recast (\ref{cme_current}) and (\ref{eq_CSE}) in terms
of the RH and LH currents $\vec{\mathbf J}_{R/L}$, as follows:
\begin{eqnarray} \label{eq_CME_LR}
\vec{\mathbf J}_{R/L} = \frac{\vec{\mathbf J} \pm \vec{\mathbf J}_{5}}{2} =  \pm \sigma_{R/L}   \vec{\mathbf B}   \,\, .
\end{eqnarray}
with $\sigma_{R/L}=\frac{Qe}{4\pi^2}\mu_{R/L}$. The above has the simple interoperation as the CME separately for the
purely right-handed and purely left-handed Weyl fermions: note the
sign difference in the RH/LH cases. It reveals that the CME and the
CSE are two sides of the same coin, which is why their conductivities 
 are both determined from chiral anomaly coefficient. It though should be emphasized that possible radiative corrections to the CME and CSE coefficients could occur when the pertinent 
 gauge fields are treated dynamically (see e.g. recent discussions in \cite{Jensen:2013vta,Gorbar:2013upa}).

\subsection{The Chiral Electric Separation Effect}

Now in view of the transport effects in
Eqs.~(\ref{cme_current},\ref{eq_Ohm},\ref{eq_CSE}) that we have discussed
thus far, one may realize a possibly missing phenomenon: can axial
current be generated in response to an external probe of electric
field? This question has been answered
recently~\cite{Huang:2013iia,Jiang:2014ura} and a new effect, called
the Chiral Electric Separation Effect (CESE), has been found:
\begin{eqnarray} \label{eq_CESE}
\vec{\mathbf J}_5 = \sigma_{\chi e}   \vec{\mathbf E}   \,\, .
\end{eqnarray}
Again, normally this is {\it forbidden} by the symmetry argument:
$\vec{\mathbf J}_5$ is a ${\cal P}$-even axial vector quantity while
the $\vec{\mathbf E}$ is a ${\cal P}$-odd vector quantity. The above
CESE is an anomalous transport process that becomes possible only in a
chirally imbalanced environment with $\mu_5\neq 0$. However, different
from the CME, the CESE does not originate from the chiral anomaly but
rather is connected to the usual conducting phenomenon in the electric
field. Its coefficient $\sigma_{\chi e} = \chi_e (Qe)\mu \mu_5$, called CESE conductivity,
does depend on the specific dynamical system, and has been computed
for weakly coupled QED and QCD
plasma~\cite{Huang:2013iia,Jiang:2014ura} as well as for a certain
strongly coupled holographic system~\cite{Pu:2014cwa}.

\begin{figure}[!hbt]
\begin{center}
\includegraphics[width=0.8\textwidth]{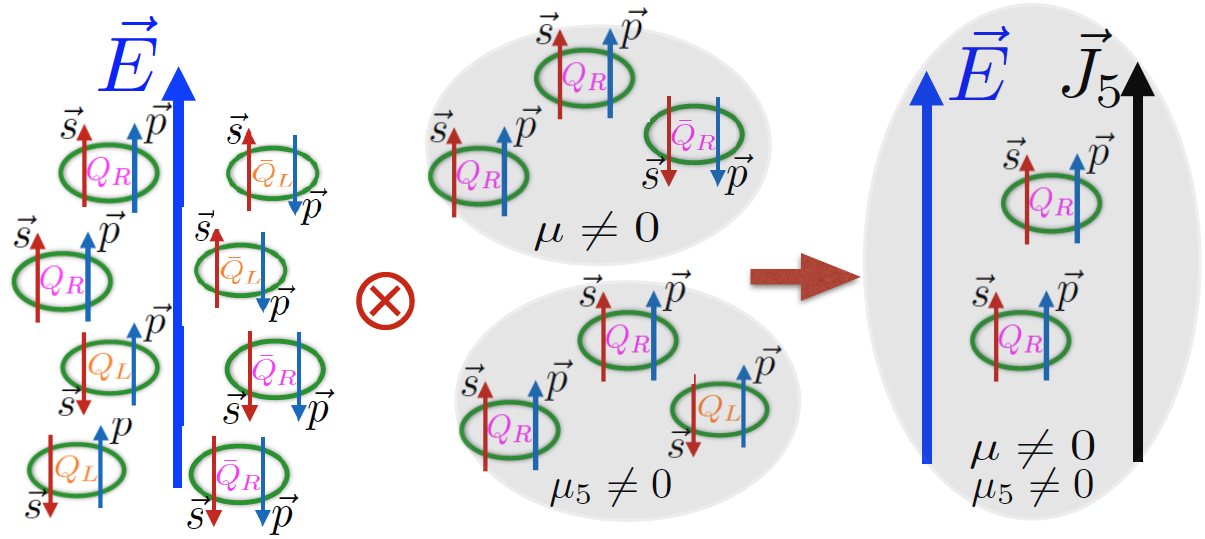}
\end{center}
\caption{(Color online) Illustration of the Chiral Electric Separation
  Effect. To be specific, the illustration is for one kind of massless
  (positively charged) quarks and (negatively charged) antiquarks, and
  for the case of $\mu>0$ and $\mu_5>0$. Changing the sign of either
  $\mu$ or $\mu_5$, the current $\vec{\mathbf J}_5$ will flip
  direction.}
\label{fig_CESE}
\end{figure}

Intuitively the CESE may be understood in the following way, as
illustrated in Fig.~\ref{fig_CESE}.  The electric field leads to the
usual conducting currents, which implies that the positively charged
quarks have their momenta preferably aligned along the $\vec{\mathbf
  E}$ field direction, while the negatively charged anti-quarks have
their momenta oppositely aligned, i.e. $\langle \vec p \rangle \propto
(Qe) \vec{\mathbf E}$. Given a nonzero $\mu_5$ (e.g. considering
$\mu_5>0$) there will be more RH particles than LH particles, with net
RH quarks moving along $\vec{\mathbf E}$ while net RH antiquarks
moving against $\vec{\mathbf E}$. Provided a further nonzero $\mu$
(e.g. considering $\mu>0$) there will then be more RH quarks than
antiquarks: this net amount of RH quarks move along $\vec{\mathbf E}$
and contribute to an axial current $\vec{\mathbf J}_5 \propto (\mu
\mu_5) (Qe) \vec{\mathbf E}$. When either $\mu$ or $\mu_5$ vanishes,
this current ceases to exist owing to cancellations.

It is useful to recast Ohm's law (Eq.~(\ref{eq_Ohm})) and the CESE
(Eq.~(\ref{eq_CESE})) into the RH/LH formulation:
\begin{eqnarray}
\vec{\mathbf J}_{R/L} =   \left[  \frac{\sigma}{2} \pm  \frac{\sigma_{\chi e}}{2} \right]\,\vec{\mathbf E}  =   \left[  \frac{\sigma}{2} \pm  \frac{{\chi_e}(Qe)}{8}  \left(\mu_R^2 - \mu_L^2 \right) \right]\,\vec{\mathbf E}   \,\, .
\end{eqnarray}
The above can be interpreted as the $\vec{\mathbf E}$-induced
conduction currents separately for RH and LH particles, and they
differ from each other when there is an imbalance $|\mu_R| \neq
|\mu_L|$.

Finally the various transport effects, Ohm's law (Eq.~(\ref{eq_Ohm})),
the CME (Eq.~(\ref{cme_current})), the CSE (Eq.~(\ref{eq_CSE})) and the
CESE (Eq.~(\ref{eq_CESE})), can be nicely summarized as follows:
\begin{eqnarray}
\left(\begin{array}{c} \vec{\mathbf J} \\ \vec{\mathbf
    J}_5 \end{array}\right) = \left(\begin{array}{cc} \sigma &
  \sigma_5   \\ \sigma_{\chi e}   & \sigma_s
\end{array}\right) \left(\begin{array}{c}
\vec{\mathbf E} \\ \vec{\mathbf B}
\end{array}\right). \label{eq_vaeb}
\end{eqnarray}
This therefore completes our discussions on the generation of currents
in a chiral quark-gluon plasma in response to external electromagnetic
fields.

\subsection{The Chiral Vortical Effect}

The anomalous transport effects can also occur when a system of chiral
fermions is undergoing a global rotation. Such a fluid rotation can be
quantified by a {\em vorticity} $\vec{\mathbf \omega}
=\frac{1}{2}\vec\nabla \times \vec{\mathbf v}$, where $ \vec{\mathbf
  v}$ is the flow velocity field. Interesting analogy may be drawn
between the fluid rotation and electromagnetic fields as first
emphasized in \cite{Kharzeev:2007tn}: $\vec{\mathbf v}$ is analogous
to vector gauge potential $\vec {\mathbf A}$, and the vorticity
$\vec{\mathbf \omega}$ is then similar to the magnetic field
$\vec{\mathbf B}=\nabla \times \vec{\mathbf A}$. Consider a charged
particle moving in a circle perpendicular to a constant $\vec{\mathbf
  B}$ field, the quantum mechanical effect gives rise to a phase
factor $e^{i (Qe) \Phi_B/\hbar}$ (with $\Phi_B$ the magnetic flux
through the circle).  Similarly when such a particle moves in a circle
perpendicular to a constant $\vec{\mathbf \omega}$ field, it aquires a
phase factor $e^{i L/\hbar}$ (with $L$ the corresponding angular
momentum). Given such similarity, it is therefore natural to expect
that vorticity-driven effects similar to the CME and the CSE may
occur.

\begin{figure}[!hbt]
\begin{center}
\includegraphics[width=0.8\textwidth]{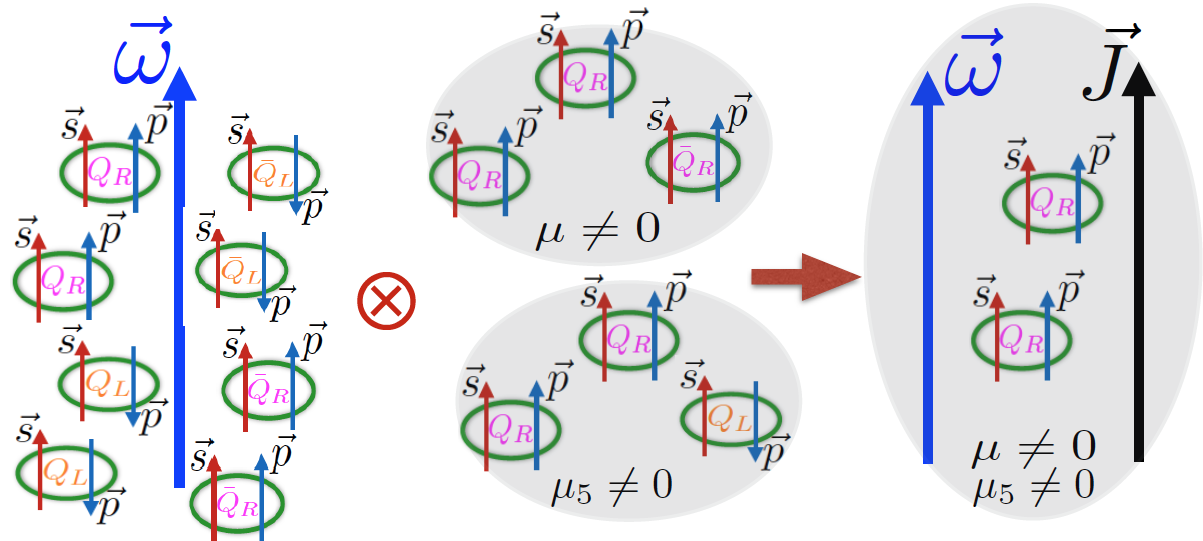}
\end{center}
\caption{(Color online) Illustration of the Chiral Vortical Effect. To
  be specific, the illustration is for one kind of massless quarks and
  antiquarks, and for the case of $\mu>0$ and $\mu_5>0$. Changing the
  sign of either $\mu$ or $\mu_5$, the current $\vec{\mathbf J}$ will
  flip direction.}
\label{fig_CVE}
\end{figure}

Such vortical effect was quantitatively identified first in
holographic
models~\cite{Banerjee:2008th,Erdmenger:2008rm,Torabian:2009qk} and
later understood in the anomalous hydrodynamic
framework~\cite{Son:2009tf}. For given vorticity $\vec \omega$, the
Chiral Vortical Effect (CVE) quantifies the generation of a vector
current ${\vec {\mathbf J}} $ along the vorticity direction:
\begin{eqnarray} \label{eq_CVE}
&&\vec{\mathbf J} =\frac{1}{\pi^2}\mu_5 \mu \vec{\mathbf \omega} \,\, .
\end{eqnarray} 
While the CME (Eq.~(\ref{cme_current})) is driven by $\vec{\mathbf B}$, the
above CVE is driven by $\mu \vec{\mathbf \omega}$ in a chiral medium
with $\mu_5\neq 0$.  Intuitively the above CVE may be understood in
the following way, as illustrated in Fig.~\ref{fig_CVE}.  In the
presence of a global rotation, the underlying fermions experience an
effective interaction of the form $\sim - \vec \omega \cdot \vec S$ in
their local rest frame, with $\vec S$ the spin of fermions. This
causes a spin polarization effect (as indeed found in other context
\cite{Liang:2004ph,Becattini:2007sr}), namely the fermions will have
their spins preferably aligned with $\vec{\mathbf \omega}$.  We
emphasize that such spin polarization $\langle \vec s \rangle \propto
\vec{\mathbf \omega}$ is {\it charge-blind}, which is different from
the magnetic polarization.  Given a nonzero $\mu_5$ (e.g. considering
$\mu_5>0$) there will be more RH particles than LH particles, with net
RH particles (both quarks and antiquarks) moving along $\vec{\mathbf
  \omega}$ due to $\langle \vec p \rangle \propto \langle \vec s
\rangle \propto \vec{\mathbf \omega}$. Provided a further nonzero
$\mu$ (e.g. considering $\mu>0$) there will then be more RH quarks
than antiquarks: this net amount of RH quarks move along $\vec{\mathbf
  \omega}$ and contribute to a vector current $\vec{\mathbf J} \propto
(\mu \mu_5) \vec{\mathbf \omega}$. When either $\mu=0$ or $\mu_5=0$,
this current ceases to exist owing to cancellations.

In fact similarly to the CSE, an axial current can be generated as
well under a global rotation:
 \begin{eqnarray} \label{eq_CVAE}
&&\vec{\mathbf J}_5=\left [
     \frac{1}{6}T^2+\frac{1}{2\pi^2}(\mu^2+\mu^2_5) \right ]
   \vec{\omega} \,\, .
\end{eqnarray}
Again one can rewrite the vortical effects (Eq.~(\ref{eq_CVE}) and
Eq.~(\ref{eq_CVAE})) in terms of chiral currents $\vec{\mathbf
  J}_{R/L} $ as follows:
\begin{eqnarray} \label{eq_CVE_LR}
&&\vec{\mathbf J}_{R/L}=\pm \left( \frac{1}{12}T^2 + \frac{1}{4\pi^2}\mu^2_{R/L} \right) \vec{\omega} \,\, .%\\
\end{eqnarray}
Clearly the above can be interpreted as the CVE separately for RH/LH
particles. The coefficient $1/{4\pi^2}$ in front of the chemical
potential term is dictated by the chiral anomaly, similarly to the
$\sigma_5/2=(Qe)/(4\pi^2)$ in the CME case.  It has been suggested~\cite{Landsteiner:2011iq,Landsteiner:2013aba} that the $T^2$ term originates from gravitational anomaly. 
Possible corrections to the coefficient of $T^2$ terms have been discussed in \cite{Golkar:2012kb,Hou:2012xg,Kalaydzhyan:2014bfa}.

\subsection{The Collective Excitations}

 While for the intuitive illustrations in preceding discussions we
 have relied upon individual particle pictures, the various anomalous
 transport effects are actually about the behavior of macroscopic
 (i.e. thermodynamic and hydrodynamic) densities and currents,
 irrespective of whether the underlying systems may allow a
 quasiparticle description or not.  A very nontrivial feature of these
 effects, is that they couple together the vector and axial
 densities/currents in the presence of electromagnetic fields or a
 fluid rotation. It is natural to wonder if certain collective modes
 may arise from mutually induced vector/axial density
 fluctuations. Let us recall the well-known example in hydrodynamics
 where the fluctuations of energy density and pressure mutually induce
 one another and form propagating collective modes i.e. the sound
 waves.  Indeed a number of robust gapless excitations in a chiral
 fluid system such as the Chiral Magnetic Wave
 (CMW)~\cite{Kharzeev:2010gd,CMW_calc_Burnier} and Chiral Vortical
 Wave (CVW)~\cite{Jiang:2015cva} have been found, which we discuss
 below.

\subsubsection{The Chiral Magnetic Wave}

\begin{figure}[!hbt]
\begin{center}
\includegraphics[width=0.7\textwidth]{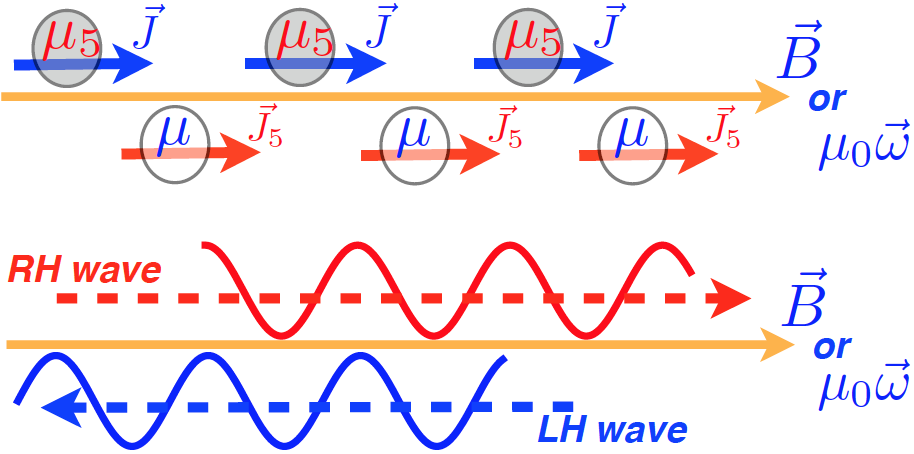}
\end{center}
\caption{(Color online) Illustration of the Chiral Magnetic Wave and
  the Chiral Vortical Wave. }
\label{fig_Wave}
\end{figure}

Let us first consider  small fluctuations in the vector and axial
densities of a QGP in an external magnetic field $\vec{\mathbf
  B}$. (For simplicity we assume here a neutral QGP with zero
background densities but the discussion otherwise would be essentially
similar.)  Suppose a nonzero axial density fluctuation $\delta J_5^0$
occurs, and it implies a locally nonzero $\mu_5 \propto (\delta
J)_5^0$, which induces a CME vector current $\vec{\mathbf J}$ via
Eq.~(\ref{cme_current}).  Such a current will transport vector charges
along the $\vec {\mathbf B}$ direction and thus cause the nearby
vector density to fluctuate off equilibrium. A nonzero vector density
fluctuation $\delta J^0$, in turn, implies a locally nonzero $\mu
\propto \delta J^0$, which induces a CSE axial current $\vec{\mathbf
  J}_5$ via Eq.~(\ref{eq_CSE}). Such a current will transport axial
charges along the $\vec {\mathbf B}$ direction and further cause the
nearby axial density to fluctuate off equilibrium. In this way, the
vector and axial density fluctuations mutually induce each other, and
their evolutions are entangled nontrivially through the CME and the
CSE. As a result, these density fluctuations disseminate with time
along the external $\vec{\mathbf B}$ direction to far locations, and
thus a propagating wave forms: see Fig.~\ref{fig_Wave} (upper panel)
for an illustration of this phenomenon. This collective mode is the
Chiral Magnetic Wave~\cite{Kharzeev:2010gd}.

Mathematically the wave equations to describe the CMW can be derived
by combining the RH/LH form of the CME and the CSE in
Eq.~(\ref{eq_CME_LR}) with the continuity equations for RH/LH currents
$\partial_t J_{R/L}^0 + \nabla\cdot \vec{\mathbf J}_{R/L} = 0$:
\begin{eqnarray} \label{eq_CMW}
\left(\partial_0 \pm \frac{(Q e) }{(4\pi^2) \chi} \vec{\mathbf B} \cdot \nabla   \right )  \, \delta J^0_{R/L} =  \left(\partial_0 \pm {v_B} \partial_{\hat{\mathbf B}} \right ) \, \delta J^0_{R/L}=0 . 
\end{eqnarray} 
An expansion of the fluctuations via Fourier modes with frequency $\nu$ and wave-vector $k \hat{\mathbf B}$ then gives  
\begin{eqnarray}  \label{eq_dispersion}
\nu \mp v_B  k = 0,
\end{eqnarray}
where we can identify the propagation speed of the wave, $v_B \equiv
\frac{(Qe) B }{(4\pi^2) \chi}$.  The parameter $\chi$ is the
thermodynamic susceptibility that connects density with chemical
potential, i.e. $\chi_{R/L}=\partial J^0_{R/L} / \partial \mu_{R/L}$
(which can be easily related to the usual vector and axial
susceptibilities $\chi=\partial J^0 / \partial \mu$ and
$\chi_5=\partial J^0_5 / \partial \mu_5$).  The above wave equations
reveal the nature of the CMW in a transparent way: it consists of
actually {\em two chiral gapless modes} traveling at the same speed
$v_B$, with the RH wave that transports RH density and current in the direction 
parallel to the $\vec{\mathbf B}$ direction as well as the LH wave
that transports LH density and current in the direction antiparallel to
$\vec{\mathbf B}$. This is illustrated in Fig.~\ref{fig_Wave} (lower
panel).

A more general analysis~\cite{Huang:2013iia} of various possible
collective modes on top of a possible non-neutral-background QGP
(i.e. with nonzero $\mu$ and/or $\mu_5$) in external electric as well
as magnetic fields can be done by starting from the response relations
(Eq.~(\ref{eq_vaeb})), combining it with continuity equations, and
linearizing the equations for small fluctuations on top of background
densities. Two general wave modes are found in~\cite{Huang:2013iia},
with rather complicated structures. These however reduce to simple
wave modes with clear physical interpretations in a number of special
settings: (1) in the case of pure magnetic field $\vec{\mathbf B}$ the
two modes become (slightly generalized) the CMW; (2) in the case of
pure electric field $\vec{\mathbf E}$ and nonzero background vector
density, the two modes become a new type of {\em Chiral Electric Wave
  (CEW)} arising from CESE and propagating in parallel/antiparallel to
the $\vec{\mathbf E}$ field; (3) in the case of pure electric field
$\vec{\mathbf E}$ and nonzero background axial density, the two modes
turn into a vector density wave and an axial density wave. In general,
when there are both $\vec E$ and $\vec B$ fields and both types of
background densities, these physically distinct modes are intertwined
into the more complex collective excitations.  
In collisions of symmetric nuclei (e.g. AuAu, CuCu, or PbPb) there is nonzero electric field locally 
but on average no net electric field (in contrast to the nonzero net magnetic field), so one may 
expect that the possible mixing effects are unimportant in those collisions.  

Note that in realistic systems there would be dissipative effects such
as electric conductance as well as diffusion of charge
densities. These would change the wave dispersion relations
Eq.~(\ref{eq_dispersion}) by contributing imaginary terms to the
frequency. In the case of the CMW, such contributions take the form
$-i(Qe \sigma/2 + D_L k^2 + D_\perp k_\perp^2)$, where $\sigma$ is the
electric conductivity, and $D_L$ and $D_\perp$ are the diffusion
constants along and perpendicular to the $\vec{\mathbf B}$ direction,
respectively. 
The effect of diffusion would somewhat reduce the CMW signal in heavy ion collisions, even though with reasonable values of diffusion coefficients  this reduction effect is not large as shown in \cite{CMW_calc_Burnier}. 

\subsubsection{The Chiral Vortical Wave}

By virtue of the close analogue between magnetic-field-driven effects
and rotation-driven effects, it is tempting to explore whether certain
interesting collective excitations may stem from the CVE-entailed
interplay between vector and axial density fluctuations in a rotating
chiral fluid. Indeed, it has been recently found that there are
vorticity-induced collective excitations called the Chiral Vortical
Wave (CVW). To see how they arise, let us consider a rotating QGP with
a homogeneous background vector density $\mu_0\neq 0$. In this case
with $\mu_0 \vec{\bf \omega}$ playing the role of $\vec{\mathbf B}$ in the CMW,
a nonzero axial density fluctuation $\delta J_5^0$ induces a CVE
vector current $\vec{\mathbf J}$ via Eq.~(\ref{eq_CVE}), while a
vector density fluctuation also induces a CVE axial current
$\vec{\mathbf J}_5$ via Eq.~(\ref{eq_CVAE}).  In this way, the vector
and axial density fluctuations mutually induce each other and form
propagating waves along the rotation axis i.e. the $\vec{\mathbf
  \omega}$ direction: again see Fig.~\ref{fig_Wave} (upper panel) for
an illustration of this phenomenon. This collective mode is the Chiral
Vortical Wave~\cite{Jiang:2015cva}.

To derive the wave equations for the CVW, it is convenient to start
from the RH/LH CVE in Eq.~(\ref{eq_CVE_LR}) and consider small
fluctuations of RH or LH densities on top of a uniform equilibrium
background (assuming no fluctuations in temperature). By combining the
linearized version of Eq.~(\ref{eq_CVE_LR}) with continuity equations
$\partial_t J_{R/L}^0 + \nabla\cdot \vec{\mathbf J}_{R/L} = 0$, one
can easily derive the following wave equations:
\begin{eqnarray} \label{eq_CMW}
\left(\partial_0 \pm \frac{\mu_{0}}{(2\pi^2) \chi_{\mu_{0}}} \vec{\mathbf \omega} \cdot \nabla   \right )  \, \delta J^0_{R/L} =  \left(\partial_0 \pm {v_\omega} \partial_{\hat{\mathbf \omega}} \right ) \, \delta J^0_{R/L}=0.  
\end{eqnarray} 
An expansion of the fluctuations via Fourier modes with frequency
$\nu$ and wave-vector $k \hat{\mathbf \omega}$ then gives
\begin{eqnarray}  \label{eq_dispersion}
\nu \mp v_\omega  k = 0,
\end{eqnarray}
where we can identify the propagation speed of the wave, $v_\omega
\equiv \frac{\mu_0 \omega }{(2\pi^2)\chi_{\mu_0}}$. (To be precise the
$\mu_0$ here is the background RH or LH density, and the
$\chi_{\mu_0}$ is the corresponding susceptibility
$\chi_{R/L}=\partial J^0_{R/L} / \partial \mu_{R/L}$ evaluated at
$\mu_{R/L}=\mu_0$.  Similar to the CMW case, the above equations imply
that the CVW also consists of {\em two chiral gapless modes} traveling
at the same speed $v_\omega$, with the RH wave that transports RH
density and current in parallel to the $\vec{\mathbf \omega}$
direction as well as the LH wave that transports LH density and
current in antiparallel to the $\vec{\mathbf \omega}$. This is also
illustrated in Fig.~\ref{fig_Wave} (lower panel).

It is important to note that because the CVE current (unlike the CME one) depends quadratically on the chemical potentials, the dynamics of the corresponding collective excitations in systems with vorticity is in general nonlinear. For example, in ``hot" systems with temperature much larger than the chemical potentials, the collective excitations are described \cite{Basar:2013iaa} by the Burgers-Hopf equation that is known to describe rich nonlinear dynamics, including solitons and shock waves.   The mixing of the chiral magnetic and chiral vortical waves has recently been addressed in \cite{Chernodub:2015gxa,Frenklakh:2015fzc}.  Possible mixing of these with other collective excitations, such as a Chiral Heat Wave, has also been discussed in \cite{Chernodub:2015gxa}.

%%% Local Variables: 
%%% mode: latex
%%% TeX-master: "master"
%%% End: 

\section{Anomalous Chiral Effects in Heavy Ion Collisions}
\label{sec:3}

So far we have discussed the physical ideas of various anomalous
chiral effects. It is of fundamental interest to look for experimental
manifestations of such effects in systems from novel
semi-metals~\cite{Li:2014bha,Basar:2013iaa,Ong,Kharzeev:2012dc,Gorbar:2013dha}
to hot dense QCD
matter~\cite{Kharzeev:2007jp,Kharzeev:2004ey,Kharzeev:2007tn,CMW_calc_Burnier,Kharzeev:2010gr,Huang:2013iia,Jiang:2015cva,Gorbar:2011ya,Ma:2015isa}. In
the rest of this review we focus on the search of anomalous chiral
effects in the quark-gluon plasma created in heavy-ion collision
experiments at the Relativistic Heavy Ion Collider (RHIC) as well as
the Large Hadron Collider (LHC).  In preparation for later discussions
on experimental search, a number of important phenomenological aspects
need to be addressed here. Emphasis will be put on three examples, the
Chiral Magnetic Effect, the Chiral Magnetic Wave, and the Chiral
Vortical Effect, which have been extensively studied both
phenomenologically and experimentally.

\subsection{The magnetic field and vorticity}

To induce effects like the CME and the CMW, strong electromagnetic
fields need to be present in the system. There are indeed such fields,
originated from the highly charged ions (e.g. Au nucleus with $Z=79$
at RHIC and Pb nucleus with $Z=82$ at LHC) that move at nearly the
speed of light. An elementary estimate can be done as follows: 
%$eB \sim \frac{\gamma\, \alpha_{EM} Z }{b^2}$ 
 $eB \sim \gamma\, \alpha_{EM} Z /b^2$
where $\alpha_{EM}\simeq
1/137$, 
$b$ is the impact parameter, and $\gamma=(\sqrt{s}/2)/M_N$ is
the Lorentz factor (with $(\sqrt{s}/2)$ the energy per nucleon in the
beam and $M_N$ the nucleon mass). Upon plugging in numbers for RHIC
collisions one immediately recognizes that $eB \sim 1/(1 fm)^2 \sim
(m_\pi)^2$ which is on the typical hadronic interaction scale and
which represents the strongest electromagnetic fields accessible to
human. A lot of computations have been done to quantify such
electromagnetic fields on the event-by-event basis (see
e.g.~\cite{Bzdak:2011yy,Deng:2012pc,Bloczynski:2012en}). Their spatial
distribution as well as the dependence on colliding nuclei, centrality
and beam energy have been studied.

An important feature of the magnetic field $\vec {\bf B}$ in heavy-ion collisions
 is its azimuthal orientation on
the transverse plane.  Fig.~\ref{fig:Overlap} schematically depicts
the transverse plane for a collision of two heavy ions.  Following
common practice we label the beam axis as $\hat{z}$, the impact
parameter direction as $\hat{x}$, and the out-of-plane direction as
$\hat{y}$, with $\hat{x}-\hat{z}$ as the reaction plane and
$\hat{x}-\hat{y}$ as the transverse plane.  Event-by-event simulations
have demonstrated that the geometric orientation of $\vec {\bf B}$ is
approximately in parallel to the out-of-plane direction
$\hat{y}$. This is extremely useful: it implies that any signal (such
as a CME current) along $\vec {\bf B}$ will also be in the out-of-plane
direction, with the latter information being experimentally
accessible. 

Another important aspect of such magnetic field $\vec {\bf B}$ is its duration in QCD fluid, see
e.g.~\cite{Tuchin:2013ie,McLerran:2013hla,Guo:2015nsa,Gursoy:2014aka}. This represents 
one of the major remaining sources of uncertainty in theoretical calculations. 
The time dependence of $\vec {\bf B}$
  after the impact of the two nuclei would
crucially depend upon whether/when/how a conducting medium may form and lead to 
a much elongated lifetime of magnetic field.   Such uncertainty can be reduced by the study of directed flow of charged hadrons away from mid-rapidity, as  proposed recently in ~\cite{Gursoy:2014aka}. The magnetic field is also expected to contribute to the photon and dilepton production through the ``magneto-sono-luminescence": the conversion of phonons into real or virtual photons in a magnetic background ~\cite{Basar:2012bp,Basar:2014swa}.  There would also be nontrivial interplay between a dynamically evolving magnetic field and the CME itself (see e.g.~\cite{Manuel:2015zpa}). 
 To ultimately resolve this issue, one needs to treat both the field and
the medium dynamically and efforts are underway to develop such a
chiral magnetohydrodynamic simulation.

The vortical effects, on the other hand, are to be induced by the
global rotation of the QGP in heavy-ion collisions. In a general
non-central collision, there is obviously a nonzero global angular
momentum $\vec
{\bf L}$~\cite{Liang:2004ph,Becattini:2007sr,vorticity}. While the majority
of this angular momentum is carried away by the spectator nucleons,
recent simulations~\cite{vorticity,Baznat:2015eca} do show that a considerable
fraction (about $10\sim 20\%$) of $\vec {\bf L}$ remains in the QGP in the
collision zone and is approximately conserved in time. This could
imply a relatively long time duration of the vortical effects. It is
important to emphasize that this angular momentum is also pointing
approximately in the out-of-plane direction. Attempts on the
computation of local vorticity $\omega$ and its space-time
distribution have also been
made~\cite{vorticity,Baznat:2015eca,Becattini:2015ska,Becattini:2013vja,Florchinger:2011qf,Gao:2014coa}. The
vorticity is a more subtle quantity owing to different ways of
defining it. Furthermore the vorticity can receive nonzero local contributions from bulk flow
 that are not related to the global rotation: for example with a pure radial flow field $\vec {\bf v}_\perp \propto f(\vec {\bf r}_\perp) \hat{\bf r}_\perp$, the vorticity $\omega=\vec\bigtriangledown\times \vec {\bf v} /2$ is nonzero locally but vanishes upon average over space. The best
approach to address vorticity quantitatively, again, is to develop 3D
hydrodynamic simulations that incorporate a built-in global rotation.

\subsection{The initial conditions}

As is evident from the formulae for the various effects, given an
external ``driving force'' ($\vec {\bf B}$ or $\vec {\bf \omega}$), the next
crucial elements are the vector/axial charge densities that
``trigger'' the anomalous transport. In the context of heavy-ion
collisions, the issue concerns the initial conditions for the various
charge densities. For example, in order for the CME to occur, one
needs nonzero initial axial charge density present in the system, for
the CMW one needs nonzero initial vector charge density, and for the
CVE one needs both types of initial charge densities.

In heavy-ion collisions such initial charge densities naturally arise
from fluctuations. The axial charge density may be generated from a
number of
sources~\cite{Kharzeev:2007jp,Kharzeev:2004ey,Kharzeev:2007tn,Kharzeev:2001ev,Iatrakis:2014dka,Fukushima:2010vw}:
the topological fluctuations of the gluonic sector (via instanton and
sphaleron transitions), the chromomagnetic flux tubes with nonzero
local $\vec{\bf E}\cdot \vec{\bf  B}$ in the initial glasma, as well as simple
fluctuations in the quark sector. The obvious way of acquiring the
vector charge density is from ``deposition'' in the collision zone by
the initial colliding nuclei which possess large baryonic, electric,
and isospin charges. 

\begin{figure}
\begin{center}
\includegraphics[width=14cm]{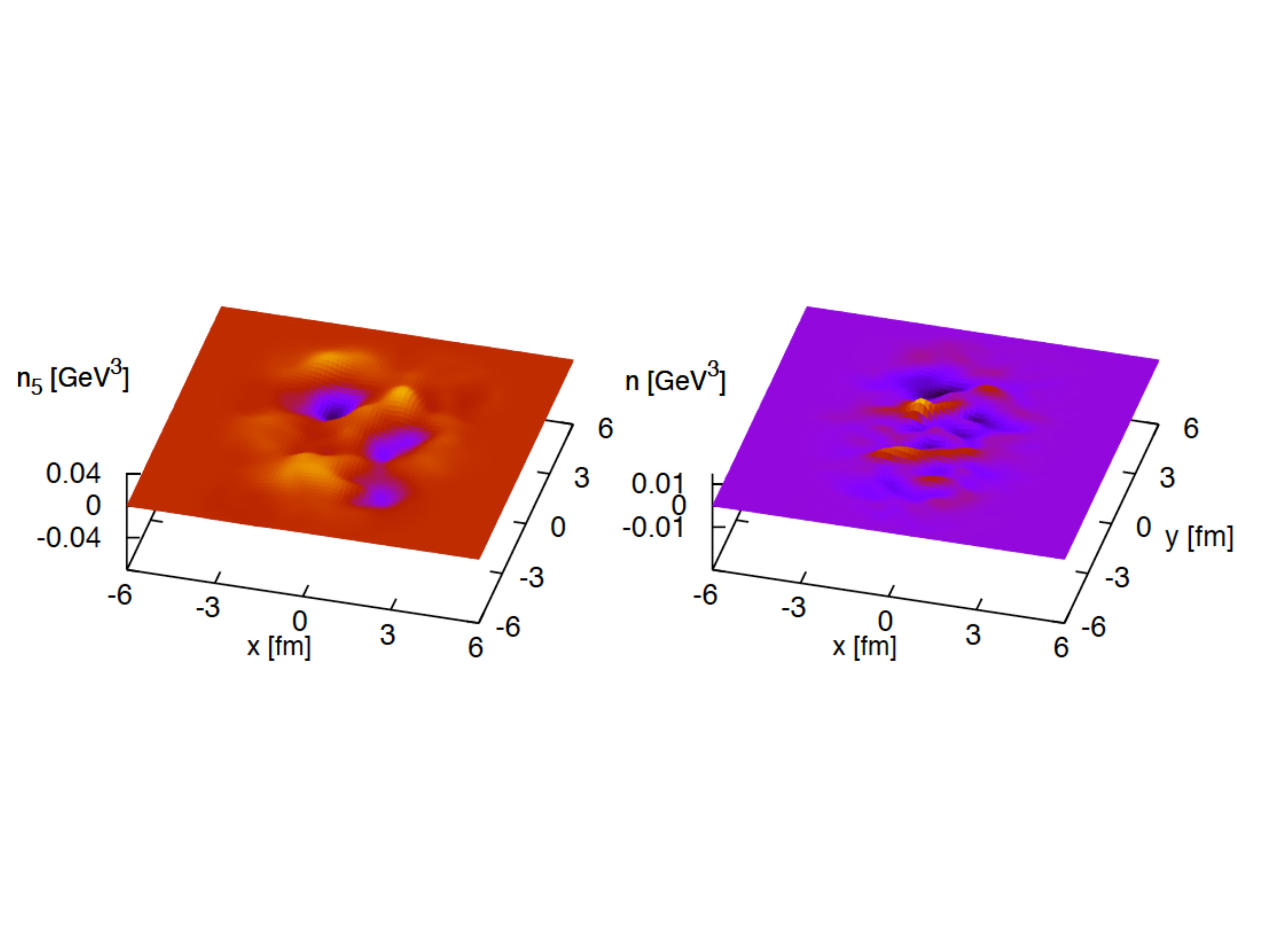}
\end{center}
\vspace{-3cm}
\caption{Distributions of the chiral (left) and electric (right) charge densities in the transverse plane at mid-rapidity and a proper time $\tau = 1.5$fm/c of a Au+Au collision event at $\sqrt{s_{\rm NN}}=200$ GeV as computed  in anomalous hydrodynamics; from \cite{Hirono:2014oda}.}
\label{axial_electric}
\end{figure}

To quantitatively model and constrain the charge
initial conditions is crucial for the search of anomalous effects and
could be quite challenging. A fully quantitative theoretical approach to describe charge asymmetries requires the use of relativistic hydrodynamics that includes the terms arising from the chiral anomaly, supplemented by the initial conditions describing topological fluctuations at the early stage of a heavy collision. A study of that kind has recently been performed in \cite{Hirono:2014oda} where the initial condition is provided by the fluctuating longitudinal ``glasma" fields with 
${{\vec {\bf E}}}^a {{\vec{\bf B}}}^a \neq 0$ \cite{Kharzeev:2001ev,Lappi:2006fp,Kharzeev:2005iz}. The snapshot of the resulting chiral and electric charge densities in the QCD fluid is shown in Fig. \ref{axial_electric}.

\subsection{Chiral Magnetic Effect}

\begin{figure}[h]
\begin{minipage}[c]{0.48\textwidth}
%\center
\includegraphics[width=\textwidth]{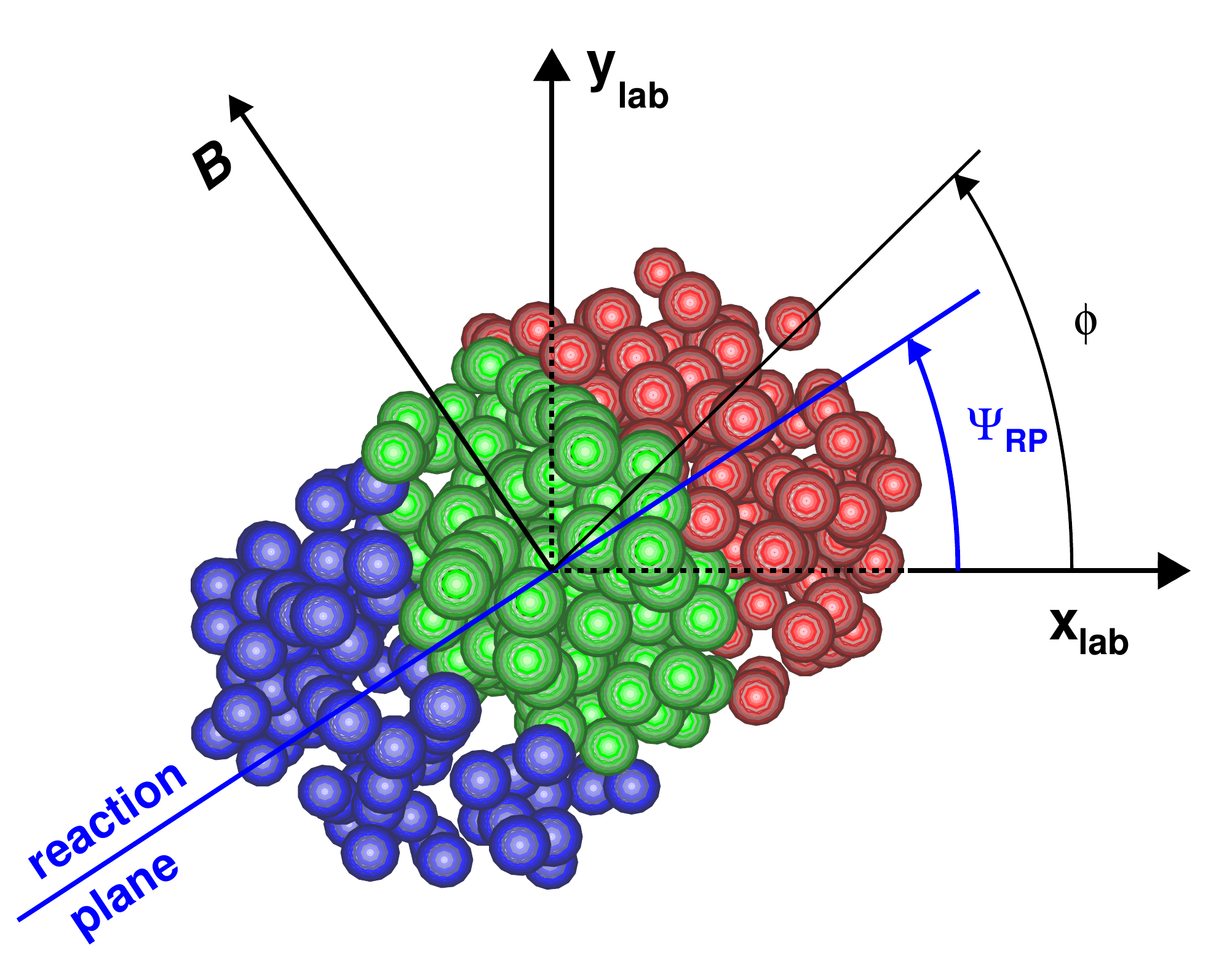}
  \caption{Schematic depiction of the transverse plane for a collision
    of two heavy ions (the left one emerging from and the right one
    going into the page)~\cite{STAR_LPV_BES}.  Particles are produced
    in the overlap region (green-colored nucleons). }
\label{fig:Overlap}
\end{minipage}
\hspace{0.4cm}
\begin{minipage}[c]{0.48\textwidth}
%\center
\includegraphics[width=\textwidth]{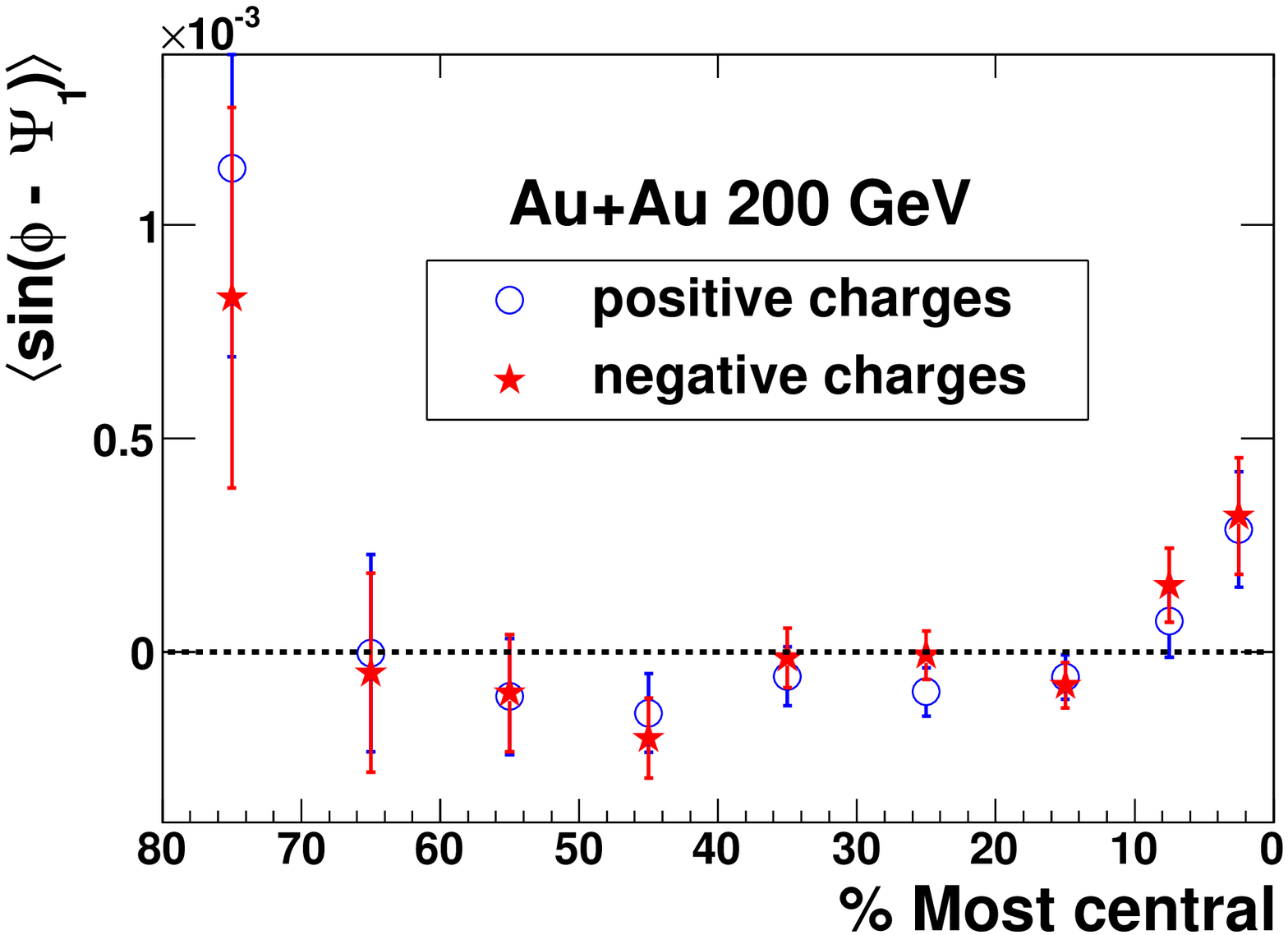}
  \caption{$\langle \sin(\phi-\Psi_1) \rangle$ for positive and
    negative charges versus centrality for Au+Au collisions at
    $\sqrt{s_{\rm NN}}=200$ GeV~\cite{STAR_LPV3}.  }
\label{fig:a1star}
\end{minipage}
\vspace{-0.2cm}
\end{figure}

We now discuss the possible signal of the CME in heavy-ion
collisions. Given the field $\vec{\bf B}$ and initial axial charge $\mu_5$,
the CME current $\vec{\bf J}$ is induced along  $\vec{\bf B}$ (thus in the
out-of-plane direction) with its sign depending upon $\mu_5$. An
electric charge current $\vec{\bf J}_Q$ forms as a result of all
contributing quark-level CME currents (see Eq.~(\ref{eq_Q_B})).  This
charge current transports positive charges toward one pole of the QGP
fireball and negative charges toward the opposite pole, thus forming a
dipole moment in the charge distribution of the QGP.  This effect can
be incorporated into the hadron production at freeze-out through a
nontrivial electric charge chemical potential of the form $\sim \mu_e
\sin(\phi_s-\Psi_{\rm RP})$ (where $\phi_s$ is the spatial azimuthal
angle and $\Psi_{\rm RP}$ is the reaction plane angle). Its
consequence can be demonstrated via the Cooper-Frye procedure for
produced final hadron's spectra:
\begin{eqnarray}
\frac{dN_{\pm}}{d\phi} \propto \int_{source} e^{-p^\mu u_\mu} e^{\pm
  (\mu_e/T_f) \sin(\phi_s-\Psi_{\rm RP})}.
\end{eqnarray} 
Here we have suppressed other kinetic variables and focused on the
azimuthal angle distribution, and for simplicity have used the
Boltzmann approximation with the freeze-out temperature $T_f$.  The
strong radial flow (hidden in flow velocity field $u_\mu$) will
collimate the azimuthal angle $\phi$ of emitted hadron's momentum with
the spatial angle $\phi_s$ of the local emission cell in the source,
and thus the out-of-plane dipole in the chemical potential will
``translate'' into a charge-dependent dipole term in the emitted
hadron distributions.  Using the parameterization of the paricle
  azimuthal distribution in a form~\cite{Voloshin:2004vk}: 
\begin{eqnarray}
\frac{dN_{\pm}}{d\phi} \propto 1 + 2v_1\cos(\phi-\Psi_{\rm RP}) + 
2v_2\cos[2(\phi-\Psi_{\rm RP})] + ... + 2a_{\pm}\sin(\phi-\Psi_{\rm RP}) + ...,
\end{eqnarray} 
where  
$v_1$ and $v_2$ are coefficients accounting for the so-called directed
and elliptic flow  \cite{flow}, one finds that
  $a_+ = - a_- \propto \mu_e \propto \mu_5 |\vec{\bf B}|$.  There is
however an important complication: the $\mu_5$ arising from
fluctuations will take different signs from event to event, and on
event average this dipole term vanishes, so a direct measurement of
this ${\cal P}$-odd effect is not possible. 
Indeed a non-zero value of $a_\pm$ would manifest  global parity
  violation which should not occur in QCD.  Fig.~\ref{fig:a1star}
  presents the STAR measurements of $\mean{a_\pm}$ with the $1^{\rm st}$ harmonic event
      plane reconstructed from spectator neutrons~\cite{STAR_LPV3}.
     These results indicate no
      significant charge dependence in all centrality intervals, where
      the typical difference between positive and negative charges is
      less than $10^{-4}$.

What can be measured is the event-by-event  correlations of $a_\pm$,
that is, a term  $\mean{ a_\alpha a_\beta}$ 
 where $\alpha$ and $\beta$ represent electric charge $+$
or $-$. This however comes with the price of dealing with now ${\cal P}$-even
observables that become vulnerable to background effects which
  could dominate the measurements.
  One way  of suppression of these
  background effects, proposed by Voloshin~\cite{Voloshin:2004vk}, is
to make a subtraction between the desired out-of-plane correlation and
the in-plane correlation:
\bea
\hspace*{-2cm}
\gamma \equiv \langle \cos(\phi_\alpha +\phi_\beta -2\Psi_{\rm RP}) \rangle &=
& \langle \cos\Delta \phi_\alpha\, \cos\Delta \phi_\beta \rangle
- \langle \sin\Delta \phi_\alpha\,\sin\Delta \phi_\beta \rangle
\label{eq:obs1}
\\ & =& [\langle v_{1,\alpha}v_{1,\beta} \rangle + B_{\rm IN}] -
   [\langle a_{\alpha} a_{\beta} \rangle + B_{\rm OUT}] \nonumber \\&
   \approx& - \langle a_{\alpha} a_{\beta} \rangle + [ B_{\rm IN} -
     B_{\rm OUT}], \nonumber \eea 
where $\Delta\phi = (\phi-\Psi_{\rm RP})$, and the averaging is done
over all particles in an event and over all events.  $B_{\rm IN}$ and
$B_{\rm OUT}$ represent contributions from ${\cal P}$-even background
processes.  
 This method allows to scale down the contribution from background
  correlations by approximately a factor of $v_2$ , since the difference between $B_{\rm IN}$ and $B_{\rm OUT}$ 
  must be proportional to elliptic flow. In the originally studied
  case of the so-called ``flowing clusters''~\cite{Voloshin:2004vk,STAR_LPV2}
\be
\frac{B_{\rm IN}-B_{\rm OUT}}{B_{\rm IN}+B_{\rm OUT}}=
v_{2,cl}\frac{\mean{\cos(\phi_\alpha+\phi_\beta-2\phi_{cl})}}
{\mean{\cos(\phi_\alpha-\phi_\beta)}},
\ee
where $\phi_{cl}$ is the cluster emission azimuthal angle, and
$\phi_\alpha$ and $\phi_\beta$ and the azimthal angle of two decay
products. 
A useful way to help decipher the remaining backgrounds is by examining the 
above $\gamma$ correlator together with another correlator $\delta \equiv \langle \cos(\phi_\alpha - \phi_\beta)
\rangle = \langle \cos \Delta \phi_\alpha\, \cos \Delta \phi_\beta
\rangle + \langle\sin\Delta \phi_\alpha\, \sin\Delta \phi_\beta
\rangle$ (see e.g.~\cite{Bzdak:2009fc}), from which one can separate the in-plane and out-of-plane projected correlations.

To demonstrate  further how such
elliptic-flow-induced backgrounds may contribute to the correlators,
let us also examine the well studied example of the transverse momentum
conservation (TMC) effect (see detailed discussions in
e.g.~\cite{Liao:2010nv,v2N,Pratt:2010gy}).  The TMC leads to the
following pertinent two-particle correlation term:
\begin{eqnarray}
f_2 (\phi_\alpha,\phi_\beta) \propto ... + f_1(\phi_\alpha - \Psi_{\rm
  RP}) f_1(\phi_\beta - \Psi_{\rm RP}) \left[ F\, \left(
  \frac{p^\alpha_x p^\beta_x}{ \langle p_x^2 \rangle_F} +
  \frac{p^\alpha_y p^\beta_y}{\langle p_y^2 \rangle_F} \right) \right
],
\end{eqnarray}
where the coefficient $F$ represents the strength of this correlation
term,  the $f_1(\phi-\Psi_{\rm RP})$
is the measured single particle distribution of the form $f_1 \propto
1+ 2\langle v_2 \rangle_\Omega \cos2(\phi-\Psi_{\rm RP})+ ...$,   
$p_{x}=p_T\cos(\phi-\Psi_{\rm RP})$ and
$p_{y}=p_T\sin(\phi-\Psi_{\rm RP})$.  It is worth emphasizing that
$\langle\rangle_F$ denotes an average of all produced particles in the
full phase space; the actual measurements
will be only in a fraction of the full space, which we denote by
$\langle\rangle_\Omega$.  
 Assuming for simplicity $v_2(\pt)=const$, 
to the linear order of the small quantity $v_2$
we have $\langle p_x^2 \rangle \approx \langle p_T^2 \rangle
(1+\langle v_2 \rangle_F)/2$ and $\langle p_y^2 \rangle \approx \langle
p_T^2 \rangle (1- \langle v_2 \rangle_F)/2$.  
It is then not difficult to find that such background effect
would make the following leading contributions to the observables
$\gamma$ and $\delta$:
\begin{eqnarray}
\gamma \to \kappa \langle v_2 \rangle_\Omega F  \,\, , \,\, \delta \to F,
\end{eqnarray}
where the coefficient $\kappa\approx 2 - \langle v_2 \rangle_F /
\langle v_2 \rangle_\Omega$ would become unity in the ideal full
acceptance case. Another extensively studied elliptic-flow-induced
background, the positive-negative charge correlation from local charge
conservation (LCC) effect~\cite{Pratt:2010zn,Schlichting:2010qia}, has a
similar characteristic structure as the above. Note such a structure
is quite different from that of the CME, which gives $\gamma \to -H $
and $\delta \to H$ (with $H= \langle a_\alpha a_\beta\rangle$ the
signal strength). These observations have motivated the following
decomposition analysis~\cite{STAR_LPV_BES} that can help obtain 
 a qualitative estimates of 
the CME signal and flow backgrounds:
\begin{eqnarray}
\gamma = \kappa \langle v_2 \rangle_\Omega F - H \,\, , \,\, \delta = F + H.
\label{eq:gd}
\end{eqnarray}  
 Note that the coefficient $\kappa$ in the above expression
  depends on particle charge combination and particle transverse
  momentum. It may also depend on centrality and collision energy,
reflecting slightly
  different particle production mechanism in different conditions. 
Further discussions will follow in the next Section on
 experimental results.

 \begin{figure}[h]
\begin{minipage}[c]{0.48\textwidth}
%\center
\includegraphics[width=1.2\textwidth]{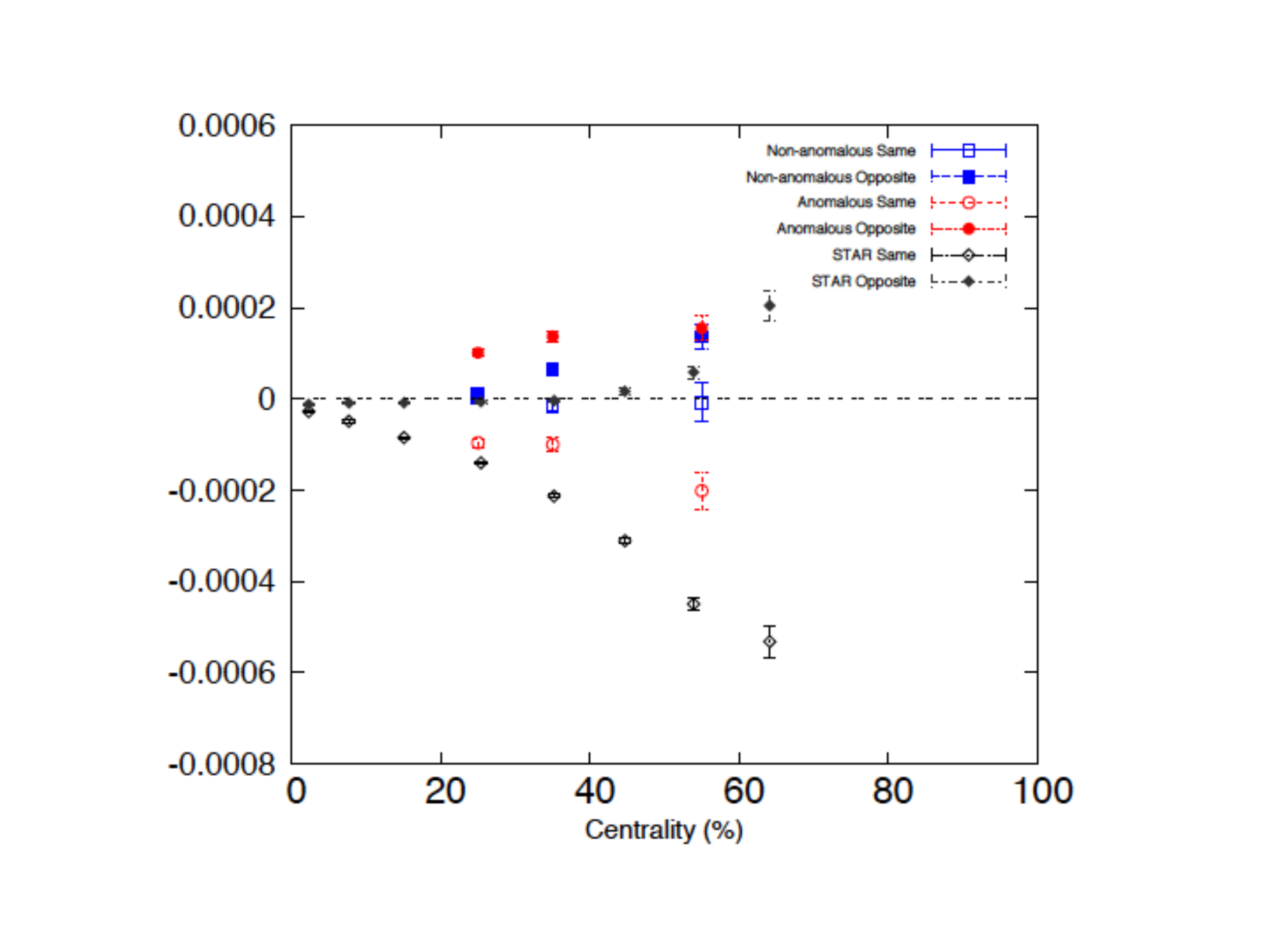}
\vspace{-1cm}
\caption{The three-point correlator $\gamma$ for same and opposite charge hadrons as computed using anomalous hydrodynamics compared to the results from STAR Collaboration in Au+Au collisions at $\sqrt{s_{\rm NN}}=200$ GeV  (from \cite{Hirono:2014oda}).}
\label{an_hyd_star}
\end{minipage}
\hspace{0.4cm}
\begin{minipage}[c]{0.48\textwidth}
%\center
\includegraphics[width=\textwidth]{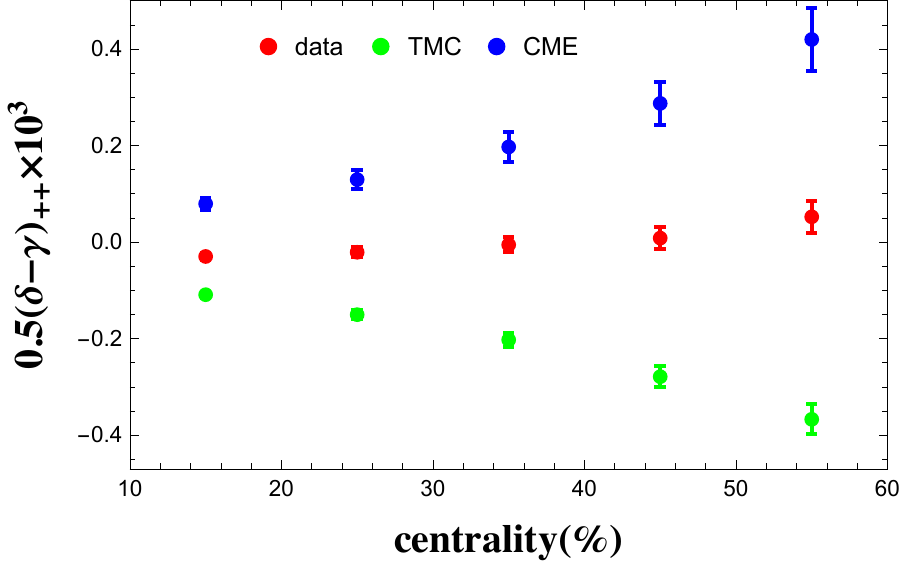}
  \caption{Anomalous hydrodynamic computation of CME and TMC contributions to correlator $ (\delta_{++}-\gamma_{++})/2$   for different centrality in Au+Au collisions at $\sqrt{s_{\rm NN}}=200$ GeV  (from \cite{Yin:2015fca}).  }
\label{fig:a1}
\end{minipage}
\vspace{-0.2cm}
\end{figure}

 On the theoretical side it is vital to develop anomalous hydrodynamic 
 simulations that quantify the CME signals with realistic initial
 conditions as well as account for background
 contributions. Significant steps toward this goal have been taken
 recently. In \cite{Hirono:2014oda}, the authors have performed the
 first event-by-event simulations of the CME in the anomalous
 hydrodynamic framework with   initial conditions (shown in Fig. \ref{axial_electric}) from glasma flux
 tubes. The computed (with and without anomalous terms) charge asymmetries are shown in Fig. \ref{an_hyd_star} in comparison with the STAR experimental results \cite{STAR_LPV1}. One can see that i) chiral anomaly has a big effect on charge asymmetries and ii) the results of the computation including the anomaly (and thus the CME) agree with the data within a factor of two, possibly leaving room for some background contributions.  
 In \cite{Yin:2015fca}, the authors have made the first attempt
 to consistently quantify contributions to observed charge
 correlations from both the CME and background effects in one and same
 framework that integrates anomalous hydro with data-validated bulk
 viscous hydro simulations. The results, shown in Fig. \ref{fig:a1}, demonstrate that the same-charge correlation data by STAR can be described quantitatively with CME and TMC together, computed with realistic magnetic field lifetime and initial axial charge density.  Both studies have conveyed the same
 message that the existence of the CME contribution is not only
 consistent with but also appears necessary for explaining the
 data. Further important progress is anticipated to come along this
 line of the quantitative CME study.

\subsection{Chiral Magnetic Wave}

Now we discuss the phenomenology of the Chiral Magnetic Wave. Consider
the QGP created in heavy-ion collisions in the presence of the
out-of-plane magnetic field $\vec{\bf B}$. Suppose the QGP has a finite
initial vector density $\rho_V^0$ concentrated around the center of
the fireball, it will trigger a right-handed CMW wave traveling toward
one pole of the fireball and a left-handed CMW wave traveling toward
the opposite pole. Such CMW evolution will transport both axial and
vector charges and result in an axial charge dipole and a vector
charge quadrupole along $\vec{\bf B}$.  Specifically considering the
electric charge distribution, the CMW will induce an electric
quadrupole moment in the QGP, with the two (out-of-plane) poles
acquiring additional positive charges and the ``equator'' acquiring
additional negative charges. As first proposed and quantified in
~\cite{CMW_calc_Burnier}, the formation of this charge quadrupole
lifts the degeneracy between the elliptic flows of positive and
negative pions leading to a splitting $\Delta v_2 = v_2^{\pi-} -
v_2^{\pi+}$ that linearly grows with the fireball charge asymmetry
$A_{\rm ch} = (N_+ - N_-)/(N_+ + N_-)$ where $N_{\pm}$ are numbers of produced positive/negative charged hadrons. 
  
Similarly to our analysis in the CME case, let us incorporate the
CMW-induced charge quadrupole into the hadron production at freeze-out
through a nontrivial electric charge chemical potential of the form
$\sim \mu_0 [1 - 2 \tilde{\mu_2} \cos(2\phi_s-2\Psi_{\rm RP})]$ with
the small quantity $\mu_0 << T_f$ characterizing the background charge
density $\bar{\rho}_e$. The coefficient $\tilde{\mu}_2$ represents the
relative ratio between the quadrupole and the ``monopole''
(i.e. isotropic) components in the charge distribution, $\tilde{\mu}_2
\simeq q_e/\bar{\rho}_e$.  Its consequence can be demonstrated via the
Cooper-Frye procedure for produced final hadron's spectra:
\begin{eqnarray}
\frac{dN_{\pm}}{d\phi} \propto \int_{source} e^{-p^\mu u_\mu} e^{\pm
  \frac{ \mu_0}{T_f} [1 - 2 \tilde{\mu}_2 \cos(2\phi_s-2\Psi_{\rm
      RP})]} \simeq \int_{source} e^{-p^\mu u_\mu} \left[1 \pm \frac{
    \mu_0}{T_f} [1 - 2 \tilde{\mu}_2 \cos(2\phi_s-2\Psi_{\rm RP})]
  \right].
\end{eqnarray} 
Here we have suppressed other kinetic variables and focused on the
azimuthal angle distribution, and for simplicity have used the
Boltzmann approximation.  The strong radial flow (hidden in flow
velocity field $u_\mu$) will collimate the azimuthal angle $\phi$ of
emitted hadron's momentum with the spatial angle $\phi_s$ of the local
emission cell in the source, and thus the out-of-plane quadrupole in
the chemical potential will ``translate'' into the following emitted
hadron distributions:
\begin{eqnarray}
\frac{dN_{\pm}}{d\phi} &\propto & N_0 [1+ 2 v^{\rm base}_{2, \pm}
  \cos(2\phi-2\Psi_{\rm RP}) + ...] \left[ 1 \pm A_{\rm ch} [1 - r
    \cos(2\phi-2\Psi_{\rm RP}) ]\right] \nonumber \\ &\simeq &
N_0(1\pm A_{\rm ch}) \left[ 1+ 2 v^{\rm base}_{2, \pm}
  \cos(2\phi-2\Psi_{\rm RP}) \mp r A_{ch} \cos(2\phi-2\Psi_{\rm RP}) +
  ... \right].
\end{eqnarray}  
In the above, the $v^{\rm base}_{2, \pm}$ is the elliptic flow of
these positive/negative hadrons at zero charge asymmetry $\mu_0=0$
(thus without any CMW effect). The charge asymmetry $A_{\rm ch}=
\langle \mu_0/T_f \rangle \propto\int_{source} e^{-p^\mu u_\mu}
(\mu_0/T_f)$ and the slope parameter $r=2\langle \tilde{\mu}_2
\rangle$. From the above one can read off the elliptic flow
coefficient:
\begin{eqnarray}
v_2^{\pm} \simeq  v^{\rm base}_{2, \pm} \mp r A_{\rm ch}/2.  
\label{eq:slope}
\end{eqnarray} 
Such a flow splitting can be measured either for relatively lower beam
energy collisions where typical events have nonzero $A_{\rm ch}$ or
for high energy collision events that are binned and selected
according the events' charge asymmetry. Indeed both types of
measurements have been reported~\cite{STAR_CMW} and the predicted
splitting patterns have been quantitatively verified. Several
hydrodynamic-based simulations of CMW-induced flow splitting have been
done for computing the slope parameter
$r$~\cite{Hongo:2013cqa,Taghavi:2013ena,Yee2014} and possible origins
of the finite intercept $\Delta v^{\rm base}_{2}$ (at zero charge
asymmetry) have also been investigated~\cite{Stephanov:2013tga}. A
number of potential non-CMW
effects~\cite{Dunlop:2011cf,Xu:2012gf,Steinheimer:2012bn,Bzdak2013,Hatta:2015hca}
that may contribute to the observed flow splitting were proposed and
studied but so far there has been no compelling alternative
interpretation of data.  The important next step would be an anomalous
hydrodynamic simulation framework incorporating dynamical magnetic
fields as well as realistic charge initial conditions and evolutions
to fully quantify the CMW signals for data validation.

\subsection{Chiral Vortical Effect}

The phenomenology of the CVE in heavy-ion collisions is quite similar to that of CME. Provided nonzero average rotation $\vec {\bf \omega}$ of the QGP along the out-of-plane direction (in non-central collisions) together with nonzero background vector charge density (specifically considering baryon density $\mu_B$ from initial deposition), the CVE current $\vec{\bf J} $ can be generated by initial axial charge $\mu_5$ along the out-of-plane direction according to Eq.~(\ref{eq_CVE}). Therefore just as in the case of the CME, the (quark-level) CVE current leads to a separation of quarks and their anti-quarks across the reaction plane, with more quarks transported to one pole of the QGP fireball and more anti-quarks to the other pole. The CVE current, for example, can thus manifest itself through the baryonic charge separation (in analogy to the electric charge separation of the CME) $\frac{dN_{B/\bar{B}}}{d\phi} \propto ... + a_{B/\bar{B}} \sin(\phi-\Psi_{\rm RP})$, 
 which can be measured through baryon-number-dependent azimuthal correlations $\gamma_{BB/{\bar{B}\bar{B}}}$ and $\gamma_{B\bar{B}}$~\cite{Kharzeev:2010gr}. 

Such similarity between CME and CVE currents, both arising at quark level, actually brings a complication: the two effects, if both occurring, would ``mix up'' and can not be easily separated from each other. But this may turn out to be an opportunity, as proposed in Ref. \cite{Kharzeev:2010gr},  as different hadron-level observables come from different combinations of quark-level contributions. By examining the specific patterns of hadronic observables, one could identify the contributions from the CME and the CVE and thus verify both effects.  To see how this works, let us consider two observables, the baryon current $\vec{\bf J}_B$ and the electric charge current  $\vec{\bf J}_E$, with contributions from both the CME and the CVE. At quark level, each flavor develops a current given by: 
\begin{eqnarray}
\vec{\bf J}_f = \frac{N_c \mu_5}{2\pi^2} \left[ Q_f (e\vec{\bf B}) +   2 B_f (\mu_B \vec{\bf \omega}) \right].
\end{eqnarray}
It may be noted that a nonzero background electric charge density could also contribute a term $2Q_f (\mu_e \vec{\bf \omega})$ to the vortical effect which nevertheless might be small due to cancellation when summing over quark flavors. 
From Eq.~(\ref{eq_Q_B}),  one can construct the observable baryon and charge currents by summing over flavor. 
The results would however depend on the extent to which the strange quark current mass is negligible at relevant scale.  If one assumes  contributions from only two light flavors, the results are
\begin{eqnarray}
\vec{\bf J}_Q^{\, 2f} = \frac{N_c \mu_5}{2\pi^2} \left[ \frac{5}{9} (e\vec{\bf B}) +   \frac{2}{9} (\mu_B \vec{\bf \omega})  \right] \,\, , \,\, 
\vec{\bf J}_B^{\, 2f} = \frac{N_c \mu_5}{2\pi^2} \left[ \frac{1}{9} (e\vec{\bf B}) +   \frac{4}{9} (\mu_B \vec{\bf \omega})  \right].
\end{eqnarray} 
 If one assumes  contributions from all three flavors, then the results are
\begin{eqnarray}
\vec{\bf J}_Q^{\, 3f} = \frac{N_c \mu_5}{2\pi^2} \left[ \frac{2}{3} (e\vec{\bf B}) +  0\times (\mu_B \vec{\bf \omega})  \right] \,\, , \,\, 
\vec{\bf J}_B^{\, 3f} = \frac{N_c \mu_5}{2\pi^2} \left[ 0 \times (e\vec{\bf B}) +   \frac{2}{3} (\mu_B \vec{\bf \omega})  \right]. 
\end{eqnarray} 
Therefore, depending on the relative strength of the CME versus the
CVE and depending on strange flavor contributions, the ratio of charge
and baryon separation can vary significantly.  Furthermore one can
also construct the strangeness current as well as various other hadronic
currents which all have different patterns. By measuring these
observables one can hopefully separate the contributions from the CME
and the CVE, and decipher the extent to which strange flavor becomes
chiral in the QGP.  One can also study how the patterns of these
currents change with beam energy which shall considerably shift the
relative strength between the CME and the CVE.

%%% Local Variables: 
%%% mode: latex
%%% TeX-master: "master"
%%% End: 

%===============================================================
\section{Experimental Results}
\label{sec:3}

Since the first STAR publications~\cite{STAR_LPV2,STAR_LPV1}
presenting results qualitatively consistent with the CME  expectations, the search for the CME has been actively
pursued by several   major collaborations studying heavy-ion collisions, including 
STAR~\cite{STAR_LPV3,STAR_LPV_BES,STAR_LPV4,STAR_LPV_UU},
PHENIX~\cite{PHENIX_LPV1} and ALICE
collaborations~\cite{ALICE_LPV}. In addition, the correlation
observables sensitive to the CMW~\cite{STAR_CMW,Qi-ye,Belmont:2014lta} and
the CVE~\cite{Lambda_CVE} have been also studied.  In the absence of
detailed quantitative predictions for the anomalous chiral effects,
the experimental program is mostly driven by search for qualitative
features that might be difficult or impossible to explain by any
other ``background'' physics. Thus, the CME predicts electric
charge separation along the direction of the magnetic field, the CVE
similarly predicts  separation of  baryon number along the
direction of the orbital angular momentum, and there are no obvious reasons for
such a separation based on non-anomalous physics.  As the direction of
the magnetic field and the orbital angular momentum on average coincide and
are perpendicular to the reaction plane, the searches are for the
separation of the electric charge and baryon number with respect to
the reaction plane. Due to the fluctuating nature of the phenomena, the
direction of the separation also fluctuates, such that on average the
effect is zero. One can observe the charge separation only by means of
correlations, looking for correlated emission of particles with the same electric
charge or baryon number into similar azimuthal directions.

A ``drawback'' of the correlation observables is that they are \P- and
\CP-even, unlike the anomalous effects lying in their origin. This
opens a possibility for a background contribution that is totally
unrelated to the effects under study. These background effects are
often dominant, and a special care must be taken to suppress them. For
example, one ``obvious'' correlation to examine for the CME effect would
be to check if the particles of the same charge are preferentially
emitted in the same direction, while particles of the opposite charges 
are emitted in the opposite directions . But there exist many other
reasons for the particles to be correlated in azimuth: jets, radial
flow, resonance decays. The magnitudes of those background
correlations in general are larger than that due to the CME. These
background correlations have to be suppressed in order to gain access
to those possibly due to the CME.

An approach often used to suppress the background correlations is
based on the observation that background effects are mostly independent
from the orientation of the reaction plane, while the CME and other
effects are strongly correlated with the direction of the magnetic
field and thus with the reaction plane orientation. Then the
difference in ``projections'' of the correlations onto the reaction
plane and onto the plane perpendicular to the reaction plane would be
mostly free of the background and more sensitive to the
CME~\cite{Voloshin:2004vk}. In this way the background contribution
can be suppressed to a level close to the magnitude of the elliptic
flow $v_2$~\cite{Voloshin:2004vk}.  The same trick could be taken
further to test the remaining effect ---  by studying the ``projections'
onto the higher harmonic event planes one can try to figure out if the
signal is due to modulation in the background caused by anisotropic
flow or if the signal exhibits something special for the direction of
the magnetic field.

The search for the CMW, at first might look ``simpler'' compared with
those for the CME and the CVE -- while on average the net effect is still
zero, the predicted splitting in elliptic flow of positive and
negative particles uniquely depends on the charge asymmetry (net
vector charge) of the (sub)system. The effect still might be studied
only by many-particle correlations, three at minimum, with one
particle to be the probe, the second particle for an estimate of the
reaction plane, and the third one for an estimate of the system's net
vector (e.g. electric) charge.

Note that the analyses involving the reaction plane require
measurements of at least three particle correlations (with one particle
used for the determination of the reaction plane). For events with low
multiplicities (e.g. in very peripheral collisions) such correlations could  in principle have a noticeable
contribution from direct three particle decays or jets, and  special
precaution should be taken to suppress such a contribution.

In the following we discuss the experimental results on the charge
dependent correlations for the search of different anomalous
chiral effects, as well as measurements sensitive to the background. We also 
review the pertinent results from Bean Energy Scan at RHIC as well as the analysis of U+U collisions. The role of the local charge conservation combined with anisotropic flow in forming the background signal, the only real possible source of the charge-dependent background
identified so far, is discussed in a separate subsection.

Once the basic measurements of a possible signal of this or other
anomalous effect are performed and found to be qualitatively
consistent with expectations (and, remarkably they all are), the next
steps include cross comparison of different observables as well as variation of
conditions sensitively affecting the possible background or signal. Some of
these studies have been performed, while many more are still awaiting experimental efforts. We discuss the latter in the subsection on future measurements.

%----------------------------------------------------------------
\subsection{Chiral Magnetic Effect}
\label{subsec:CME}

\begin{figure}[h]
\begin{minipage}[c]{0.48\textwidth}
%\center
\includegraphics[width=\textwidth]{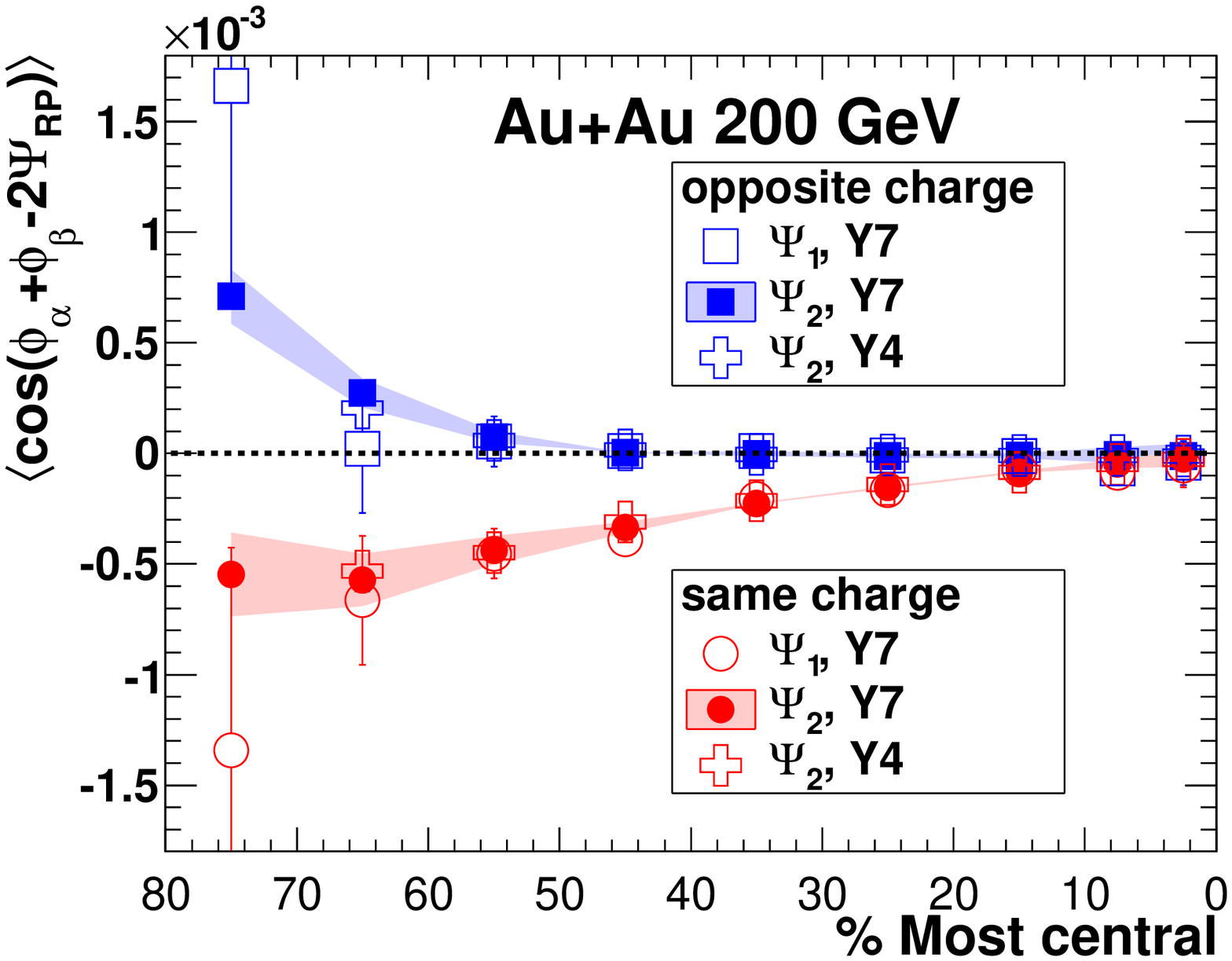}
  \caption{Three-point correlator, $\gamma$, measured with $1^{\rm
      st}$ and $2^{\rm nd}$ harmonic event planes versus centrality
    for Au+Au collisions at $\sqrt{s_{\rm NN}}=200$
    GeV~\cite{STAR_LPV3}.  Shown with crosses are STAR previous
    results from the 2004 RHIC run (Y4)~\cite{STAR_LPV1,STAR_LPV2}.  }
\label{fig:gamma}
\end{minipage}
\hspace{0.4cm}
\begin{minipage}[c]{0.48\textwidth}
%\center
\includegraphics[width=\textwidth]{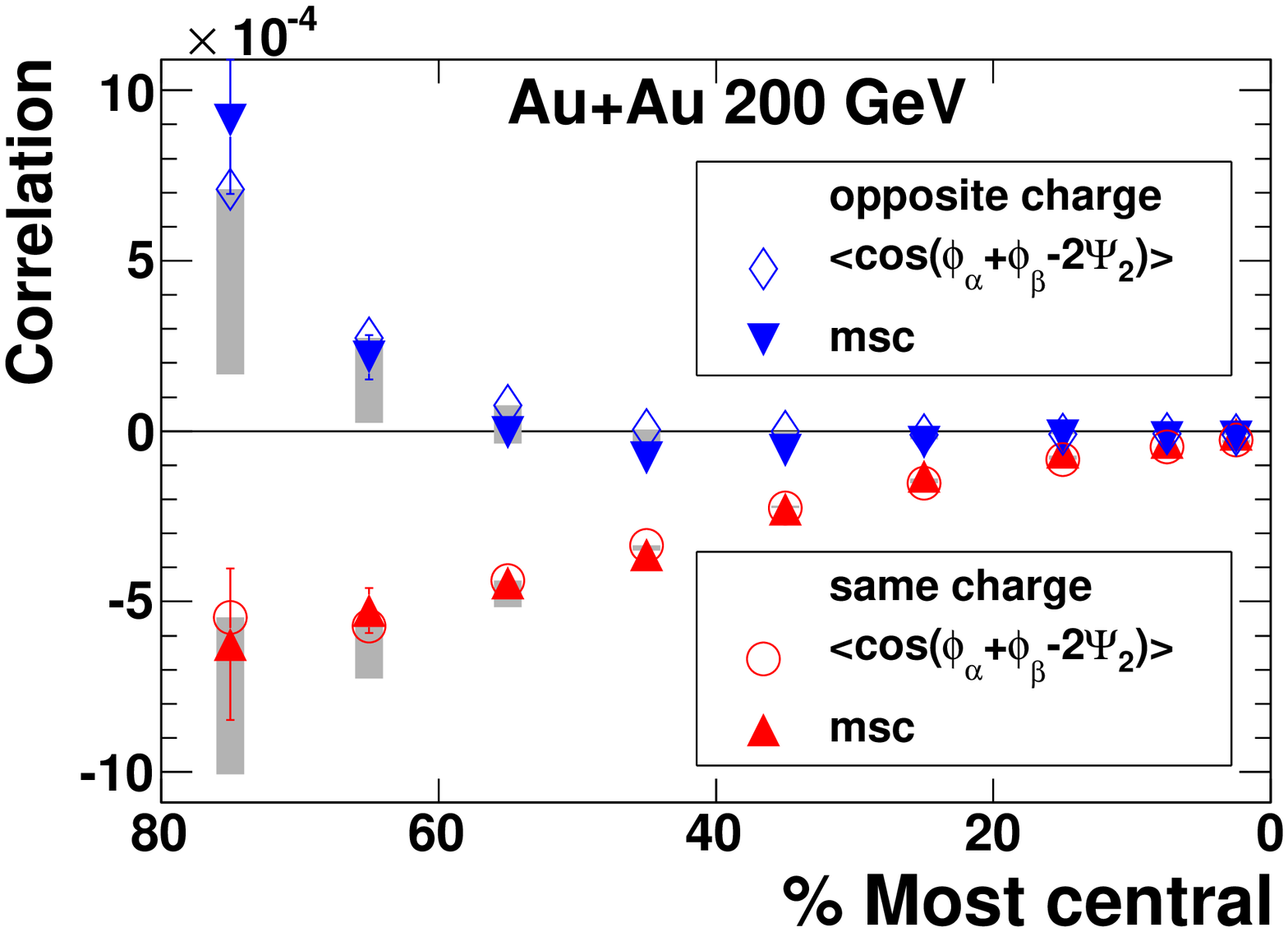}
  \caption{Modulated sign correlations (msc) compared to the
    three-point correlator versus centrality for Au+Au collisions at
    $\sqrt{s_{\rm NN}}=200$ GeV~\cite{STAR_LPV3}. The grey bars
    reflect the conditions of $\Delta p_T > 0.15$ GeV/$c$ and $\Delta
    \eta > 0.15$ applied to $\gamma$.  }
\label{fig:msc}
\end{minipage}
\vspace{-0.2cm}
\end{figure}

The STAR Collaboration first measured the $\gamma$ correlator for
Au+Au (shown with crosses in Fig.~\ref{fig:gamma}) and Cu+Cu
collisions at 62.4 and 200 GeV with data from the 2004/2005 RHIC
runs~\cite{STAR_LPV2,STAR_LPV1}.  The $2^{\rm nd}$ harmonic event
plane used in the correlator was reconstructed with the STAR Time
Projection Chamber (TPC)~\cite{STAR_TPC}.  The opposite charge
($\gamma_{\rm OS}$) and the same charge ($\gamma_{\rm SS}$)
correlations display the ``right" ordering, supporting the picture of
the CME.  Similar $\gamma$ results for 200 GeV Au+Au were observed by
the PHENIX Collaboration~\cite{PHENIX_LPV1}.  PHENIX also employed a
multiparticle charge-sensitive correlator, $C_c(\Delta
S)$~\cite{PHENIX_LPV2}, and their preliminary results showed a concave
$C_c(\Delta S)$ distribution~\cite{PHENIX_LPV1}, also evidencing the
charge separation effect.  To study the background from conventional
physics, Au+Au collisions were simulated with heavy-ion event
generators MEVSIM~\cite{MEVSIM}, UrQMD~\cite{URQMD}, and
HIJING~\cite{HIJING} (with and without an elliptic flow afterburner
implemented).  MEVSIM only includes correlations due to resonance
decays and an overall elliptic flow pattern.  UrQMD and HIJING are much more realistic and comprehensive simulation models of the collision, and they include correlations from
many different physical processes. No generator gives qualitative
agreement with data.

STAR has also analyzed Au+Au collision at 62 GeV as well as Cu+Cu
collisions at 200 and 62 GeV~\cite{STAR_LPV2,STAR_LPV1}. All the
results have been found to be in qualitative expectation with CME. The
opposite charge correlations in Cu+Cu collisions are stronger than
those in Au+Au, possibly reflecting the suppression of the
correlations among oppositely moving particles in a larger system.
STAR also presented $\pt$ and $\deta$ dependencies of the signal.  The
signal has a $\deta$ width of about one unit of rapidity, consistent
with small \P-odd domains. The signal is found to increases with the pair average  transverse
momentum, and it was later 
shown~\cite{Bzdak:2009fc} that the radial expansion can explain such a feature.

The charge-separation signal was cross-checked with data from the 2007
RHIC run (shown in Fig.~\ref{fig:gamma})~\cite{STAR_LPV3}.  The
$\gamma$ correlations from these data were measured with respect to both the $1^{\rm
  st}$ harmonic plane (of spectators at large rapidity) and the $2^{\rm nd}$ harmonic event planes at mid-rapidity.  Using the ZDC-SMD
first harmonic event plane determined by spectator neutrons ensures
that the signal is not coming from three-particle background
correlations, and is due to genuine correlations to the reaction
plane.  Another test was carried out by replacing one of the two
charged particles in $\gamma$ with a neutral particle, e.g. $K_S^0$,
and the results show no separation between $K_S^0-h^+$ and
$K_S^0-h^-$~\cite{Lambda_CVE}.  Thus the charge separation observed in
the $\gamma$ correlation between two charged particles is indeed due
to the electric charge.
To suppress the contribution from femtoscopic correlations, the
conditions of $\Delta p_T > 0.15$ GeV/$c$ and $\Delta \eta > 0.15$
were applied to the three-point correlator, shown with the grey bars
in Fig.~\ref{fig:msc}.  Excluding pairs with low relative momenta
significantly reduces the positive contributions to opposite charge
correlations in peripheral collisions, but the difference between same-
and opposite-charge correlations remains largely unchanged and
consistent with CME expectations.

The $\gamma$ correlator weights different azimuthal regions of charge
separation differently, e.g.~oppositely charged pairs emitted
azimuthally at $90^\circ$ from the event plane (maximally
out-of-plane) are weighted more heavily than those emitted only a few
degrees from the event plane (minimally out-of-plane).  It is a good
test to modify the $\gamma$ correlator such that all azimuthal regions
of charge separation are weighted identically.  This may be done by
first rewriting Eq.~(\ref{eq:obs1}) as
\be \langle
\cos(\phi_{\alpha}+\phi_{\beta}-2\Psi_{\rm RP}) \rangle = \langle
(M_{\alpha}M_{\beta}S_{\alpha}S_{\beta})_{\rm IN} \rangle - \langle
(M_{\alpha}M_{\beta}S_{\alpha}S_{\beta})_{\rm OUT} \rangle,
\label{eq:MMSS}
\ee
where $M$ and $S$ stand for the absolute magnitude ($0\leq M \leq 1$)
and sign ($\pm 1$) of the sine or cosine function, respectively.  IN
represents the cosine part of Eq.~(\ref{eq:obs1}) (in-plane) and OUT
represents the sine part (out-of-plane). A modulated sign correlation
(msc) is obtained by reducing the $\gamma$
correlator~\cite{STAR_LPV3}:
\be {\rm msc} \equiv \left(\frac{\pi}{4}\right)^2\left({\langle
  S_{\alpha}S_{\beta} \rangle_{\rm IN}-\langle
  S_{\alpha}S_{\beta}\rangle_{\rm OUT}}\right).
\label{eq:msc}
\ee
The modulated sign correlations are compared with the three-point
correlator for Au+Au collisions at 200 GeV in Fig.~\ref{fig:msc}.  It
is evident that the msc is able to reproduce the same trend as the
three-point correlator although their magnitudes differ slightly. STAR
also carried out another approach called the charge multiplicity
asymmetry correlation, whose methodology is similar to the msc, and
yielded very similar results~\cite{STAR_LPV4}.

\begin{figure}[h]
\includegraphics[width=\textwidth]{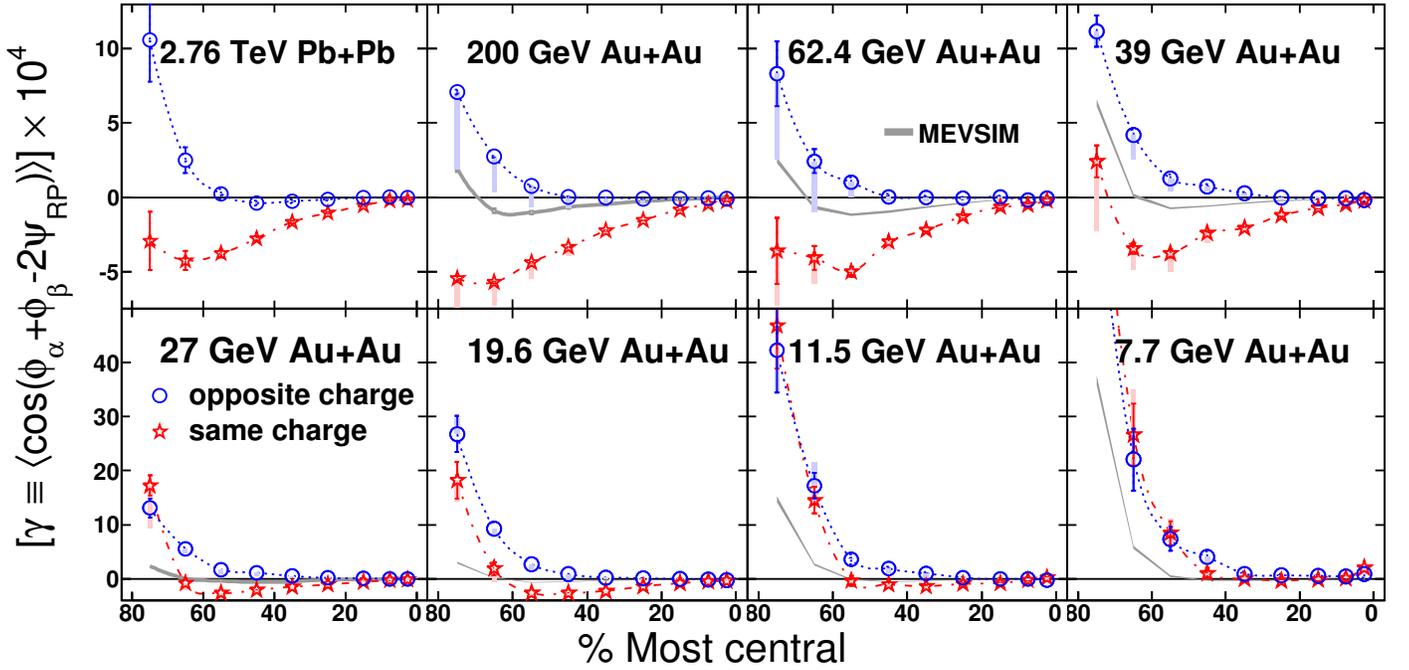}
  \caption{Three-point correlator as a function of centrality for
    Au+Au collisions at 7.7-200 GeV~\cite{STAR_LPV_BES}, and for Pb+Pb
    collisions at 2.76 TeV~\cite{ALICE_LPV}.  Note that the vertical
    scales are different for different rows.  The systematic errors
    (grey bars) bear the same meaning as in Fig.~\ref{fig:msc}.
    Charge independent results from the model calculations of
    MEVSIM~\cite{MEVSIM} are shown as grey curves.}
\label{fig:LPV_BES}
\end{figure}

A further understanding of the origin of the observed charge
separation could be achieved with a study of the beam-energy
dependence of the $\gamma$ correlation.  The charge separation effect
depends strongly on the formation of the quark-gluon plasma and chiral
symmetry restoration~\cite{Kharzeev:2007jp}, and the signal can be
greatly suppressed or completely absent at   low collision energies where a QGP has significantly shortened lifetime or not even formed. Taking into account that the life-time
of the strong magnetic field is larger at smaller collision energies,
this could lead to an almost threshold effect: with decreasing collision energy, 
the signal might slowly increase with an abrupt drop thereafter.
Unfortunately, the exact energy dependence of the chiral magnetic
effect is not calculated yet.

The question of the collision-energy dependence of the $\gamma$
correlator has been addressed during the recent RHIC Beam Energy Scan.
Figure~\ref{fig:LPV_BES} presents $\gamma_{\rm OS}$ and $\gamma_{\rm
  SS}$ correlators as a function of centrality for Au+Au collisions at
$\sqrt{s_{NN}} = 7.7 - 200$ GeV measured by STAR~\cite{STAR_LPV_BES},
and for Pb+Pb collisions at 2.76 TeV by ALICE~\cite{ALICE_LPV}.  The
difference between $\gamma_{\rm OS}$ and $\gamma_{\rm SS}$ seems to
vanish at low collision energies, again qualitatively in agreement
with expectations for the CME. At most collision energies, the
difference between $\gamma_{\rm OS}$ and $\gamma_{\rm SS}$ is still
present with the ``right" ordering, manifesting extra
charge-separation fluctuations perpendicular to the reaction plane.
With decreased beam energy, both $\gamma_{\rm OS}$ and $\gamma_{\rm
  SS}$ tend to rise up starting from peripheral collisions.  This
feature seems to be charge independent, and can be explained by
momentum conservation and elliptic flow~\cite{STAR_LPV3}.  Momentum
conservation forces all produced particles, regardless of charge, to
separate from each other, while collective flow works in the opposite
sense. For peripheral collisions, the multiplicity ($N$) is small, and
momentum conservation dominates. The lower beam energy, the smaller
$N$, and the higher $\gamma_{\rm OS}$ and $\gamma_{\rm SS}$.  For more
central collisions where the multiplicity is large enough, this type
of ${\cal P}$-even background can be estimated with
$-v_2/N$~\cite{STAR_LPV3,v2N}.  In Fig.~\ref{fig:LPV_BES}, we also
show the MEVSIM~\cite{MEVSIM} model calculations with implementation
of elliptic flow and momentum conservation, which qualitatively
describe the beam-energy dependence of the charge-independent
background.

%-------------------------------------------------
\subsubsection{CME background studies}

The ambiguity in the interpretation of experimental results comes from
a possible background of (the reaction plane dependent) correlations
not related to the CME. As illustrated in Fig.~3 of
Ref~\cite{STAR_LPV_BES}, the two-particle correlator, $\delta \equiv
\langle \cos(\phi_\alpha -\phi_\beta) \rangle$, which in the absence
of any other correlations except the CME should be proportional to
$\langle a_{\alpha} a_{\beta} \rangle$, shows the ``wrong"
ordering.  That indicates the existence of an overwhelming background
in $\delta$ over any possible CME effect. In $\gamma$ correlator those
background correlations are strongly suppressed (at the level of
$v_2$) but still might be significant. The fact that no event
generator can explain the data says that either the experimental
results are indeed due to the CME, or that all existing event generators
do not include all the possible physics. There exist several attempts
to identify the physics which might be responsible for the
experimental observations.
The most notable in this respect is the
paper~\cite{Schlichting:2010qia} where the authors show that the
difference between the same- and opposite-charge correlations as
measured by STAR can be explained within a Blast Wave model that
includes charge conservation along with radial and elliptic flow with
parameters tuned to the data.

Local charge conservation (LCC) assumes that the pairs of opposite charges
are created very close in space at the late stage of the system
evolution with developed anisotropic flow. Radial boost of the pair
due to transverse expansion leads to particle collimation in azimuth
and pseudorapidity~\cite{Voloshin:2003ud,Voloshin:2004th}. Then, due
to elliptic flow, opposite-charge pairs became stronger correlated
in-plane than out-of-plane, which causes splitting in value of
$\gamma$ correlator between same- and opposite-charge
pairs~\cite{Schlichting:2010qia} as observed in the data. While
in~\cite{Schlichting:2010qia} the authors were able to describe the
data rather closely, there exist many questions to this particular
analysis. Firstly we note that the local charge conservation (LCC) mechanism leads to strong
correlation between opposite charge pairs, while experimentally
$\gamma_{+-}$ is very close to zero. There should be another
charge-independent correlation mechanism (momentum conservation is
often discussed for that) which almost exactly compensates for the
effect of the LCC. We also note that the parameters of the Blast Wave
model in~\cite{Schlichting:2010qia} were {\em tuned} to the charge
balance function in $B(\dphi)$, the detailed shape of
which by itself could be influenced by the CME.  
 Further study of the background effects within this
  approach would be very useful.

\begin{figure}[h]
\begin{minipage}[c]{0.48\textwidth}
%\center
\includegraphics[width=\textwidth]{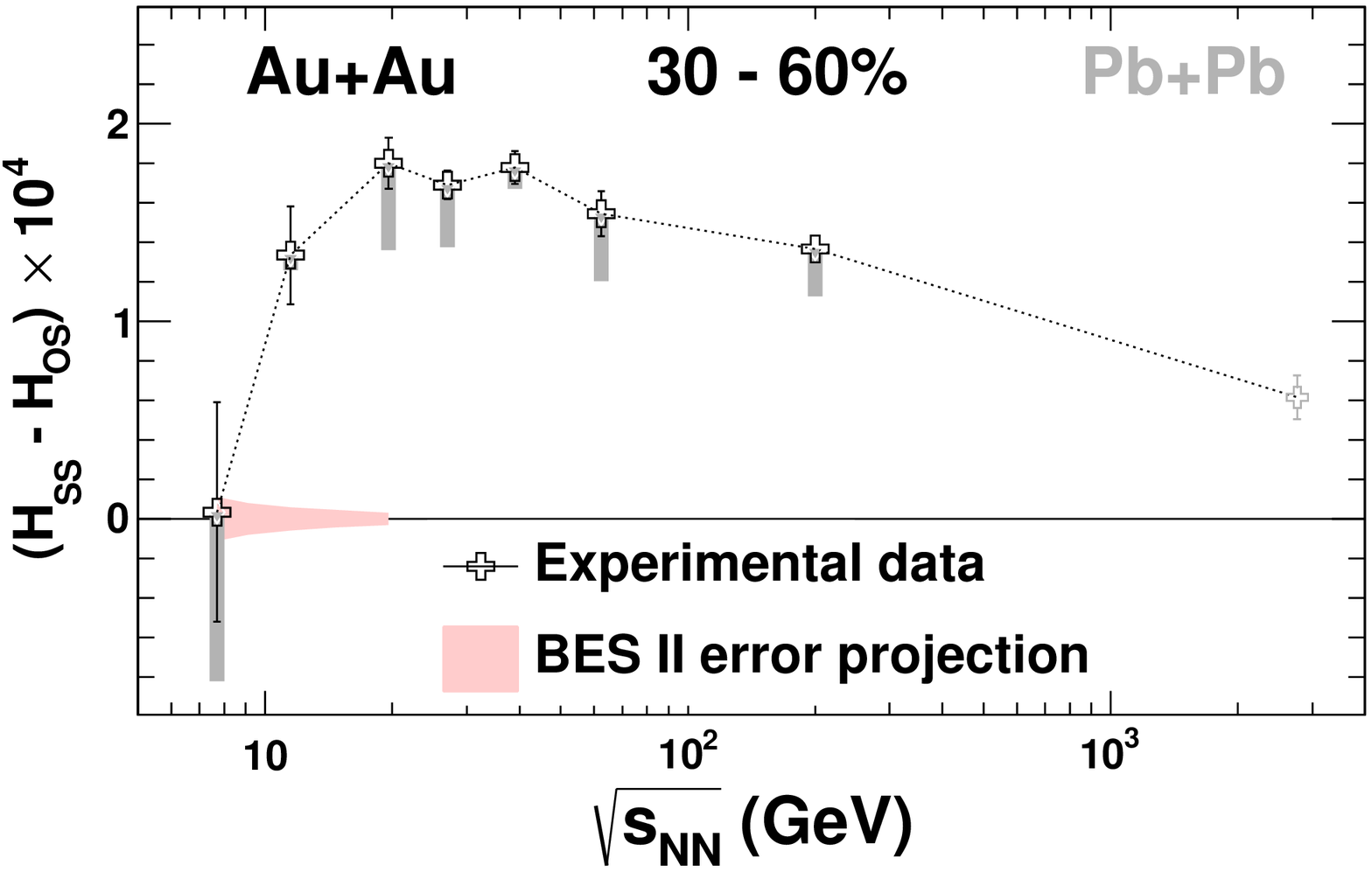}
  \caption{The modified correlator
    ($H^{\kappa=1}_{SS}-H^{\kappa=1}_{OS}$) as a function of beam
    energy for $30-60\%$ Au+Au (Pb+Pb)
    collisions~\cite{STAR_LPV_BES,ALICE_LPV}.  The systematic errors
    (grey bars) bear the same meaning as in Fig.~\ref{fig:msc}.}
\label{fig:H_BES}
\end{minipage}
\hspace{0.4cm}
\begin{minipage}[c]{0.48\textwidth}
%\center
\includegraphics[width=\textwidth]{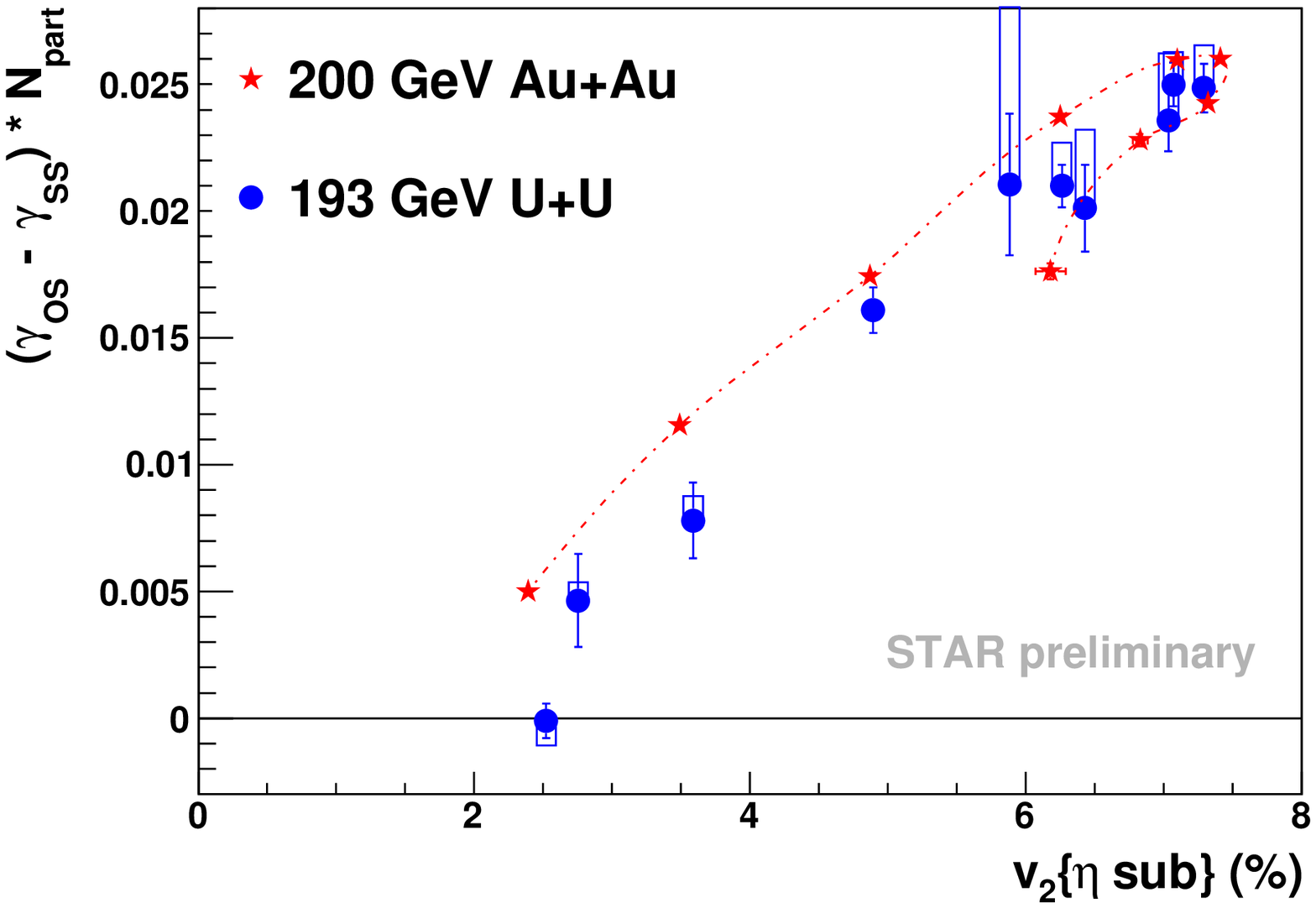}
  \caption{$(\gamma_{\rm OS}-\gamma_{\rm SS}) \times N_{part}$ vs
    $v_2$ for Au+Au collisions at 200 GeV and U+U collisions at 193
    GeV~\cite{STAR_LPV_UU}.  The open box represents the systematic
    uncertainty due to the tracking capability under the high
    luminosity in RHIC year 2012.  }
\label{fig:LPV_UU}
\end{minipage}
\vspace{-0.2cm}
\end{figure}

To single out background contributions, one can take the approximation in Eq.~(\ref{eq:gd}) and solve for $H$, the
CME signal strength,
\be H^\kappa = (\kappa v_2
\delta - \gamma)/(1+\kappa v_2).  \ee
According to Ref.~\cite{Flow_CME}, $\kappa$ is close to, but deviates from, unity owing to the finite detector acceptance.
Figure~\ref{fig:H_BES} shows $(H^{\kappa=1}_{\rm SS}-H^{\kappa=1}_{\rm
  OS})$ as a function of beam energy for $30-60\%$ Au+Au (Pb+Pb)
collisions~\cite{STAR_LPV_BES,ALICE_LPV}. In the case of unity
$\kappa$, $(H_{\rm SS} - H_{\rm OS})$ demonstrates a weak energy
dependence above 19.6 GeV, and tends to diminish from 19.6 to 7.7 GeV,
though the statistical errors are large for 7.7 GeV.  This may be
explained by the probable dominance of hadronic interactions over
partonic ones at low energies.  
 As   discussed in the previous Section, the parameter
  $\kappa$ in reality can take rather different values for different
  sources of background correlations that may exhibit varied dependence on collision energy and centrality.  A more definitive conclusion may be
reached with a more accurate  and reliable estimation of $\kappa$ and with higher
statistics at lower energies in the proposed phase II of the RHIC Beam
Energy Scan program, as illustrated by the shaded band in
Fig.~\ref{fig:H_BES}.

The prolate shape of the uranium nuclei could yield a sizable initial
eccentricity (and large elliptic flow) even in fully overlapping U+U
collisions, where the magnetic field is minimal.  Such collisions
would be dominated by the background effects and can be used as an
additional test for the nature of the charge dependent
correlations~\cite{Voloshin:2010ut}. Figure~\ref{fig:LPV_UU} shows
$(\gamma_{\rm OS} - \gamma_{\rm SS})$ multiplied by the number of
participants, $N_{\rm part}$, vs $v_2$ for different centralities in
193 GeV U+U and 200 GeV Au+Au collisions~\cite{STAR_LPV_UU}.  $N_{\rm
  part}$ was used to compensate for the dilution effect, which is due
to multiple sources involved in the collision~\cite{STAR_LPV2} .  In
both U+U and Au+Au, the signal roughly increases with $v_2$.  The
centrality trigger in U+U collisions helps to select the most central events for disentangling the
background contribution from the signal, since the magnetic field will
be greatly suppressed and the measurement will be dominated by the
$v_2$-related background.  As a result, in $0$-$1\%$ most central U+U
collisions the signal disappears as expected by the CME, while $v_2$
is still $\sim 2.5\%$. This demonstrates the smallness of the
background, and presents a challenge to the LCC interpretation. A phenomenological study for extrapolating 
both signal and background from AuAu to 
UU collisions was done in~\cite{Bloczynski:2013mca}.

\begin{figure}[htbp]
\begin{center}
\includegraphics[width=0.48\textwidth]{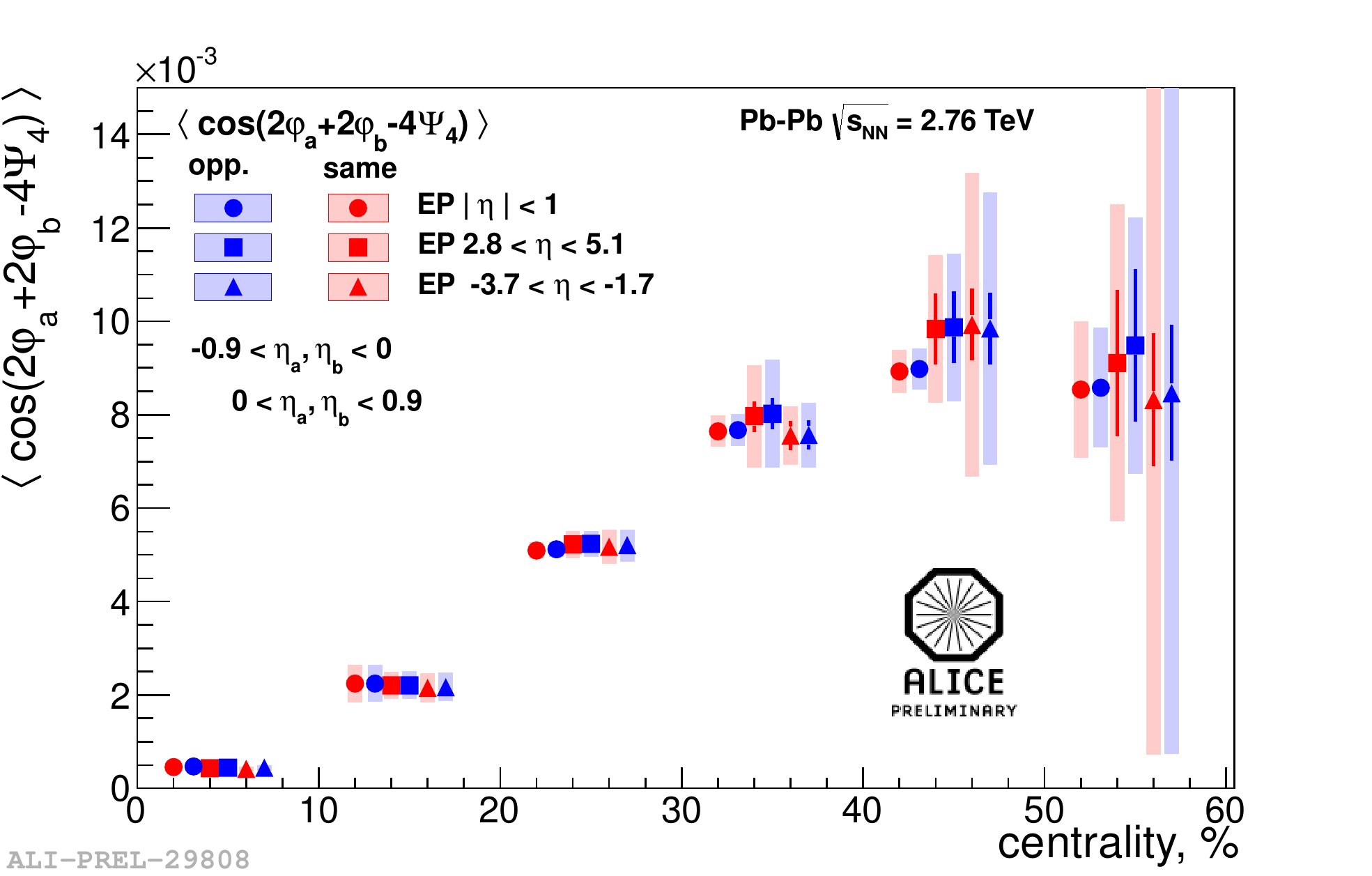}
\includegraphics[width=0.48\textwidth]{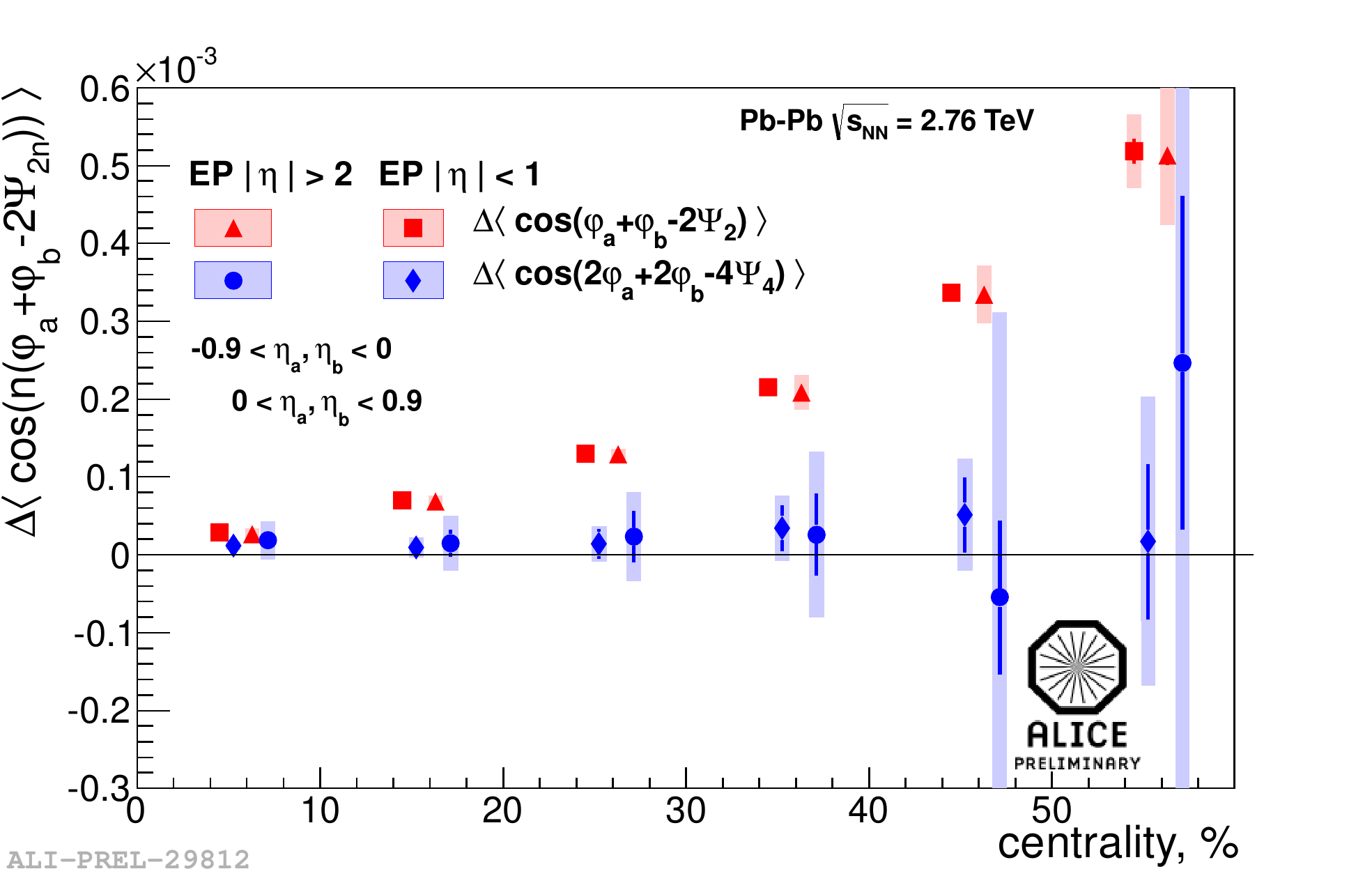}
\end{center}
\vspace*{-4mm}
\caption{(Left panel) Same-charge and opposite-charge pair  correlations relative to the
fourth harmonic event plane as a function of centrality. (Right panel)
Comparison of the charge-dependent parts in correlations with respect
to the second and fourth harmonic event planes. }
\label{fig:jm}
\end{figure}

Another measurement that can clarify the origin of the charge
dependent correlations and the role of the LCC
was suggested in~\cite{Voloshin:2011mx}: the correlations measured
with respect to the fourth harmonic event plane, $
\mean{\cos(2\phi_\alpha+2\phi_\beta-4\Psi_4)}$, should not contain any
contribution from the CME, but it should include the effect of
the LCC. The correlations due to the LCC in this case are expected to be somewhat smaller in
magnitude as the fourth harmonic flow is not that strong as the
elliptic flow. The preliminary results of such measurements are
presented in Fig.~\ref{fig:jm}~\cite{Voloshin:2012fv} with the charge-dependent part shown in the right panel. The correlations
relative to the fourth harmonic event plane are very weak and
suggestive of small contribution from the LCC, but
the detailed blast wave simulation has to be performed to draw more
definite conclusion from this measurement.

%---------------------------------------------------------------
\subsection{Chiral Magnetic Wave}

The chiral magnetic wave (CMW) manifests itself in a finite electric
quadrupole moment of the collision system, when the ``poles" of the
produced fireball acquire, depending on the net charge of the system,
excess of either positive or negative charge, with the opposite charge
concentrated on the ``equator''.  This effect, if present, will lead
to charge-dependent elliptic flow.  Taking pions as an example, the
CMW will lead to~\cite{CMW_calc_Burnier}
\be v_2(\pi^\pm) = v_2^{\rm
 base}(\pi^\pm) \mp r A_{\rm ch}/2,
\label{eq:v2_slope}
\ee 
where $v_2^{\rm base}(\pi^\pm)$ is the presumably charge independent,
``baseline'', elliptic flow, $A_{\rm ch} = (N_+ - N_-)/(N_+ + N_-)$ is
the charge asymmetry of the event, and $r$ is the coefficient
reflecting the strength of the CMW.  As $\langle A_{\rm ch} \rangle$
is always positive in heavy-ion collisions, the $A_{\rm ch}$-integrated $v_2$ of $\pi^-$
($\pi^+$) should be above (below) the baseline owing to the CMW.
However, the baseline $v_2$ may be different for $\pi^+$ and $\pi^-$
because of several other physics
mechanisms~\cite{Dunlop:2011cf,Xu:2012gf}, so it is less ambiguous to
study the CMW via the $A_{\rm ch}$ dependence of pion $v_2$ than via
the $A_{\rm ch}$-integrated $v_2$.

\begin{figure}[H]
\center
\includegraphics[width=0.8\textwidth]{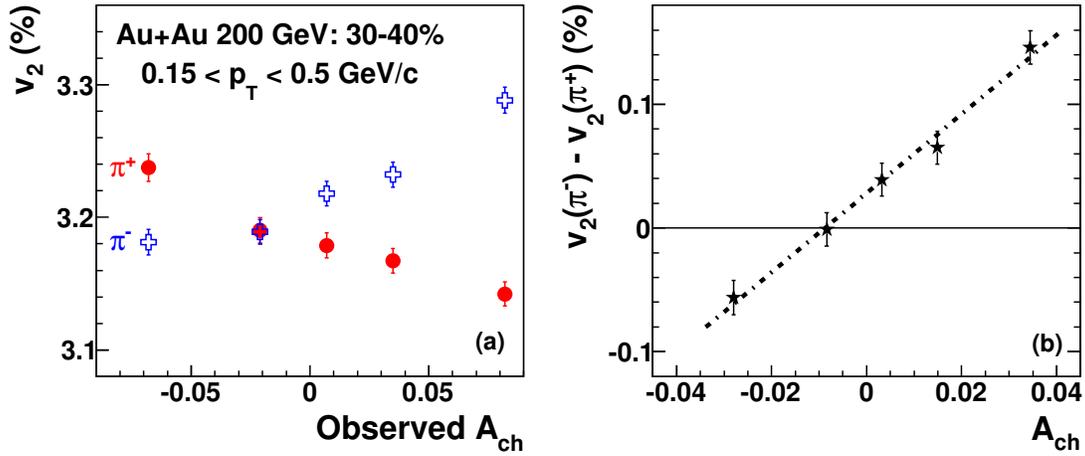}
  \caption{(a) pion $v_2$ as a function of observed charge asymmetry
    and (b) $v_2$ difference between $\pi^-$ and $\pi^+$ as a function
    of charge asymmetry with the tracking efficiency correction, for
    $30$-$40\%$ Au+Au collisions at 200 GeV~\cite{STAR_CMW}.}
\label{fig:v2_A}
\end{figure}

Taking $30$-$40\%$ 200 GeV Au+Au as an example, pion $v_2$ is shown as
a function of $A_{\rm ch}$ in panel (a) of
Fig.~\ref{fig:v2_A}~\cite{STAR_CMW}. $\pi^-$ $v_2$ increases with
$A_{\rm ch}$ while $\pi^+$ $v_2$ decreases with a similar magnitude of
the slope.  Note that $v_2$ was integrated over a narrow low $p_T$
range ($0.15 < p_T < 0.5$ GeV/$c$) to focus on the soft physics of the
CMW. Such a $p_T$ selection also ensures that the $\langle p_T \rangle$
is independent of $A_{\rm ch}$ and is the same for $\pi^+$ and
$\pi^-$, so that the $v_2$ splitting is not a trivial effect due to
the $\langle p_T \rangle$ variation.  The $v_2$ difference between
$\pi^-$ and $\pi^+$ is fitted with a straight line in panel (b). The
slope parameter $r$ is positive, qualitatively consistent with the
expectation of the CMW picture.  The fit function is non-zero at
$\langle A_{\rm ch} \rangle$ (i.e. the event-average value in given centrality class), indicating the $A_{\rm ch}$-integrated
$v_2$ for $\pi^-$ and $\pi^+$ are different, which was also observed
in~\cite{BESv2_PID}.

\begin{figure}[h]
\includegraphics[width=\textwidth]{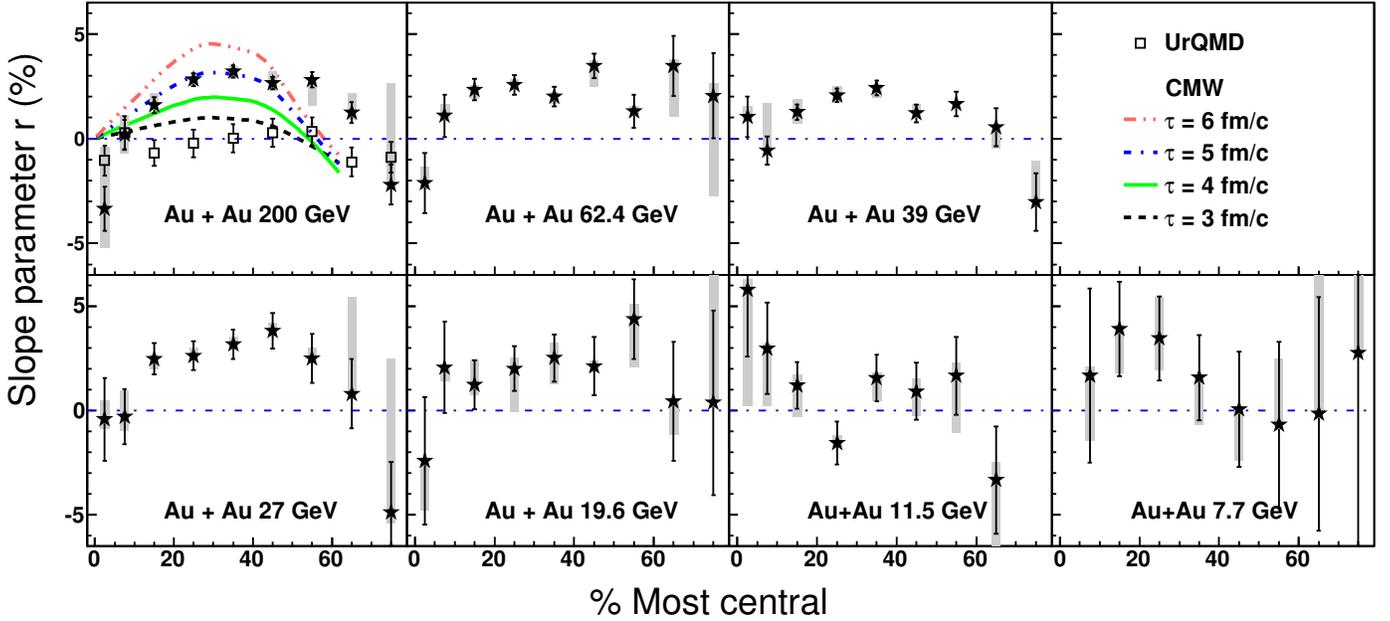}
  \caption{The slope parameter $r$ as a function of centrality for
    Au+Au collisions at 7.7-200 GeV~\cite{STAR_CMW}.  The grey bands
    include the systematic errors due to the track selection cut, the tracking
    efficiency and the $p_T$ range of particles involved in the event
    plane determination.  For comparison, we also show the UrQMD
    calculations~\cite{URQMD} and the calculations of the
    CMW~\cite{CMW_calc_Burnier} with different magnetic field
    duration times.  }
\label{fig:CMW_BES}
\end{figure}

The same procedure as above was followed to retrieve the slope
parameter $r$ as a function of centrality for Au+Au collisions at 200,
62.4, 39, 27, 19.6, 11.5 and 7.7 GeV, as shown in
Fig.~\ref{fig:CMW_BES}~\cite{STAR_CMW}.  A similar rise-and-fall trend
is observed in the centrality dependence of the slope parameter for
all the beam energies except 11.5 and 7.7 GeV, where the slopes are
consistent with zero with large statistical uncertainties.  It was
argued~\cite{Dunlop:2011cf} that at lower beam energies the $A_{\rm
  ch}$-integrated $v_2$ difference between particles and
anti-particles can be explained by the effect of quark transport from
the projectile nucleons to mid-rapidity, assuming that the $v_2$ of
transported quarks is larger than that of produced ones. The same
model, however, when used to study $v_2(\pi^-) - v_2(\pi^+)$ as a
function of $A_{\rm ch}$, suggested a negative slope~\cite{Campbell},
which is in contradiction with data.
Charge dependence of the elliptic flow on the event charge asymmetry
was confirmed by preliminary ALICE results for Pb+Pb collisions at
2.76 TeV~\cite{Belmont:2014lta}.

Recently a more realistic implementation of the CMW~\cite{Yee2014}
confirmed that the CMW contribution to $r$ is sizable, and that
the centrality dependence of $r$ is qualitatively similar to the data.
A quantitative comparison between data and theory requires further
work on both sides to match the kinematic regions used in the
analyses.

One drawback of the measurement of $v_2(A_{\rm ch})$ is that the
observed $A_{\rm ch}$ requires a correction factor due to the finite
detector tracking efficiency, as well as dependence on a particular
experimental acceptance.  A novel
correlator that is independent of
efficiency 
was proposed  in the following cumulant form~\cite{Voloshin:2014gja}:
\be \langle \langle \cos[n(\phi_1-\phi_2)]q_3 \rangle
\rangle = \langle \cos[n(\phi_1-\phi_2)]q_3 \rangle - \langle
\cos[n(\phi_1-\phi_2)] \rangle \langle q_3 \rangle_1.
\label{DAQ}
\ee
Here $\phi_1$ and $\phi_2$ are the azimuthal angles of particles 1 and
2, and $q_3$ is the charge ($\pm1$) of particle 3. $\mean{q_3}_1$ is
the average charge of particle ``3'' under condition of observing a
particle ``1'' of a particular charge (whereas the particle 2 is all inclusive regardless of charge).   The $\cos[n(\phi_1-\phi_2)]$ part was estimated by
ALICE~\cite{Belmont:2014lta} using the cumulant method and denoted as
$c_n\{2\}$.  In the absence of charge dependent correlations, the
correlator should be equal to zero.  Note that when the charge of the
third particle is averaged over all particles in the event (in a
specified kinematic region), the mean is equal to the charge
asymmetry, i.e. $\langle q_3 \rangle = A_{\rm ch}$.

\begin{figure}[h]
\begin{minipage}[c]{0.48\textwidth}
%\center
\includegraphics[width=\textwidth]{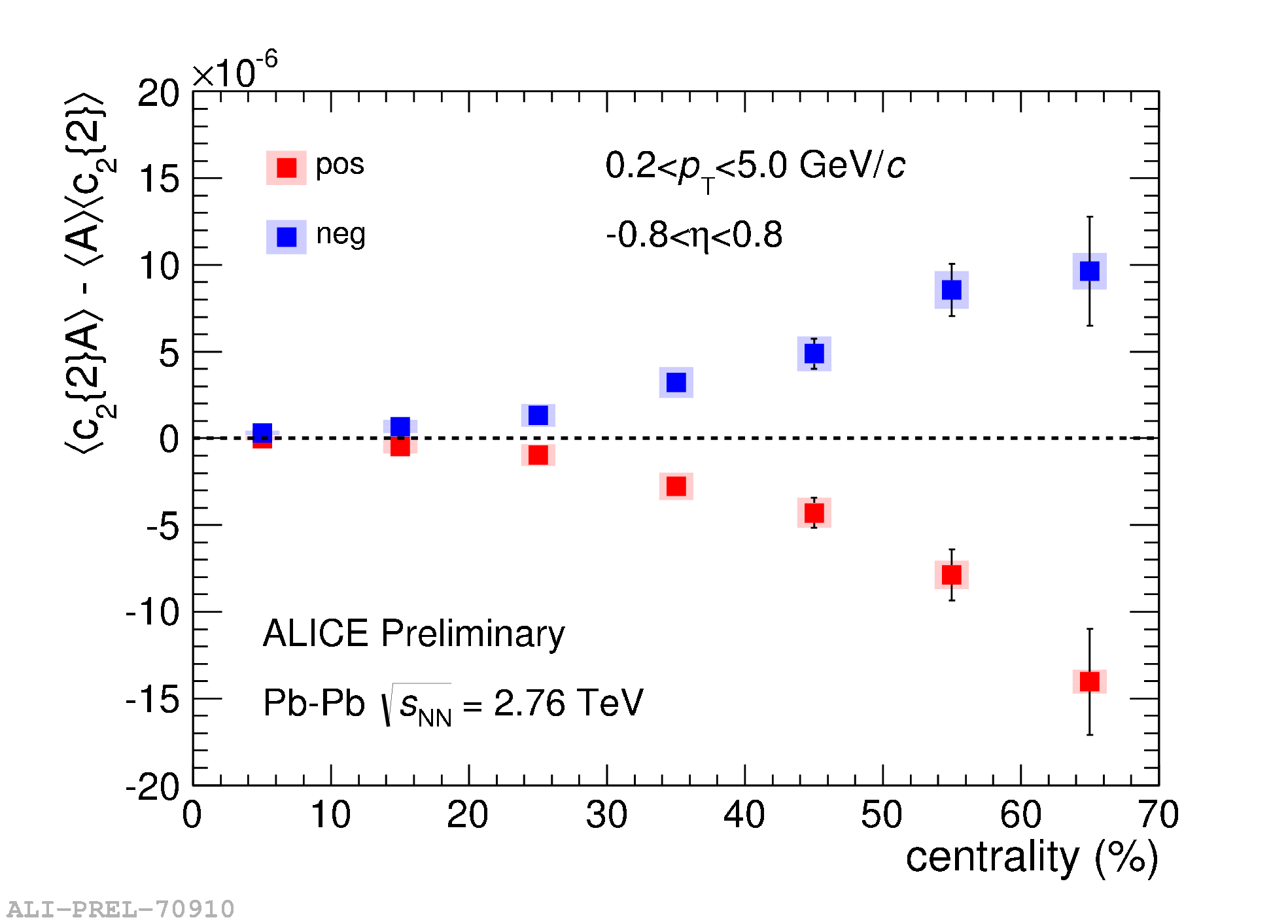}
  \caption{The correlation of $(\langle c_2\{2\}A_{\rm ch} \rangle -
    \langle A_{\rm ch} \rangle \langle c_2\{2\} \rangle)$ as a
    function of centrality in Pb+Pb collisions at 2.76 TeV~\cite{Belmont:2014lta}.  }
\label{fig:threepart_DAQ22}
\end{minipage}
\hspace{0.4cm}
\begin{minipage}[c]{0.48\textwidth}
%\center
\includegraphics[width=\textwidth]{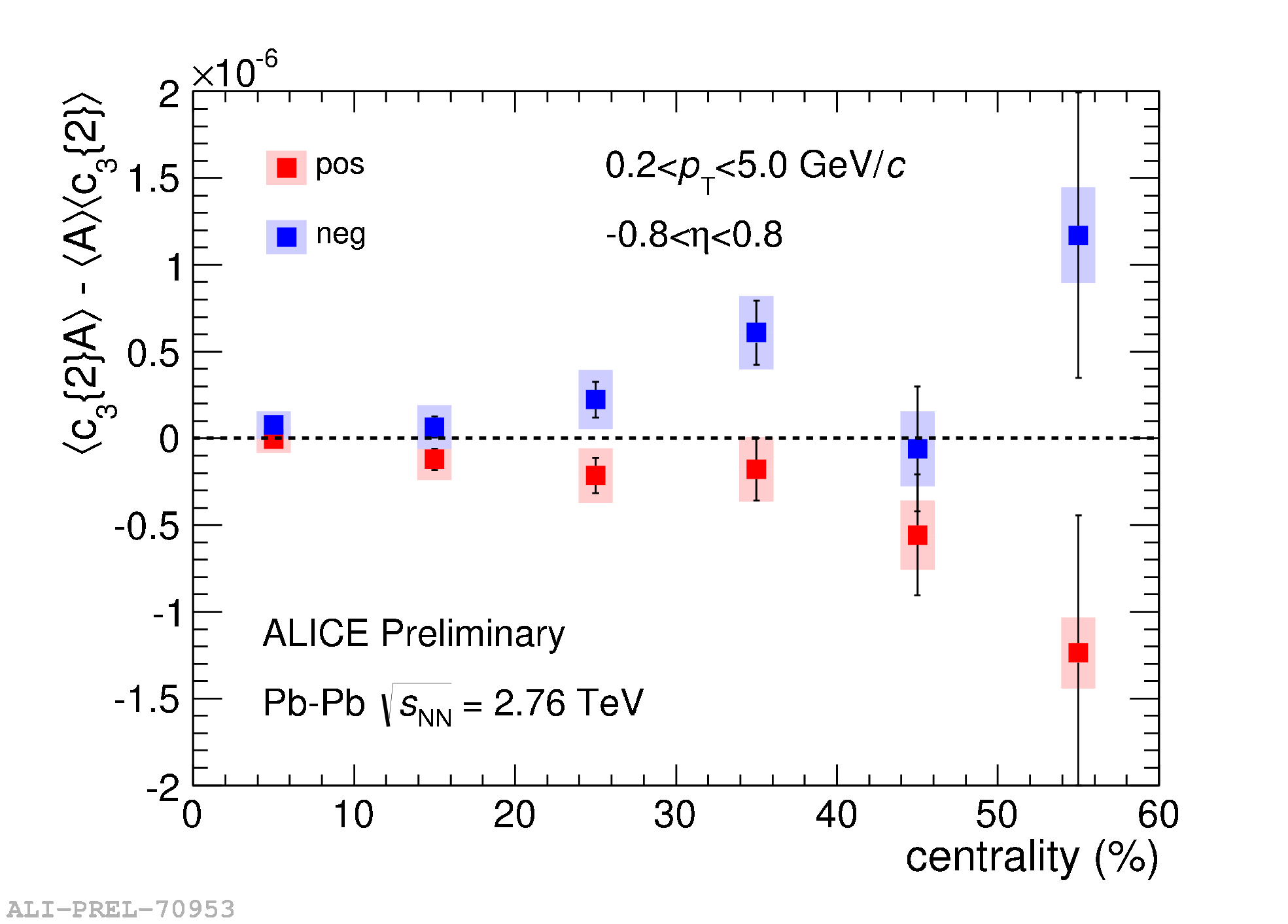}
  \caption{The correlation of $(\langle c_3\{2\}A_{\rm ch} \rangle -
    \langle A_{\rm ch} \rangle \langle c_3\{2\} \rangle)$ as a
    function of centrality in Pb+Pb collisions at 2.76 TeV~\cite{Belmont:2014lta}.  }
\label{fig:threepart_DAQ32}
\end{minipage}
\vspace{-0.2cm}
\end{figure}

Figs.~\ref{fig:threepart_DAQ22} and ~\ref{fig:threepart_DAQ32} show
the three-particle correlator (as in Eq.~(\ref{DAQ})) for the second and
third harmonic, respectively, as a function of centrality in Pb+Pb
collisions at 2.76 TeV~\cite{Lambda_CVE}.  For the second
harmonic, the correlation is charge dependent, with an ordering that
supports the CMW picture. The third harmonic correlator is much
  weaker (see discussion of the background effects below).

\begin{figure}[h]
\begin{minipage}[c]{0.48\textwidth}
%\center
\includegraphics[width=\textwidth]{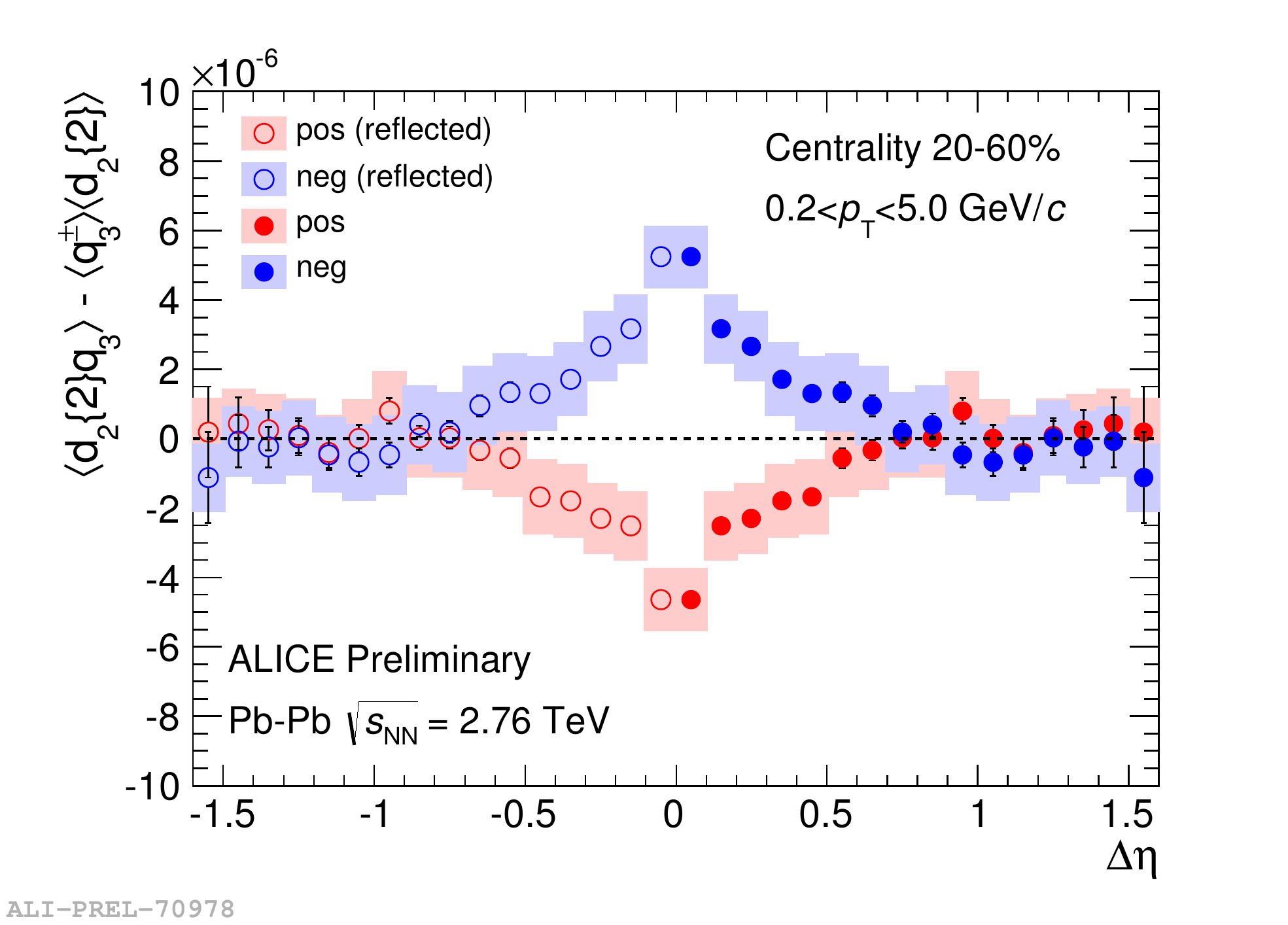}
  \caption{
Second harmonic differential three particle correlator for 20--60\%
    centrality
 Pb+Pb collisions at 2.76 TeV~\cite{Belmont:2014lta}.  
}
\label{fig:3pcd22}
\end{minipage}
\hspace{0.4cm}
\begin{minipage}[c]{0.48\textwidth}
%\center
\includegraphics[width=\textwidth]{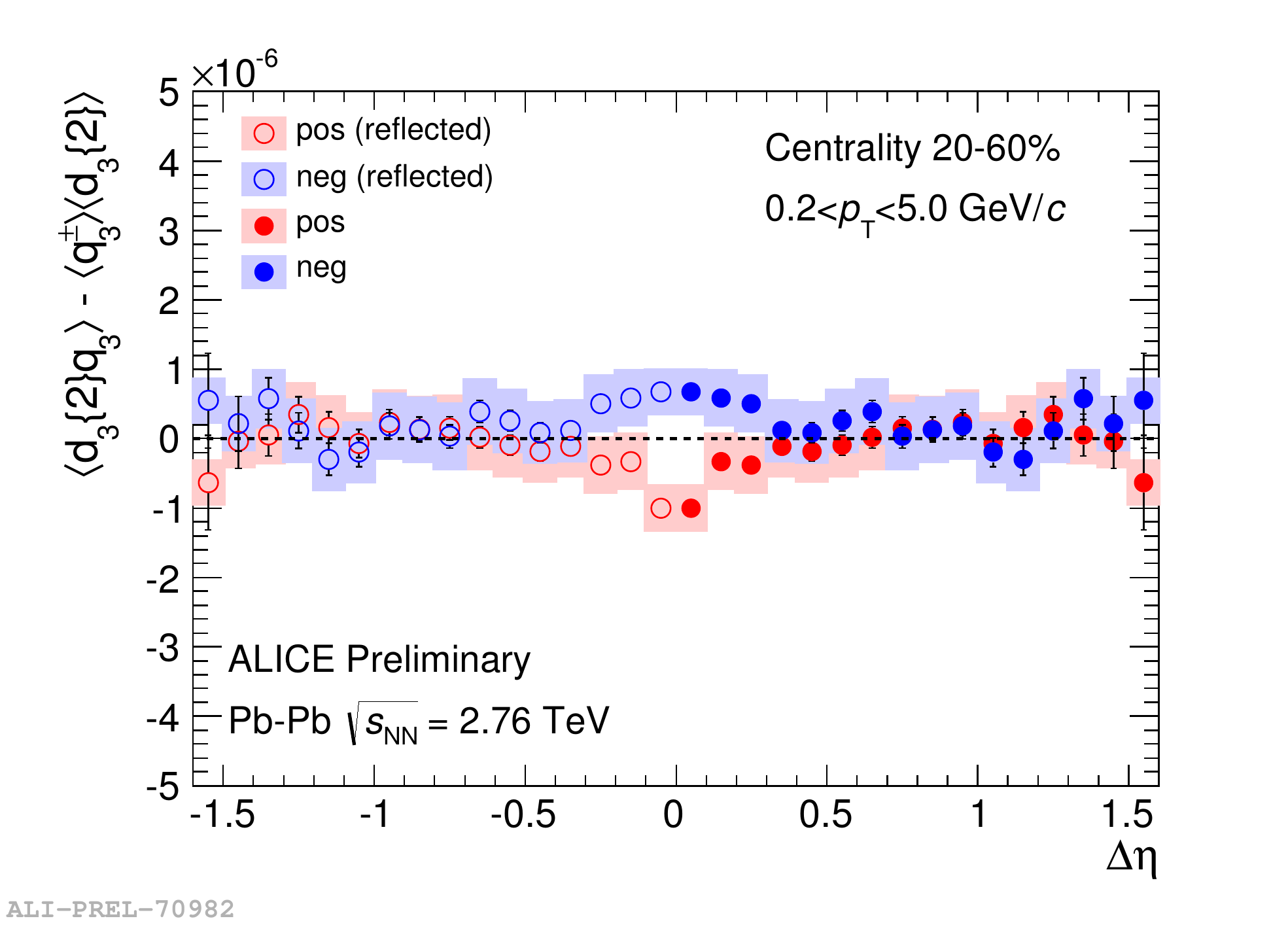}
  \caption{Third harmonic differential three particle correlator for 20--60\%
    centrality
 Pb+Pb collisions at 2.76 TeV~\cite{Belmont:2014lta}.  }
\label{fig:3pcd32}
\end{minipage}
\vspace{-0.2cm}
\end{figure}

The main advantage of the new 3-particle correlator is a possibility
of differential studies, e.g. as a function of pseudorapidity
difference between particles ``1'' and ``3''. Such measurements, see
Figs.~\ref{fig:3pcd22} and~\ref{fig:3pcd32} for ALICE preliminary
results~\cite{Belmont:2014lta}, should be significantly more informative
about the nature of the correlations and will help to identify the
background effects.

%-----------------------------------------------------
\subsubsection{CMW background studies}

Note that more recent theoretical calculations of the CMW effect are
rather controversial -- they range from a very modest signal that
would be difficult to detect~\cite{Hongo:2013cqa,Taghavi:2013ena} to a
``full strength'' effect~\cite{Yee2014} explaining the entire
signal measured by STAR.  To check if the observed slope parameters
come from conventional physics, the same analysis of
UrQMD~\cite{URQMD} events was carried out.  For Au+Au collisions at
200 GeV, the slopes extracted from UrQMD events are consistent with
zero for the $10$-$70\%$ centrality range, where the signal from the
real data is prominent, see Fig.~\ref{fig:CMW_BES}.  Similarly, the
AMPT event generator~\cite{AMPT} also yields slopes consistent with
zero (not shown here).

Naively, it is very difficult to imagine any background mechanisms
that would lead to the dependence observed by STAR and there exist a
very limited number of possible explanations proposed so far.  Perhaps
the most interesting of these explanations is the result of Bzdak and
Bozek~\cite{Bzdak2013} who tried to incorporate the effect of the
LCC in the hydrodynamic calculations.
Their calculations show that the LCC can indeed be responsible for at
least a large part of the observed effect. Unfortunately it is
difficult to trace from the final result the real mechanism -- how and
why the LCC actually contributes to this observable.  

The mechanism of such LCC effect was elaborated only  recently  with the help of differential three
particle correlator in Ref.~\cite{Voloshin:2014gja}, where the effect
was studied in detail using the Blast Wave model.  It was found that there exists   an
interesting interplay of two effects: a stronger correlation of
balancing charges in-plane compared with out-of-plane, and the
statistical ``dilution'' of the correlation due to uncorrelated
background. 

If there is a considerable LCC effect, then it should also be seen in higher harmonic correlators,
for which the effects of the CME and the CMW should be
minimal~\cite{Voloshin:2011mx,Bzdak2013}.   
Fig.~\ref{fig:threepart_DAQ32} demonstrates that the correlations for
the third harmonic bear the similar ordering as those for the second
harmonic, but the correlation strength is about 10 times smaller.  The third harmonic differential correlator in Fig.~\ref{fig:3pcd32} 
shows a signal about 3 times smaller than the
second harmonic one. 
%Both these results in principle are consistent with expectation in the strength of the signals due to the difference in $v_3$ and $v_2$.  
The corresponding slope parameters for the  third harmonic flow splitting 
were reported by STAR to be consistent with zero~\cite{Qi-ye}, which
would thus suggest the smallness of the LCC effect.

%Note that the correlator becomes negative at $\Delta\eta
%\gtrsim 0.6$ -- 0.7.  It then approaches zero owing to too few balancing
%particles at large rapidity separation.

%\begin{figure}
%\begin{center}
%\includegraphics*[width=0.4\textwidth]{correlator.pdf}
%\hspace{0.1\textwidth}
%%\includegraphics*[width=0.4\textwidth]{mpt.pdf}
%\vspace*{-1mm}
%\caption{
% The correlator $\corr$ $(n=2)$ as a function of $\Delta\eta$ from Blast Wave model. 
%Right: Mean $\pt$ of the balancing charge particle.}
%\label{fig:corr}
%\end{center}
%\end{figure}

Another interesting point made in Ref.~\cite{Bzdak2013} is that the LCC at freeze-out,
when convoluted with the characteristic shape of $v_2(\eta)$ and
$v_2(p_T)$, may provide a qualitative explanation for the finite $v_2$
slope observed from data.  A realistic estimate of the contribution of
this mechanism, however, turns out to be smaller than the measurement by an order
of magnitude~\cite{STAR_CMW}. 

In a very recent hydrodynamic study~\cite{Hatta:2015hca}, it was suggested that simple viscous transport of charges combined with certain specific initial conditions might lead to a sizable contribution to the observed $v_2$ splitting of charged pions. In order for the results of pion splitting in line with data, the authors had to assume a crucial relation between isospin chemical potential and the electric charge asymmetry, which seems rather shaky, to say the very least. Furthermore certain predictions of this model (e.g. splitting for kaons) appear not in line with current experimental information.  Clearly whether such idea works or not, would need to be thoroughly vetted by realistic viscous hydrodynamic simulations. But all that said, this study poses a very important question that has to be answered: to make a firm case for the observation of anomalous charge transport via CMW,  the normal (viscous hydrodynamical) transport of charges should be quantitatively understood.

%---------------------------------------------------
\subsection{Chiral Vortical Effect}

The Chiral Vortical Effect (CVE) is in some sense
  similar to the CME: its experimental manifestation is the
baryon-number separation, instead of the electric-charge separation,
perpendicular to the reaction plane.  As a result, the three-point
correlator, $\gamma$, needs to be studied between two (anti)baryons.
However, if both particles are (anti)protons that carry also electric
charges, there will be an ambiguity in the possible signal arising
from the CME.  The study of the $\gamma$ correlator with an
electrically neutral baryon, such as $\Lambda$, will provide more
conclusive evidence of the baryon-number separation effect.

\begin{figure}[h]
\begin{minipage}[c]{0.48\textwidth}
%\center
\includegraphics[width=\textwidth]{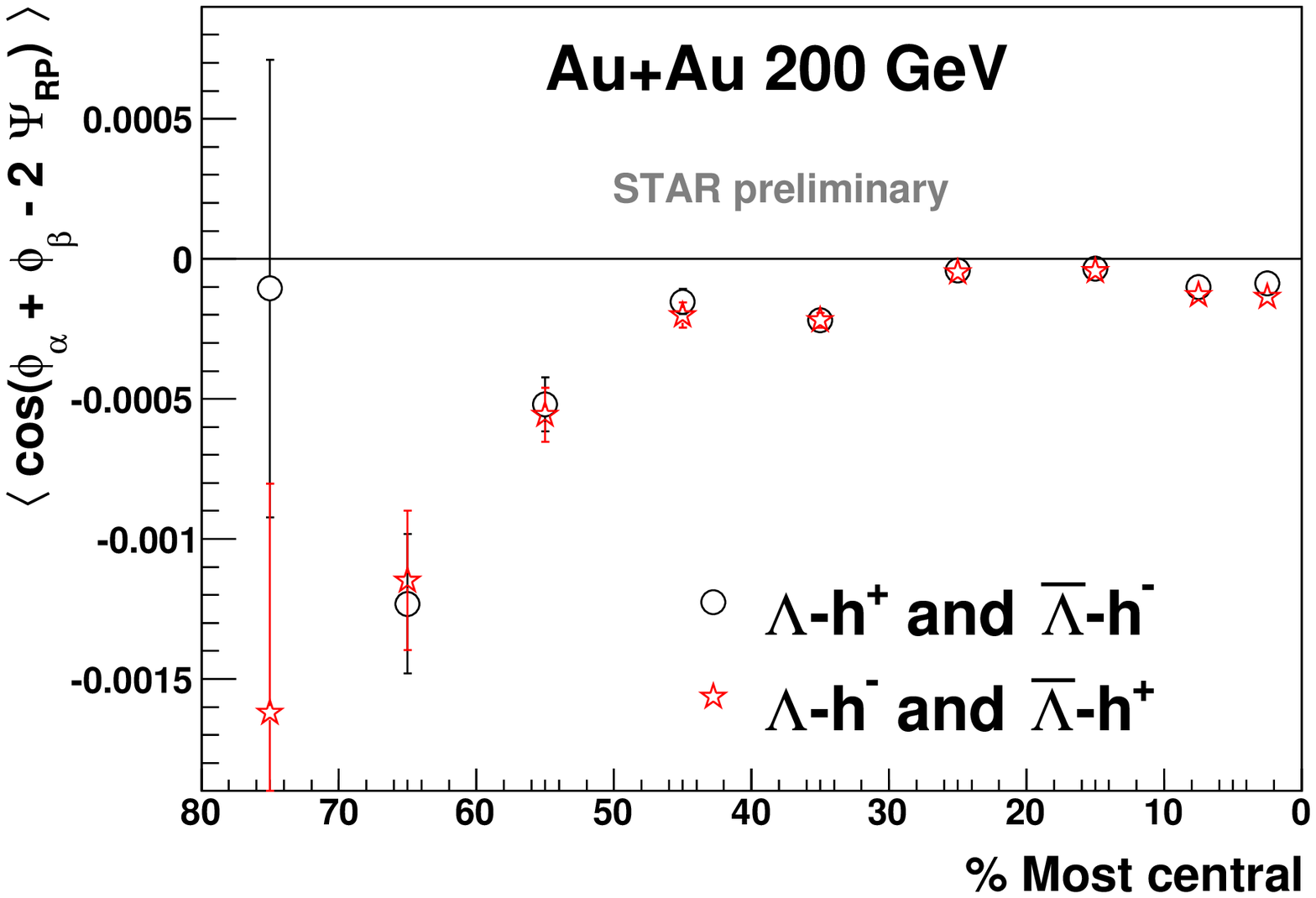}
  \caption{$\gamma$ correlation of $\Lambda$-$h^{+}$ ($\bar{\Lambda}$-$h^{-}$) and $\Lambda$-$h^{-}$ ($\bar{\Lambda}$-$h^{+}$)
 as a function of centrality in Au+Au collisions at 200 GeV~\cite{Lambda_CVE}.
    }
\label{fig:Lambda_h}
\end{minipage}
\hspace{0.4cm}
\begin{minipage}[c]{0.48\textwidth}
%\center
\includegraphics[width=\textwidth]{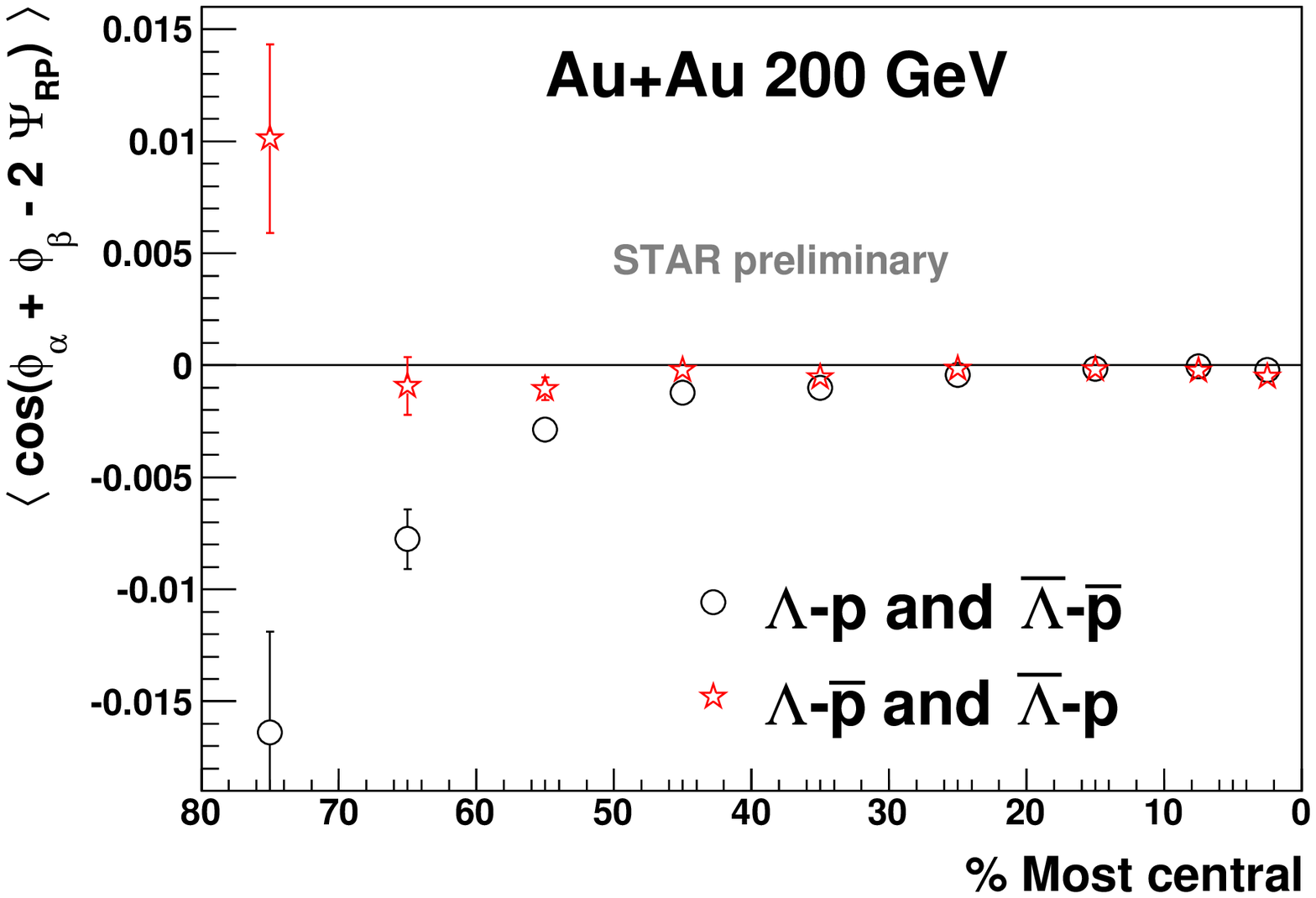}
  \caption{$\gamma$ correlation of $\Lambda$-$p$ ($\bar{\Lambda}$-$\bar{p}$) and $\Lambda$-$\bar{p}$ ($\bar{\Lambda}$-$p$)
 as a function of centrality in Au+Au collisions at 200 GeV~\cite{Lambda_CVE}.
        }
\label{fig:Lambda_P}
\end{minipage}
\vspace{-0.2cm}
\end{figure}

Although (anti)$\Lambda$s are electrically neutral, it is still a
question whether the strange quarks behave the same way as the up/down
quarks in the chiral dynamics during the collision.  If the answer is
no, then (anti)$\Lambda$s may still act like electrically charged
particles in the $\gamma$ correlation.  Fig.~\ref{fig:Lambda_h}
shows the $\gamma$ correlation of $\Lambda$-$h^{+}$
($\bar{\Lambda}$-$h^{-}$) and $\Lambda$-$h^{-}$
($\bar{\Lambda}$-$h^{+}$) as a function of centrality in Au+Au
collisions at 200 GeV~\cite{Lambda_CVE}.  Note that (anti)protons have
been excluded from the charged hadrons in the correlation to avoid any
possible CVE contribution.  Tentatively assuming $\Lambda$s
($\bar{\Lambda}$s) are positively(negatively)-charged, we find that
the ``same-charge" and ``opposite-charge" correlations are consistent
with each other, which means no charge-dependent effect.  The message
is twofold. First, from the $K_S^0-h$ correlations~\cite{Lambda_CVE}
we learn that the different behaviors of same-charge and
opposite-charge particle correlation as shown in Fig.~\ref{fig:gamma}
are really due to the electric charge, and therefore the null
charge-separation effect in $\Lambda$-$h$ indicates that
(anti)$\Lambda$s manifest no electric charges in the $\gamma$
correlation.  So the strange quarks inside the $\Lambda$ hadron seem to behave the same way
as the up/down quarks in the chiral dynamics.  Second, the
$\Lambda$-$h$ correlation provides a baseline check for the
$\Lambda$-$p$ correlation, and any possible signal in the latter
should not come from the CME contribution.

Fig.~\ref{fig:Lambda_P} shows $\gamma$ correlation of $\Lambda$-$p$
($\bar{\Lambda}$-$\bar{p}$) and $\Lambda$-$\bar{p}$
($\bar{\Lambda}$-$p$) as a function of centrality in Au+Au collisions
at 200 GeV~\cite{Lambda_CVE}.  The same-baryon-number correlation has
a different behavior from the opposite-baryon-number correlation from
mid-central to peripheral collisions. This baryon-number separation
with respect to the event plane is consistent with the expectation
from the CVE. More investigations into the
background contribution are needed. For example, in analog with the
LCC, there could be the Local Baryon-number
Conservation that plays a similar role as the LCC when coupled with the
collective flow.  The magnitudes of the $\Lambda$-$p$ correlations are
much larger than those of the $h$-$h$ correlations.  This is partially
because the $\langle p_T \rangle$ of baryons is higher than that of
mesons, and the correlation strength increases with the average $p_T$
of the two particles in the correlation.  A future differential
measurement vs the average $p_T$ and further correlations between
identified particles may provide a better comparison of the
correlation strength between the CME- and CVE-related correlations.

%==============================================================
\subsection{Future experimental studies}

Experimental observation of the charge dependent correlations
consistent with the theoretical expectations for several chiral
anomalous effects, if confirmed, can be a beginning of an exciting
program --
% Of
%course, one would first need to confirm that the observed correlations
%are indeed related to these effects.  
%If confirmed, 
%It will open the
%door for 
{\em direct} experimental study of the effects from
non-perturbative sector of QCD.
Below we briefly discuss several future experimental measurements 
aimed at more detailed study of the observed signals as well as 
understanding the background effects.

%---------------------------------------------------------
\begin{itemize}

%------------------------------------------------------------
\item {\sl Beam Energy Scan II.} 
%------------------------------------------------------------
The results from the BES I indicate that the signal likely
disappears at lower energies. Unfortunately the statistical
uncertainties are still about a factor of three too high to make a definite
conclusion. The upcoming BES II should resolve this question. 
%---------------------------------------------------------
\item {\sl Collisions of isobaric nuclei.}
%-----------------------------------------------------------
The signal dependence on the strength of the magnetic field
can be verified with collisions of isobaric nuclei (the same mass but
different charges), such as
$^{96}_{44}Ru$ and $^{96}_{40}Zr$. In such collisions the background
effects (that depend mostly on the elliptic flow) should not change,
while the magnetic field, proportional to
the electric charge of the nuclei, would vary by about $10\%$, resulting in
a $\sim 20\%$ change in the CME signal.  
%Such variations should be readily
%measurable.  The collisions of $^{96}_{44}Ru$ and $^{96}_{40}Zr$
%isotopes have been successfully used at
%GSI~\cite{Hong:2001tm,Hong:2003jk} in a study of baryon stopping.
Collisions of isobaric nuclei at RHIC 
would also help in understanding baryon stopping, initial velocity fields, and
the origin of directed flow.
%---------------------------------------------------------
\item {\sl Higher harmonic correlators.}
%-----------------------------------------------------------
The fluctuations in the initial conditions result in nonzero higher
harmonic flow ($n > 2$).  The background correlations, if measured 
relative to the higher harmonic 
event planes, while smaller in magnitude
(according to higher harmonic flow), should be finite. However the correlations
caused by the magnetic field should be highly suppressed, if not equal to zero.  
Several such measurements have
been discussed above, but they will 
be significantly improved with higher statistics available in the next
few years.
%---------------------------------------------------------
\item {\sl Event shape engineering (ESE).}~\cite{Schukraft:2012ah}
%-----------------------------------------------------------
Large fluctuations of anisotropic flow open another possibility to disentangle
effects associated with the magnetic field (or orbital angular momentum) from the 
background correlations. With the event shape
engineering one is able to select events corresponding to the large
or small flow while keeping the magnetic field the same. These
measurements also require higher statistics.
%-----------------------------------------------------
\item {\sl U+U collisions.} 
%---------------------------------------------------------
While the selection of body-body (large elliptic flow) and tip-tip 
(small elliptic flow) collision orientations solely based
on the measured multiplicity appears to be much more difficult than
expected~\cite{Adamczyk:2015obl}, these collisions, owing to increased flow
fluctuations, would be the best place for the application of   ESE method.

\item {\sl  Charge-dependent $\hat{Q}$-vector analysis.}  Both the CME and CMW observables pertain to  nontrivial charge distributions  in azimuthal angles (e.g. the electric charged dipole and quadrupole), induced by anomalous transport effects. Some of the identified background effects (e.g. the local charge conservation, normal viscous charge transport) could also lead to nontrivial charge azimuthal correlations when coupled with various harmonic flows. It would therefore be very useful to develop possible new measurements that could extract full information for the azimuthal charge distributions. One class of observables that can help decipher such information, are the charged multipole vectors $\hat{Q}^c_n$, defined for  the measured charged hadrons in an event as $Q^c_n\, e^{i \Psi^c_n} = \sum_i \, q_i \,  e^{i \phi_i}$ 
 where the summation runs over charged hadrons   with $q_i$ and $\phi_i$ the electric charge and azimuthal angle of the $i$-th particle~\cite{Voloshin:2008dg,Liao:2010nv}. In particular the $\hat{Q}^c_1$ and $\hat{Q}^c_2$ are directly relevant to the CME and CMW signals.  One may also think of sub-event version of this analysis or possible  improved version with multi-particle correlation. These observables are different and independent   from the usual $\hat{Q}_n$ vectors  related to the collective flow measurements. The $\hat{Q}_n$ is charge blind and includes all charges similarly while the $\hat{Q}_n^c$ takes the difference between positive and negative charges therefore yielding information on the charge distribution. With a joint $\hat{Q}_n$ and $\hat{Q}_n^c$ analysis one can study the strength and azimuthal correlations among all harmonics and charged multipoles. A systematic charged multipole vector analysis would provide extremely valuable information about the ``charge landscape'' in heavy ion collisions. 
 
%---------------------------------------------------------
\item {\sl Cross-correlation of different observables.}
%-----------------------------------------------------------
Cross-correlation and cross-comparison of different observables might
be very valuable for understanding the nature of the correlations. For
example, the baryonic charge separation due to the CVE and electric charge
separation due to the CME should be strongly correlated owing to correlations
in the direction of the magnetic field and the system orbital angular
momentum. This implies a strong correlation in baryonic -- electric charge
correlations (e.g. $\Lambda-\pi^\pm$) that can be measured
experimentally. Another example of cross comparison of different
observables can be the LCC contribution to the CME and the CMW studies. This
would require a detailed modeling of the LCC, with both
measurements to be explained by the same set of parameters. 
%------------------------------------------------------------
\item {\sl Correlations with identified particles}~\cite{Voloshin:2009hr}.  
%---------------------------------------------------
At least two directions can be pursued here, both in heavy ion and elementary $pp$ collisions. 
The various anomalous chiral effects discussed previously occur at quark level, and when combined into hadronic observables, will show specific patterns for different identified hadrons according to their electric, baryonic, and strangeness quantum numbers. Detailed measurements of correlation observables with identified particles will provide crucial verifications of data interpretation in terms of anomalous chiral effects. The neutral particles should not be affected by the magnetic field, and 
this can be used to separate the background effects.  The 
measurements of the ``quark content'' of the correlated cluster 
will also be extremely interesting. The topological cluster
decays in equal number of $q\bar{q}$-pairs of all flavors. 
The so-called 't Hooft interaction (e.g. used to
explain the difference in $\bar{d}/\bar{u}$ ratio in the nucleon ``sea'')
predicts that the topological cluster couples to equal number of
$q\bar{q}$ pairs of all flavors, which
would lead, e.g., to weaker correlations
between hadrons with $u-u$ combination than that with the $u-d$ one.
 
%------------------------------------------------------------
\item {\sl Studies of the electromagnetic fields.}
%------------------------------------------------------------
As mentioned earlier, there exist substantial uncertainties in the time evolution of the
magnetic field.  It depends strongly on the
electric conductivity of the
system~\cite{Tuchin:2013ie,Gursoy:2014aka}.  Experimental measurements
sensitive to the strength and time dependence of the strong
electromagnetic fields would be extremely valuable for a better
theoretical modeling of the CME and CMW effects, as well as for
understanding the system evolution and hadronization in general.
There have been several proposals for such
measurements~\cite{Gursoy:2014aka,Voronyuk:2014rna,Hirono:2012rt}, and
we expect the first experimental results on the charge dependent
directed flow in Cu+Au collisions to be available soon. 
Note that the electric conductivity of the plasma should strongly
depend on the quark production time. From this point of view such
measurements could serve as an important test of the ``two waves''
scenario proposed by  S.~Pratt~\cite{Pratt:2012dz,Pratt:2013xca}. 
Quark propagation in a strong magnetic field can also lead to the charge
dependence of particle elliptic flow: $v_{2,p}>v_{2,n} \ge
v_{2.\Lambda}$, $v_{2,\pi^+}=v_{2,\pi^-} = v_{2,\pi^0}$,
$v_{2,K^+}=v_{2,K^-} > v_{2,K^0}$. The recent ALICE
measurements of the kaon flow~\cite{Abelev:2014pua} appear to be consistent
with this ordering.  

\end{itemize}

%==========================================================

\section{Summary and Outlook}
\label{sec:5}

The interplay of quantum anomalies with magnetic field and vorticity induces a variety of novel non-dissipative transport phenomena in systems possessing chiral fermions. In heavy ion collisions, these phenomena provide a unique possibility to probe the topological properties of the quark-gluon plasma by measuring the charge dependence of the azimuthal distributions of the produced hadrons. This is possible because relativistic heavy ion collisions produce hot QCD matter characterized by strong fluctuations of topological charge as well as approximately chiral fermions (namely the light quarks). In addition, the colliding ions generate strong magnetic fields $eB \sim {\cal O} (10\ m_\pi^2)$. Theoretical estimates discussed in this review indicate that the chiral magnetic and chiral vortical effects lead to the event-by-event charge separation that can be observed in heavy ion collisions. Moreover, the chiral magnetic wave induced by the vector chemical potential leads to the characteristic charge dependence of the elliptic flow of hadrons controlled by the electric charge asymmetry in a given event. 

The experimental data from STAR Collaboration at Relativistic Heavy Ion Collider at BNL and 
ALICE Collaboration at Large Hadron Collider at CERN discussed in section 4 provide an evidence for the predicted effects, with magnitude consistent with theoretical estimates. 
There exist known conventional backgrounds to all of these experimental observables. However at present there is no compelling alternative explanation that can describe all of the data without invoking the anomalous chiral effects. 

Nevertheless, a lot remains to be done both in experiment and theory to substantiate the existing evidence for the anomalous chiral effects in heavy ion collisions, and we outlined the future program of these studies in section 4. The physics of anomalous transport is at the heart of QCD as  a non-Abelian gauge theory. It is crucially important to establish and quantify the anomalous chiral effects in heavy ion collisions, and we hope that this goal will be accomplished in the near future.   

\section*{Acknowledgements}
The research of DK was supported in part by the U.S. Department of
Energy under Contracts DE-FG-88ER40388 and DE-SC0012704. JL was partly supported by  the U.S. National Science Foundation (Grant No. PHY-1352368) and by the RIKEN BNL Research Center. The research of SV was supported by the U.S. Department of Energy under Award Number DE-FG02-92ER-40713.  The research of GW was supported by the  U.S. Department of Energy under Award Number DE-FG02-88ER40424.

\end{document}